\pdfoutput=1
\documentclass[a4paper,11pt]{article}
\usepackage{jheppub}
\usepackage{graphicx}
\usepackage{bm}
\usepackage{booktabs}
\usepackage{array}
\usepackage{multirow}
\usepackage{xcolor}
\usepackage{amsmath,amssymb}
\usepackage{tikz}
\newsavebox{\fitwidthbox}
\newenvironment{fitwidth}{\begin{lrbox}{\fitwidthbox}}{\end{lrbox}%
  \ifdim\wd\fitwidthbox>\linewidth\resizebox{\linewidth}{!}{\usebox{\fitwidthbox}}\else\usebox{\fitwidthbox}\fi}
\usetikzlibrary{decorations.pathmorphing,decorations.markings}
\tikzset{
  fermion/.style={postaction={decorate},decoration={markings,mark=at position 0.55 with {\arrow[scale=1.2]{latex}}}},
  gluon/.style={decorate,decoration={coil,amplitude=2.2pt,segment length=3.5pt,aspect=0.8,pre length=1pt,post length=1pt}},
  photon/.style={decorate,decoration={snake,amplitude=1.6pt,segment length=6pt,pre length=2pt,post length=2pt}},
  boson/.style={decorate,decoration={snake,amplitude=1.6pt,segment length=6pt,pre length=2pt,post length=2pt}},
  scalar/.style={dashed,thick},
  effvertex/.style={circle,draw,fill=gray!40,inner sep=0pt,minimum size=7pt},
}

\newcommand{\at}{a_t}
\newcommand{\bt}{b_t}
\newcommand{\dphijj}{\Delta\phi_{jj}}
\newcommand{\dphigrp}{\Delta\phi_{jj}^{\mathrm{grp}}}
\newcommand{\myy}{m_{\gamma\gamma}}
\newcommand{\pT}{p_{\mathrm{T}}}
\newcommand{\OOone}{\mathcal{O}_1}
\newcommand{\OOtwo}{\mathcal{O}_2}
\newcommand{\GeV}{\,\mathrm{GeV}}
\newcommand{\TeV}{\,\mathrm{TeV}}
\newcommand{\pb}{\,\mathrm{pb}}
\newcommand{\fb}{\,\mathrm{fb}}
\newcommand{\ifb}{\,\mathrm{fb}^{-1}}
\newcommand{\cHGt}{c_{H\widetilde{G}}}
\newcommand{\mgfive}{\textsc{MadGraph5\_aMC@NLO}}
\newcommand{\pythia}{\textsc{Pythia}\,8}
\newcommand{\delphes}{\textsc{Delphes}\,3}
\newcommand{\sherpa}{\textsc{Sherpa}}

\newcommand{\rhoI}{\rho_I}
\newcommand{\rhoO}{\rho_O}

\title{Beyond two jets: recovering the CP information of QCD radiation in gluon-fusion $H\to\gamma\gamma$
       with matrix-element and transformer-learned observables}

\author[1]{Pratik Kafle,}
\author[1]{Kirtimaan Mohan,}
\author[1]{Marcel Yanez,}
\author[1]{Nityaansh Parekh,}
\author[1]{Joey Huston,}
\author[1,2]{and Keping Xie}
\emailAdd{kaflepra@msu.edu}
\affiliation[1]{Department of Physics and Astronomy, Michigan State University,\\ East Lansing, Michigan 48824, USA}
\affiliation[2]{State Key Laboratory of Dark Matter Physics \& Key Laboratory for Particle Astrophysics and Cosmology (MOE) \& Shanghai Key Laboratory for Particle Physics and Cosmology,\\ Tsung-Dao Lee Institute \& School of Physics and Astronomy, Shanghai Jiao Tong University, Shanghai 201210, China\looseness=-1}

\abstract{Violation of charge-parity (CP) symmetry in the Higgs-top interaction leaves an imprint on the jets produced together
with the Higgs boson. Parity-odd observables built from the jet momenta isolate the interference between the CP-even ($\at$) and
CP-odd ($\bt$) Higgs-gluon couplings. Existing experimental analyses build such observables from a two-jet system, typically the two
jets with the highest transverse momentum. About a third of the selected signal events, however, contain a third reconstructed jet.
This raises the question of how much CP information is lost when multijet events are reduced to a two-jet description, which we
answer with the Fisher information on $\bt$ in $H\to\gamma\gamma$ plus jets production at $\sqrt{s}=13.6\TeV$. In events with a third
hard parton, the two-jet optimal observables evaluated with the two leading jets retain only 41\% of the information available in the hard process, and 27\%
when evaluated on reconstructed jets. We recover part of this information with observables that also use the additional jets: an
analytic observable based on the matrix element for Higgs boson production with three jets, and learned observables from two
transformer networks, a Particle Transformer and the Lorentz-equivariant L-GATr. In a realistic analysis with backgrounds and
systematic uncertainties, the expected 68\% CL half-width on $\bt$ at $300\ifb$ is $0.263$ with the two leading jets, $0.241$ with all
jets grouped into two, $0.227$ with the analytic observable and $0.210$ with the learned observables. An analysis based on the two
leading jets would therefore need about 56\% more integrated luminosity to match the learned observables. Pileup widens the intervals of the optimal and learned observables
by 33--36\% but leaves their ordering unchanged.}


\begin{document}
\maketitle
\flushbottom

\section{Introduction}\label{sec:intro}

Since the discovery of the Higgs boson~\cite{ATLAS:2012yve,CMS:2012qbp}, its couplings to weak gauge bosons and third-generation fermions have been measured to agree with Standard Model (SM) predictions at the level of approximately 5--10\% and 7--20\%, respectively~\cite{ATLAS:2022vkf,CMS:2022dwd}. 
These measurements probe the strength of the couplings, but their CP structure is far less constrained. Determining whether the Higgs interactions conserve charge-parity (CP) symmetry is therefore a central objective of the high-luminosity LHC (HL-LHC)~\cite{Cepeda:2019klc} and of future collider programs~\cite{deBlas:2019rxi,CERN-ESU-2025-002,P5:2023wyd}. 

The observed matter--antimatter asymmetry of the Universe requires new sources of CP violation beyond those present in the Standard Model~\cite{Planck:2018vyg,Sakharov:1967dj,Gavela:1993ts,Huet:1994jb,Morrissey:2012db}. 
Many well-motivated extensions of the SM introduce additional CP violation in the Higgs sector, either through an extended scalar sector or through higher-dimensional operators~\cite{Lee:1973iz,Branco:2011iw,th:warsaw}. 
Such effects can enter the Higgs Yukawa couplings already at tree level, so they need not be small~\cite{Weinberg:1976hu,Weinberg:1990me}. The CP structure of the Higgs couplings to fermions, in particular to the top quark, has thus become an important target of current and future collider experiments~\cite{Cepeda:2019klc,FCC:2025lpp}. 

In this work we focus on the coupling of the Higgs boson to the top quark. Allowing for a CP-violating component, we parametrize the interaction as 
\begin{equation} 
	\mathcal{L}_Y = -\frac{m_t}{v}\, H\, \bar t \left(\at+i\bt\gamma_5\right) t, 
	\label{eq:yukawa} 
\end{equation} 
where $\at$ and $\bt$ denote the CP-even and CP-odd components, respectively, and $(\at,\bt)=(1,0)$ in the SM.

This interaction modifies Higgs boson production in gluon fusion through the top-quark loop~\cite{th:ggmn} and affects associated production with top quarks, $t\bar tH$ and $tH$, already at tree level~\cite{Ellis:2013yxa}. Measurements of Higgs boson production rates therefore constrain $\bt$~\cite{Bahl:2020wee,Bahl:2022yrs}.
In the heavy-top limit, the gluon-fusion loop can be represented by the effective interaction~\cite{th:kniehl_spira,th:spira_hgg}
\begin{equation}
  \mathcal{L}_{\mathrm{eff}} = \frac{\alpha_s}{12\pi v}\,H\left[\at\,G^a_{\mu\nu}G^{a\,\mu\nu}
   + \frac{3}{2}\,\bt\,G^a_{\mu\nu}\widetilde{G}^{a\,\mu\nu}\right],
  \label{eq:leff}
\end{equation}
where $G^a_{\mu\nu}$ is the gluon field strength and $\widetilde{G}^{a\,\mu\nu}=\tfrac12\epsilon^{\mu\nu\rho\sigma}G^a_{\rho\sigma}$ its dual, with $\epsilon^{0123}=-1$ and $\gamma_5=i\gamma^0\gamma^1\gamma^2\gamma^3$ as in \mgfive{} and ref.~\cite{peskin}. 
Any heavy colored particle coupled to the Higgs boson contributes to the same effective operators, so gluon-fusion observables probe CP-odd Higgs-gluon interactions whatever their ultraviolet origin. 
Throughout this work we assume that the effective interaction of eq.~\eqref{eq:leff} arises entirely from the modified Higgs-top coupling of eq.~\eqref{eq:yukawa}, and we express all results in terms of $\bt$. Distinguishing different ultraviolet realizations of the effective operators is left for future work.

The observables studied in this paper complement other probes of the CP structure of the Higgs-top interaction. A nonzero $\bt$ induces CP-odd $HVV$ interactions only through top-quark loops, so that measurements in vector-boson fusion (VBF), $VH$ production and $H\to ZZ^*\to4\ell$ reach only $|\bt|$ of order 10 to 100~\cite{th:cms_4l_anom,th:cms_tautau_anom,ATLAS:2022vbfyycp,ATLAS:2023vbf4lcp}. 
Electric dipole moments (EDMs) of the electron~\cite{th:acme2018,th:jila_eedm} and neutron~\cite{th:nedm_2020,th:weinberg_op} give much stronger but indirect constraints. These depend on assumptions about other Higgs couplings, carry hadronic uncertainties and can be weakened by cancellations among different contributions~\cite{th:brod_haisch_zupan,th:chien_edm,th:cirigliano_tabletop}. 
A global fit to LHC data with free top and electron Yukawa couplings still permits $|\bt|\simeq0.35$ at 90\% CL, and such values remain compatible with the electron EDM if the electron Yukawa coupling is modified as well, at the cost of significant fine tuning~\cite{Bahl:2022yrs}. 
CP-odd spin correlations in top-quark pair production probe the same coupling at one loop~\cite{Erdogan:2026cbe}. We compare our expected constraints with these and other existing constraints in appendix~\ref{sec:interp}.

Production rates, however, are CP-even and depend on $\bt$ only quadratically. They constrain the magnitude of a CP-odd coupling but cannot establish CP violation or determine the sign of $\bt$. For that, one needs observables that are odd under CP and isolate the interference between the CP-even and CP-odd amplitudes. Such observables project out the terms linear in $\bt$ and are therefore sensitive to both its magnitude and its sign.

CP-odd observables have been studied extensively in $t\bar tH$ and $tH$ production~\cite{Gunion:1996xu,Demartin:2014fia,Boudjema:2015nda} and are used by both ATLAS and CMS~\cite{th:atlas_tth_cp,th:cms_tth_cp,ATLAS:2023cbt,CMS:2022dbt}. Here we focus on gluon-fusion Higgs production. In inclusive gluon fusion the interference of the CP-even and CP-odd amplitudes integrates to zero; it becomes observable only when the Higgs boson is produced together with jets.

With one jet, the interference can be accessed through the linear polarization of the gluon jet using jet-substructure observables~\cite{Yu:2022kcj}. With two jets, it appears in the correlations among the jet momenta, in particular in the azimuthal angle between the jets, $\dphijj$. The sensitivity of $\dphijj$ to the CP structure of Higgs interactions was first pointed out for VBF~\cite{th:prz}, studied in detail in refs.~\cite{th:hankele,Hagiwara:2009wt}, and then extended to the Higgs-gluon interaction induced by heavy-quark loops~\cite{th:klamke_zeppenfeld}. While early studies suggested that additional radiation could substantially degrade this sensitivity~\cite{Odagiri:2002nd}, later work showed that it largely survives parton showering and hadronization~\cite{th:delduca_hjj_2006}, as well as next-to-leading-order (NLO) QCD corrections~\cite{Campbell:2006xx}.

This program has since been extended in several directions. To reduce the effect of additional radiation, ref.~\cite{th:andersen_arnold_zeppenfeld} proposed to combine all jets on either side of the Higgs boson into two effective jets, and ref.~\cite{Englert:2012ct} studied hadronic event shapes of the full jet system. The full top- and bottom-quark mass dependence of the one-loop amplitudes was included in ref.~\cite{Campanario:2010mi}, and the sensitivity in the $H\to\tau\tau$ channel was studied in ref.~\cite{th:dolan}. Closest to this work, refs.~\cite{Campanario:2014oua,Bhattacharya:2026ucv} studied the CP properties of Higgs boson production with three jets in gluon fusion, and ref.~\cite{Campanario:2014oua} examined, with the full quark-mass dependence of the loops, whether the azimuthal correlation of a CP-mixed Higgs boson survives the emission of a third jet. These studies characterize distributions of jet angles but do not quantify the information they carry.

ATLAS has measured the signed $\dphijj$ distribution in $H\to\gamma\gamma$ events and interpreted it in terms of a CP-odd Higgs-gluon interaction~\cite{th:atlas_yy_139_diff}, following an earlier interpretation of the signed angle~\cite{Bernlochner:2018opw}, and has used the same angle to constrain the CP structure of the effective Higgs-gluon coupling in $H\to WW^*$ events with two jets~\cite{th:atlas_ww_jj36}. CMS has used matrix-element discriminants derived from $H+2$ parton amplitudes in the $H\to\tau\tau$, $H\to\gamma\gamma$ and $H\to4\ell$ channels~\cite{th:cms_tautau_anom,th:cms_4l_anom,th:cms_yy_anom}, and the matrix-element method for Higgs couplings in this process was studied in ref.~\cite{Andersen:2012kn}. These analyses constrain the CP structure of the effective Higgs-gluon coupling, which in the SM is induced by the top quark.

All existing measurements build their CP-sensitive observables from a two-jet system. This is the natural choice at leading order (LO), where the interference can be expressed entirely in terms of the two partons that accompany the Higgs boson. Higgs events at the LHC, however, often contain additional hard radiation. Once a third hard parton is present, the interference is no longer described by a single CP-odd structure but by several pseudoscalar invariants, and it is not known how much information is lost when such events are projected onto a two-jet observable. This paper asks how much CP information the additional jets carry in gluon-fusion Higgs production, and how much of it can be recovered experimentally. To answer it, we quantify with the exact per-event weights how much of the information on $\bt$ a two-jet observable loses, construct observables that use the third jet, and evaluate them in a reconstruction-level analysis with backgrounds, systematic uncertainties and pileup.

Our reference is the conventional two-jet optimal observable~\cite{th:atwood_soni,th:davier,th:diehl_nachtmann}, computed from the $gg\to Hgg$ matrix element with the two highest-$\pT$ jets, as in the ATLAS $\dphijj$ measurement and, as far as stated, in the CMS analyses. 
We investigate three ways of using the additional jets. 
First, we evaluate the two-jet optimal observable on jet groups constructed on either side of the Higgs boson, following ref.~\cite{th:andersen_arnold_zeppenfeld}. 
Second, we derive an analytic three-jet optimal observable in the heavy-top limit for the process $gg\to Hggg$. 
Third, we learn optimal observables directly from reconstructed events with transformer networks, 
the Particle Transformer (ParT)~\cite{Qu:2022mxj} and the Lorentz-equivariant geometric algebra transformer (L-GATr)~\cite{Spinner:2024hjm,Brehmer:2024yqw}. Both are trained on matrix-element weights, and their outputs are constructed to be exactly even or odd under parity.

Machine learning has been applied to the CP structure of the Higgs-gluon coupling before, with boosted decision trees (BDTs) on the two leading jets but without the continuum background or systematic uncertainties~\cite{th:bahl_hjj_ml}. In VBF and in top-associated production, CP-sensitive observables have been learned at detector level with symbolic regression~\cite{Bahl:2025jtk}, and L-GATr with a pseudoscalar channel has been used to learn and interpret optimal observables in VBF and top-quark pair production~\cite{Bahl:2026kig}. In those processes the CP information sits in the hard-process objects. In gluon fusion it sits in the jets themselves, and two thirds of it in events with a third hard parton. Our regression of likelihood ratios follows refs.~\cite{Brehmer:2018kdj,Brehmer:2018eca,Brehmer:2019xox,Cranmer:2019eaq}.

We work in the $H\to\gamma\gamma$ channel. All observables are evaluated with the same event selection, signal-versus-background (S/B) categories and statistical procedure, so that differences in sensitivity come only from the observable and its binning. We first compare the observables on the hard partons, where the matrix-element weights of each event define the exact optimal observables and hence the maximum information on $\bt$ that any observable can reach in our simulation. We then follow the loss of information through the parton shower, hadronization and detector simulation to the reconstructed events.

The main steps and results are the following.
\begin{itemize}
\item We quantify the information on $\bt$ with the Fisher information and compare every observable with the exact optimal
observables given by the matrix-element weights of each event, tracking the information from the hard partons through the parton
shower and hadronization to the reconstructed events.
At the level of the hard partons, the two-jet optimal observables evaluated with the two leading-$\pT$ jets retain only 41\% of the
information in events with a third parton, in the binning of the information ladder of section~\ref{sec:truth}.
\item We construct an analytic three-jet optimal observable from the $gg\to Hggg$ matrix element, and learned observables with two
transformer networks, ParT and L-GATr, trained by weighted regression on the matrix-element weights, with outputs that are exactly
odd or even under parity.
At the level of the hard partons and in the same binning they retain 82\% and 90\% of the information of three-parton events, respectively, the latter equal to that of the
exact ratios in the same binning.
\item We perform a realistic $H\to\gamma\gamma$ analysis, whose workflow is summarized in figure~\ref{fig:workflow}.
The signal is simulated with merged matrix elements for up to three partons and reweighted to NLO in QCD, and the analysis
includes the continuum background, S/B categories that cannot select on the sign of $\bt$, and a binned likelihood fit with
systematic uncertainties.
The expected 68\% CL half-width on $\bt$ at $300\ifb$ decreases from $0.263$ for the two leading-$\pT$ jets to $0.210$ for the learned
observables, so that an analysis based on the two leading-$\pT$ jets would require about 56\% more integrated luminosity to reach the
same sensitivity.
\item We test how stable this gain is against the systematic uncertainties, including a conservative treatment of the higher-order
QCD corrections to the azimuthal correlations of the jets, and against pileup.
The gain survives pileup, which widens the intervals of the optimal and learned observables by 33--36\%.
Information is lost at reconstruction level due to the parton shower shifting the jet momenta, as well as additional radiation
that is not captured as a reconstructed jet since it falls below the $\pT$ threshold.
\end{itemize}

Section~\ref{sec:theory} describes the process, the simulation and the event selection, section~\ref{sec:multijet} the CP-sensitive observables and the Fisher information, and section~\ref{sec:mj-learned} the transformer networks. Sections~\ref{sec:categories} and~\ref{sec:stat} give the S/B categories and the fit, section~\ref{sec:truth} follows the information from hard partons to reconstructed jets, section~\ref{sec:results} gives the expected constraints at $300$ and $3000\ifb$, and section~\ref{sec:conclusions} concludes. Appendix~\ref{sec:interp} translates the results into other parametrizations and compares them with existing constraints, and the other appendices give technical details and cross-checks.

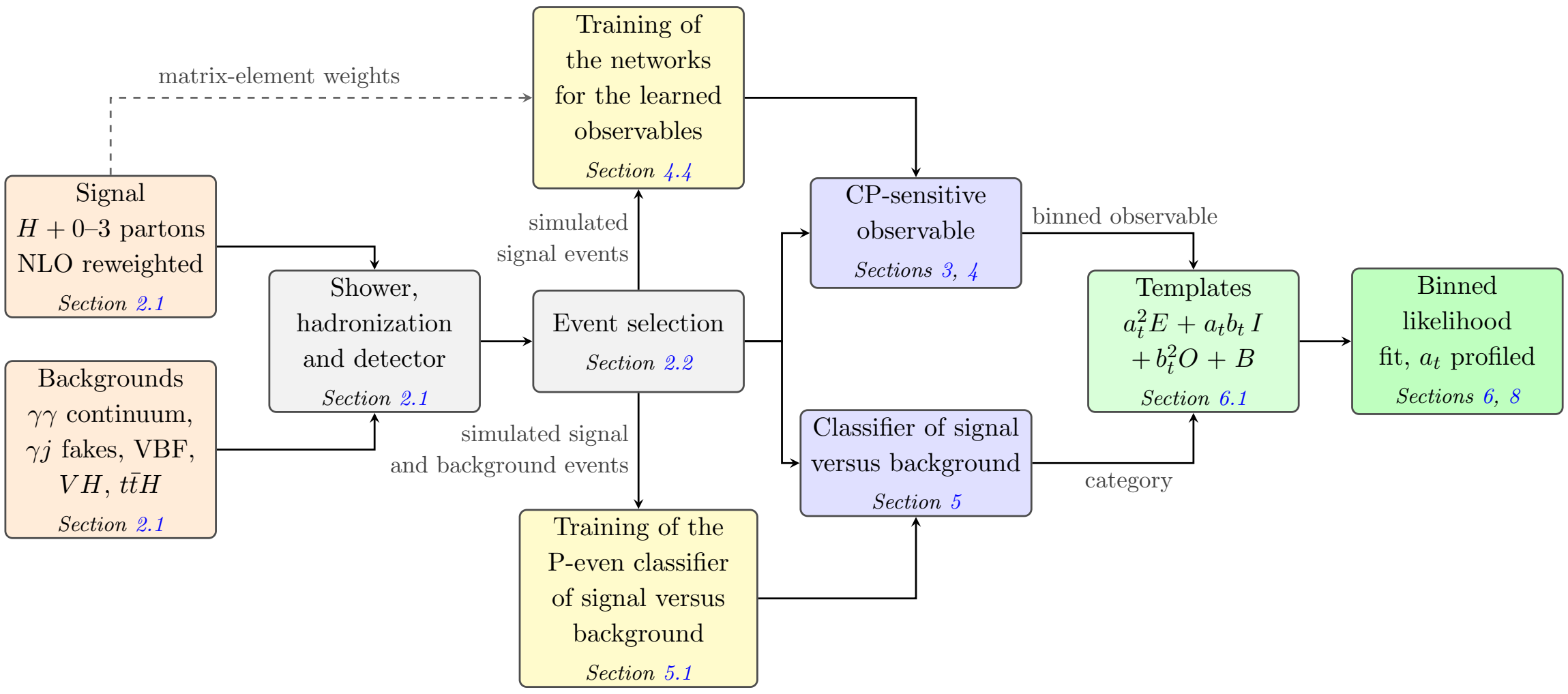
\begin{figure}[tbp]
  \centering
  \resizebox{\linewidth}{!}{%
\begin{tikzpicture}[
  box/.style={rectangle,rounded corners=3pt,draw=black!70,minimum height=15mm,text width=29mm,align=center,inner sep=3pt,
              execute at begin node={\hyphenpenalty=10000\exhyphenpenalty=10000}},
  lab/.style={font=\small,text=black!75},
  >=stealth,thick]
  \newcommand{\wfsec}[1]{\\[1pt]{\footnotesize\itshape #1}}
  \node[box,fill=orange!15] (sig) at (0,1.4) {Signal\\ $H+0$--$3$ partons\\ NLO reweighted\wfsec{Section~\ref{sec:sim}}};
  \node[box,fill=orange!15] (bkg) at (0,-1.6) {Backgrounds\\ $\gamma\gamma$ continuum, $\gamma j$ fakes, VBF, $VH$, $t\bar tH$\wfsec{Section~\ref{sec:sim}}};
  \node[box,fill=black!5] (sim) at (3.9,0) {Shower, hadronization and detector\wfsec{Section~\ref{sec:sim}}};
  \node[box,fill=black!5] (sel) at (7.8,0) {Event selection\wfsec{Section~\ref{sec:selection}}};
  \node[box,fill=yellow!25] (trn) at (7.8,3.6) {Training of the networks for the learned observables\wfsec{Section~\ref{sec:ml-training}}};
  \node[box,fill=yellow!25,text width=33mm] (trb) at (7.8,-3.8) {Training of the P-even classifier of signal versus background\wfsec{Section~\ref{sec:st-classifiers}}};
  \node[box,fill=blue!12] (cp) at (11.9,1.6) {CP-sensitive observable\wfsec{Sections~\ref{sec:multijet}, \ref{sec:mj-learned}}};
  \node[box,fill=blue!12,text width=32mm] (sb) at (11.9,-1.8) {Classifier of signal versus background\wfsec{Section~\ref{sec:categories}}};
  \node[box,fill=green!15] (tpl) at (16.0,0) {Templates\\ $\at^2E+\at\bt\,I$\\ $+\,\bt^2O+B$\wfsec{Section~\ref{sec:st-templates}}};
  \node[box,fill=green!25] (fit) at (19.9,0) {Binned likelihood fit, $\at$ profiled\wfsec{Sections~\ref{sec:stat}, \ref{sec:results}}};
  \draw[->] (sig.east) -| (sim.north);
  \draw[->] (bkg.east) -| (sim.south);
  \draw[->] (sim) -- (sel);
  \draw[->] (sel.east) -- ++(0.55,0) |- (cp.west);
  \draw[->] (sel.east) -- ++(0.55,0) |- (sb.west);
  \draw[->] (sel.north) -- node[lab,left,align=right]{simulated\\ signal events} (trn.south);
  \draw[->,dashed,black!60] (sig.north) |- node[lab,pos=0.7,above]{matrix-element weights} (trn.west);
  \draw[->] (trn.east) -| (cp.north);
  \draw[->] (sel.south) -- node[lab,left,align=right]{simulated signal\\ and background events} (trb.north);
  \draw[->] (trb.east) -| (sb.south);
  \draw[->] (cp.east) -| node[lab,pos=0.3,above]{binned observable} (tpl.north);
  \draw[->] (sb.east) -| node[lab,pos=0.3,below]{category} (tpl.south);
  \draw[->] (tpl) -- (fit);
\end{tikzpicture}}
  \caption{Schematic overview of the analysis. The signal, generated with merged matrix elements for up to three partons and
  reweighted to next-to-leading order (NLO) in QCD, and the backgrounds are passed through the parton shower, hadronization and
  detector simulation. After the event selection, an S/B classifier assigns each event to one of three categories of
  signal purity, and the CP-sensitive observable is binned in each category. The resulting templates, with the dependence on $\at$
  and $\bt$ of eq.~\eqref{eq:morph}, enter a binned likelihood fit. The networks that learn the optimal observables are trained on
  simulated signal events with the matrix-element weights of the hard partons as targets. The S/B classifier is
  trained on simulated signal and background events without the interference term, and its output is symmetrized under the
  azimuthal reflection, so that the categories cannot select on the sign of $\bt$. The section in which each step is described is
  given in each box.}
  \label{fig:workflow}
\end{figure}

\section{Signal process, simulation and event selection}\label{sec:theory}

We consider Higgs boson production in gluon fusion (ggF) in association with jets, followed by the decay $H\to\gamma\gamma$. In the heavy-top limit, the top-quark loop with the Yukawa coupling of eq.~\eqref{eq:yukawa} reduces to the effective Higgs-gluon vertex of eq.~\eqref{eq:leff}, whose strength is set by $(\at,\bt)$. Figure~\ref{fig:feyn-signal} shows representative LO diagrams for the production of a Higgs boson with two and with three jets. The process with three jets, $gg\to Hggg$, is the basis of the three-jet optimal observable of section~\ref{sec:mj-oo3}. This section describes the coupling dependence of the signal, the simulation of signal and backgrounds, and the event selection.

\begin{figure}[tbp]
  \centering
  \begin{tikzpicture}[scale=0.85, baseline=(base)]
  \coordinate (base) at (0,-0.1);
  \draw[gluon] (0,1) -- (1.1,0.55);
  \draw[gluon] (0,-1) -- (1.1,-0.55);
  \draw[fermion] (1.1,0.55) -- (2.0,0);
  \draw[fermion] (2.0,0) -- (1.1,-0.55);
  \draw[fermion] (1.1,-0.55) -- (1.1,0.55);
  \draw[scalar] (2.0,0) -- (3.0,0);
  \fill (2.0,0) circle (1.3pt);
  \node[below right] at (1.9,-0.1) {\scriptsize $\at+i\bt\gamma_5$};
  \node[left] at (1.1,0) {\scriptsize $t$};
  \node[left] at (0,1) {$g$}; \node[left] at (0,-1) {$g$}; \node[right] at (3.0,0) {$H$};
  \draw[->,>=latex] (3.75,0) -- (4.65,0) node[midway,above] {\scriptsize $m_t\to\infty$};
  \draw[gluon] (5.2,1) -- (6.1,0);
  \draw[gluon] (5.2,-1) -- (6.1,0);
  \draw[scalar] (6.1,0) -- (7.1,0);
  \node[effvertex] at (6.1,0) {};
  \node[above right] at (6.15,0.05) {\scriptsize $\at,\,\bt$};
  \node[left] at (5.2,1) {$g$}; \node[left] at (5.2,-1) {$g$}; \node[right] at (7.1,0) {$H$};
  \node at (3.6,-1.95) {(a)};
\end{tikzpicture}\hfill
\begin{tikzpicture}[scale=0.85, baseline=(base)]
  \coordinate (base) at (0,-0.1);
  \draw[gluon] (0,1.1) -- (1.1,1.1);
  \draw[gluon] (1.1,1.1) -- (2.4,1.5);
  \draw[gluon] (0,-1.1) -- (1.1,-1.1);
  \draw[gluon] (1.1,-1.1) -- (2.4,-1.5);
  \draw[gluon] (1.1,1.1) -- (1.5,0);
  \draw[gluon] (1.1,-1.1) -- (1.5,0);
  \draw[scalar] (1.5,0) -- (2.6,0);
  \node[effvertex] at (1.5,0) {};
  \node[left] at (0,1.1) {$g$}; \node[left] at (0,-1.1) {$g$};
  \node[right] at (2.4,1.5) {$g$}; \node[right] at (2.4,-1.5) {$g$}; \node[right] at (2.6,0) {$H$};
  \node at (1.3,-1.95) {(b)};
\end{tikzpicture}\hfill
\begin{tikzpicture}[scale=0.85, baseline=(base)]
  \coordinate (base) at (0,-0.1);
  \draw[fermion] (0,1.1) -- (1.1,1.1);
  \draw[fermion] (1.1,1.1) -- (2.4,1.5);
  \draw[fermion] (0,-1.1) -- (1.1,-1.1);
  \draw[fermion] (1.1,-1.1) -- (2.4,-1.5);
  \draw[gluon] (1.1,1.1) -- (1.5,0);
  \draw[gluon] (1.1,-1.1) -- (1.5,0);
  \draw[scalar] (1.5,0) -- (2.6,0);
  \node[effvertex] at (1.5,0) {};
  \node[left] at (0,1.1) {$q$}; \node[left] at (0,-1.1) {$q'$};
  \node[right] at (2.4,1.5) {$q$}; \node[right] at (2.4,-1.5) {$q'$}; \node[right] at (2.6,0) {$H$};
  \node at (1.3,-1.95) {(c)};
\end{tikzpicture}
\par\bigskip
\hspace*{\fill}%
\begin{tikzpicture}[scale=0.85, baseline=(base)]
  \coordinate (base) at (0,-0.1);
  \draw[fermion] (0,1.1) -- (1.1,1.1);
  \draw[fermion] (1.1,1.1) -- (2.4,1.5);
  \draw[gluon] (0,-1.1) -- (1.1,-1.1);
  \draw[gluon] (1.1,-1.1) -- (2.4,-1.5);
  \draw[gluon] (1.1,1.1) -- (1.5,0);
  \draw[gluon] (1.1,-1.1) -- (1.5,0);
  \draw[scalar] (1.5,0) -- (2.6,0);
  \node[effvertex] at (1.5,0) {};
  \node[left] at (0,1.1) {$q$}; \node[left] at (0,-1.1) {$g$};
  \node[right] at (2.4,1.5) {$q$}; \node[right] at (2.4,-1.5) {$g$}; \node[right] at (2.6,0) {$H$};
  \node at (1.3,-1.95) {(d)};
\end{tikzpicture}\hspace*{\fill}%
\begin{tikzpicture}[scale=0.85, baseline=(base)]
  \coordinate (base) at (0,-0.1);
  \draw[gluon] (0,1.1) -- (0.6,1.1);
  \draw[gluon] (0.6,1.1) -- (1.4,1.1);
  \draw[gluon] (0.6,1.1) -- (0.95,2.15);
  \draw[gluon] (1.4,1.1) -- (2.6,1.5);
  \draw[gluon] (0,-1.1) -- (1.4,-1.1);
  \draw[gluon] (1.4,-1.1) -- (2.6,-1.5);
  \draw[gluon] (1.4,1.1) -- (1.8,0);
  \draw[gluon] (1.4,-1.1) -- (1.8,0);
  \draw[scalar] (1.8,0) -- (2.9,0);
  \node[effvertex] at (1.8,0) {};
  \node[left] at (0,1.1) {$g$}; \node[left] at (0,-1.1) {$g$};
  \node[above] at (0.95,2.15) {$g$}; \node[right] at (2.6,1.5) {$g$}; \node[right] at (2.6,-1.5) {$g$}; \node[right] at (2.9,0) {$H$};
  \node at (1.45,-1.95) {(e)};
\end{tikzpicture}\hspace*{\fill}%
\begin{tikzpicture}[scale=0.85, baseline=(base)]
  \coordinate (base) at (0,-0.1);
  \draw[gluon] (0,1.1) -- (1.1,1.1);
  \draw[gluon] (1.1,1.1) -- (2.4,1.5);
  \draw[gluon] (0,-1.1) -- (1.1,-1.1);
  \draw[gluon] (1.1,-1.1) -- (2.4,-1.5);
  \draw[gluon] (1.1,1.1) -- (1.5,0);
  \draw[gluon] (1.1,-1.1) -- (1.5,0);
  \draw[gluon] (1.5,0) -- (2.6,0.55);
  \draw[scalar] (1.5,0) -- (2.6,-0.35);
  \node[effvertex] at (1.5,0) {};
  \node[left] at (0,1.1) {$g$}; \node[left] at (0,-1.1) {$g$};
  \node[right] at (2.4,1.5) {$g$}; \node[right] at (2.4,-1.5) {$g$}; \node[right] at (2.6,0.55) {$g$}; \node[right] at (2.6,-0.35) {$H$};
  \node at (1.3,-1.95) {(f)};
\end{tikzpicture}\hspace*{\fill}
  \par
  \caption[Signal diagrams]{Representative leading-order diagrams of the signal. (a)~In the limit $m_t\to\infty$ the top-quark loop
  reduces to the effective vertex of eq.~\eqref{eq:leff} (gray circle). (b)~$gg\to Hgg$, (c)~$qq'\to Hqq'$  and (d)~$qg\to Hqg$. (e), (f)~$gg\to Hggg$.
  The decay $H\to\gamma\gamma$ is not shown.}
  \label{fig:feyn-signal}
\end{figure}
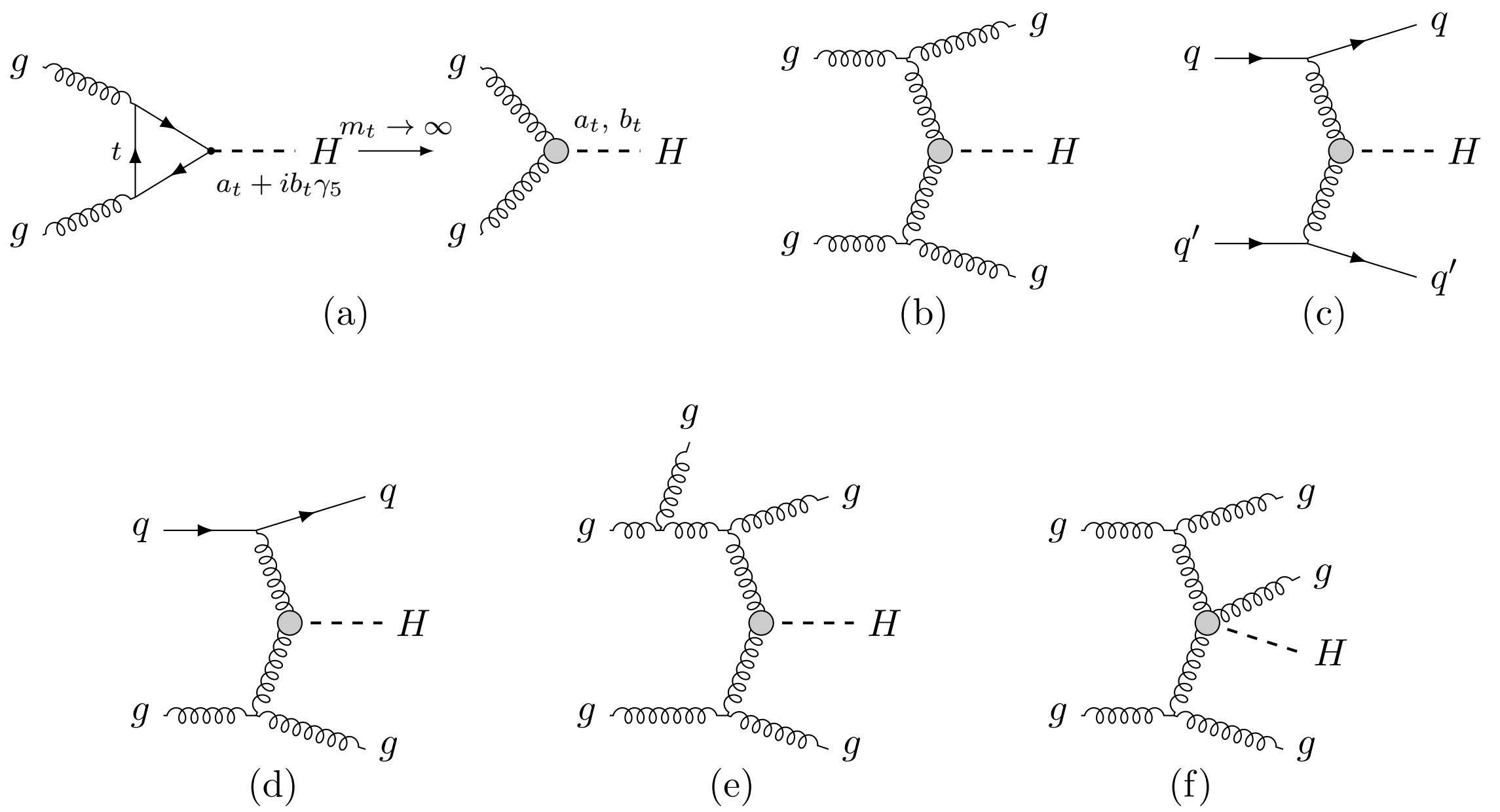

Throughout this work, $\at$ and $\bt$ denote the coefficients of the effective interaction in eq.~\eqref{eq:leff}, and we interpret our results in the heavy-top limit. Their relation to other parametrizations of the CP-odd coupling is summarized in appendix~\ref{sec:int-maps-short}. Evaluating the two-jet optimal observables with the one-loop amplitudes at the physical top-quark mass instead changes the expected interval of the two-jet analysis by less than $1\%$ (a check made with the per-helicity weights defined in section~\ref{sec:sim-signal}). The full top-mass dependence of three-parton events remains to be studied.\footnote{Beyond leading order, the coefficient of the CP-even operator receives a QCD correction, $1+\tfrac{11}{4}\,\alpha_s/\pi$ in the amplitude, from gluon exchange inside the top-quark loop~\cite{th:dawson,th:djouadi_spira_zerwas}, whereas the CP-odd coefficient is not renormalized as a consequence of the Adler--Bardeen theorem~\cite{th:adler_bardeen,th:spira_hgg}. Since we apply a common higher-order correction to all three terms of the cross section (section~\ref{sec:sim-nlo}), the ratio $\bt/\at$ extracted in this work refers to the effective interaction of eq.~\eqref{eq:leff}. The corresponding ratio of the Yukawa couplings in eq.~\eqref{eq:yukawa} is larger by the factor $1+\tfrac{11}{4}\,\alpha_s/\pi \approx 1.10$ for $\alpha_s(m_H)\approx0.11$.} 

We vary $\at$ and $\bt$ only in the ggF production vertex and keep the $H\to\gamma\gamma$ decay and all other Higgs production modes, including $t\bar tH$ and $tH$, at their SM values. A CP-odd top Yukawa coupling would modify these as well, but we leave that information out to isolate the information carried by the jets. 
The factor $3/2$ in eq.~\eqref{eq:leff} is chosen such that $\bt=1$ corresponds to a pseudoscalar coupling with the same Yukawa strength as the SM scalar coupling. A purely CP-odd interaction therefore gives a larger cross section, $\sigma(0,1)/\sigma(1,0)=(3/2)^2=2.25$ for inclusive $gg\to H$ and $2.16$ after our event selection requiring at least two jets. 
To a good approximation, the cross section after the selection is
\begin{equation} 
	\sigma_{\mathrm{sel}}(\at,\bt) \simeq \sigma_{\mathrm{sel,SM}} \left( \at^2 + 2.16\,\bt^2 \right).
	\label{eq:ring} 
\end{equation} 
There is no term linear in $\bt$ because the rate is CP-even, so a measurement of the total rate constrains only the combination $\at^2+2.16\,\bt^2$. If the observed rate agrees with the SM prediction, all points on the ellipse $\at^2+2.16\,\bt^2=1$ are equally preferred. 
The sign of $\bt$, and the distinction between a CP-odd coupling and a reduced CP-even one, must come from the kinematic distributions of the jets, through the interference of the CP-even and CP-odd amplitudes.

\subsection{Signal and background simulation}\label{sec:sim}\label{sec:sim-signal}

All samples are generated at $\sqrt{s}=13.6\TeV$. Matrix elements are computed with \mgfive{} 3.5.5~\cite{sim:mg5amc}, the parton shower, hadronization and multiple interactions are simulated with \pythia{} 8.316~\cite{sim:pythia83} using the A14 tune~\cite{sim:a14}, and the detector response is modeled with \delphes~\cite{sim:delphes}. The \mgfive{} LO samples use the NNPDF2.3 NLO parton distribution functions (PDFs)~\cite{sim:nnpdf23} through LHAPDF6~\cite{sim:lhapdf6}, and the NLO reference samples use NNPDF3.0 (appendix~\ref{app:mlm}). For comparison, all \mgfive{} samples are also reweighted event by event to the shape of the CT25 NNLO set~\cite{sim:ct25} (section~\ref{sec:st-syst}). We refer to the stable particles after the parton shower and hadronization as the particle level, and to the objects after the detector simulation as the reconstruction level. Comparisons with NLO predictions and published measurements are made in a particle-level fiducial region modeled on the ATLAS $H\to\gamma\gamma$ measurements, with two isolated photons and at least two anti-$k_t$ jets with $\pT>30\GeV$ and $|y|<4.4$.

\label{sec:sim-detector}The detector response is simulated with \delphes{}, starting from its ATLAS card and adjusted to reproduce the published Run~2 and Run~3 ATLAS performance for photons and particle-flow jets (appendix~\ref{app:detector}). Jets are reconstructed with the anti-$k_t$ algorithm with $R=0.4$~\cite{sim:antikt,sim:fastjet}, as in the ATLAS and CMS measurements of Higgs boson production with jets, and the card also provides jet-substructure variables that the S/B classifiers use to separate quark and gluon jets. The resulting diphoton mass resolution is $\sigma_{68}=1.54\GeV$, the half-width of the smallest interval containing 68\% of the signal, better than the ATLAS Run~2 value of $1.9\GeV$~\cite{th:atlas_yy_139_diff}. The overall diphoton efficiency is corrected by a factor $0.884$ to match the published Run~3 efficiency~\cite{ATLAS:2023run3hyy}. Our nominal results do not include pileup, whose effect is evaluated in section~\ref{sec:res-pileup}.

The signal is generated with \mgfive{}~\cite{sim:mg5amc} for $pp\to H+0,1,2,3$ partons in the four-flavor scheme, with the default dynamical scales of \mgfive{} and a Universal FeynRules Output (UFO) implementation of eq.~\eqref{eq:leff} in which $\at$ and $\bt$ are free parameters. Electroweak diagrams with the same final state, which correspond to VBF and hadronic $VH$ production, are removed, so that the sample contains only gluon fusion. The simulation includes all partonic channels, namely gluon--gluon, quark--gluon and quark--(anti)quark initial states with quarks and gluons in the final state.
Only in constructing the analytic optimal observables of section~\ref{sec:observables} do we approximate every parton by a gluon (appendix~\ref{app:obs-flavour}), so that this approximation affects these observables but not the simulated signal.
The approximation is not negligible, since three-parton events with at least one quark carry about four fifths of the information on $\bt$ in three-parton events, and treating them as all-gluon events leads the optimal observables to lose about a tenth of this information at the level of the hard partons.
The learned observables, which are trained on the full mixture of subprocesses, do not rely on this approximation and therefore recover the information lost by the optimal observables (section~\ref{sec:truth-flavour}).
At leading order in the heavy-top effective theory every amplitude contains exactly one effective Higgs-gluon vertex, so that $\mathcal{M}=\at\mathcal{M}_E+\bt\mathcal{M}_O$ for each helicity and color configuration, and the squared matrix element is an exact quadratic form in the couplings. Since the parton densities, the phase space, the scales and the jet matching do not depend on the couplings, the weight of each event at any coupling point follows from three basis weights,
\begin{equation}
  w(\at,\bt) = \at^2\,w_E + \at\bt\,w_I + \bt^2\,w_O ,
  \label{eq:morph}
\end{equation}
with $w_E=w(1,0)$, $w_O=w(0,1)$ and $w_I=w(1,1)-w_E-w_O$. We generate the events at $(\at,\bt)=(1,1)$, which populates the phase space of both the CP-even and the CP-odd operators, and obtain $w_E$ and $w_O$ for every event with the matrix-element reweighting of \mgfive{}~\cite{sim:mg5_reweight}, evaluating the matrix element summed over the helicities of
the external partons. We call these the helicity-summed weights. By default, the reweighting instead rescales each event by the
matrix-element ratio of the single helicity configuration assigned to it at generation. These per-helicity weights give the correct
cross sections and distributions after summing over events, but their per-event ratios depend on the unobservable helicities, so we
use them only in the cross-checks where this is stated. Because the parton shower, hadronization and detector simulation do not depend on the couplings, eq.~\eqref{eq:morph} holds event by event at every level of the simulation, and a single sample provides the prediction at any point $(\at,\bt)$. This procedure is often called morphing~\cite{ATLAS:2015rmd}.
The Higgs boson is decayed to two photons in \pythia{}, and the SM branching ratio, $\mathcal{B}(H\to\gamma\gamma)=2.27\times10^{-3}$, is applied afterwards. After jet matching, the 259 generation runs used in this analysis contain $1.05\times10^7$ events. One further run, $0.4\%$ of the sample, is not used.

Since matrix elements with different parton multiplicities are combined in a single sample, the emissions of the parton shower must not double count those of the matrix elements. This matters for the CP analysis in particular: without merging, a shower jet can replace one of the hard jets of a $gg\to Hgg$ sample and dilute or even reverse the azimuthal correlation. We use $k_T$-MLM merging~\cite{sim:mlm,sim:ktmlm,sim:mlm_comparison} with a generation cut $x_{\mathrm{qcut}}=15\GeV$ and a merging scale $Q_{\mathrm{cut}}=22.5\GeV$. We stop at three partons because our goal is the information carried by a third jet, and leave additional jets for future work. Since the merging scale lies below the analysis threshold of $30\GeV$, up to three jets above threshold are described by tree-level matrix elements, while a fourth jet comes from the parton shower.

How the information carried by the third jet depends on the parton-shower model is not evaluated in this study (section~\ref{sec:syst}). Varying $(x_{\mathrm{qcut}},Q_{\mathrm{cut}})$ between $(10,15)$ and $(20,30)\GeV$ changes the SM cross section with at least two jets by $+3.2\%$ and $-1.8\%$. The CP-dependent terms were not re-evaluated with these settings.

Of the selected signal events, counted by the number of hard partons in the generated process, $56\%$ come from three hard partons, $39\%$ from two and $5\%$ from zero or one, and the three-parton events
carry $68\%$ of the information on $\bt$ (section~\ref{sec:truth}). Events with zero or one hard parton contain two jets only through the
parton shower. Their interference term vanishes (section~\ref{sec:dphi-corr}), so they add to the rate and to the $\bt^2$ term but carry no
linear information on $\bt$.

\label{sec:sim-nlo}The merged LO sample describes the CP structure exactly at LO, but not the kinematic distributions and the rate predicted at NLO in QCD. We therefore correct it with a kinematic reweighting that does not depend on the couplings,
\begin{equation}
  w_{\mathrm{NLO}}(\at,\bt) = \left[\at^2 w_E+\at\bt\, w_I+\bt^2 w_O\right]\,r(x)\,K ,
  \label{eq:nlo}
\end{equation}
where $r(x)$ is the ratio of the NLO to the LO density in the fiducial region, learned by a gradient-boosted classifier~\cite{bkgm_xgboost} from a SM sample of gluon-fusion $H+0,1,2$ jets at NLO with FxFx merging~\cite{sim:fxfx} (appendix~\ref{app:mlm-checks}), and $K=4.145$ normalizes the sample to the gluon-fusion cross section at N$^3$LO in QCD and NLO in the electroweak coupling, $52.17\pb$~\cite{Karlberg:2024zxx}.

The classifier receives no azimuthal information, nor the variables from which it could be reconstructed. This matters because the CP information resides in the azimuthal correlations of the jets, while $r(x)$ is learned from an SM sample. A reweighting that depended on these correlations would impose the NLO corrections of the SM azimuthal pattern also on the CP-odd and interference terms, whose azimuthal patterns are different, and would thereby bias the measurement of $\bt$. With a CP-blind classifier, the same $r(x)$ multiplies all three basis weights, and the CP structure of eq.~\eqref{eq:morph} is preserved event by event. NLO samples generated directly at CP-odd and mixed coupling points (appendix~\ref{app:merging-nlo}) confirm this to $2\%$ in the fiducial rate and to about $10\%$ in the interference term. These comparisons were made with the per-helicity weights, which leave the rate and the interference moments unbiased, and in events with a third jet they test the interference only at the tree-level accuracy of the real emission.

One feature cannot be captured by a CP-blind reweighting: the SM $\cos2\dphijj$ modulation is smaller at NLO, by a factor $0.74\pm0.07$ in the fiducial region and $0.85\pm0.09$ for $|\Delta y_{jj}|>2$. This reduction enters the fit as a nuisance parameter (section~\ref{sec:stat}).
The generator settings of the LO signal, the FxFx reference sample, the classifier for $r(x)$ and the \sherpa{} continuum sample are summarized in appendix~\ref{app:mlm}.

The dominant background is the QCD production of two photons in association with jets. It is generated with \mgfive{} at LO as $pp\to\gamma\gamma+2,3$ jets, merged with the same MLM settings as the signal and without diagrams containing a Higgs boson. The LO matrix elements give about half the rate of the NLO-merged prediction used below and produce too few jets at large rapidity separation. We therefore reweight every event to the \sherpa{} 2.2~\cite{sim:sherpa22} MEPS@NLO~\cite{sim:mepsnlo} prediction, with $\gamma\gamma+0,1$ jets at NLO from OpenLoops~2~\cite{sim:openloops2}, using a per-event ratio learned at reconstruction level. This ratio uses only unsigned azimuthal information, so that it cannot introduce a parity-odd asymmetry into the background. The reweighted sample describes \sherpa{} within its statistical precision, including the signed $\dphijj$ and the optimal observables (appendix~\ref{app:bkg-rw}). The continuum is normalized to the ATLAS diphoton cross section measured at $13\TeV$~\cite{bkgm_atlas_diphoton_13tev}, which scales our \sherpa{} prediction by $1.169$ (appendix~\ref{app:bkg-norm}). In the fit with systematic uncertainties, the continuum normalization of each category is a free parameter constrained by the $\myy$ sidebands. The continuum templates are the main source of simulation-statistical uncertainty (section~\ref{sec:st-mcstat}).

Smaller backgrounds come from photon+jet events with a jet misidentified as a photon, generated as $pp\to\gamma+2,3$ jets and normalized to the photon+jet fraction measured by ATLAS~\cite{th:atlas_yy_36}, and from the purely electroweak production of $\gamma\gamma jj$, which amounts to about $1\%$ of the selected continuum. Events with two misidentified jets, about $3\%$ of the ATLAS diphoton sample~\cite{th:atlas_yy_36}, are not simulated, and their rate is absorbed by the continuum normalization of each category. Higgs boson production through VBF, $VH$ and $t\bar tH$ is simulated with SM couplings at LO (appendix~\ref{app:bkg-other}). VBF is the largest of these, about a third of the gluon-fusion yield after the selection. It produces CP-even jet correlations of its own and is therefore a background to the CP measurement, which the S/B classifier treats as a separate class (section~\ref{sec:st-classifiers}). All simulated samples, their sizes and their cross sections are listed in table~\ref{tab:sim-samples}.

\begin{table}[tbp]
  \centering
  \caption{Simulated samples. All are generated with \mgfive{} at $\sqrt{s}=13.6\TeV$ in the four-flavor scheme, showered
  with \pythia{} (A14) and passed through \delphes{}. $N_{\mathrm{LHE}}$ is the number
  of generated events (the signal templates use 259 of the 260 runs, $5.18\times10^7$ events). $\sigma_{\mathrm{gen}}$ is the generator cross section within the
  generator cuts of appendix~\ref{app:mlm} and $\sigma_{\mathrm{matched}}$ the cross section after jet matching (samples without merging have
  only $\sigma_{\mathrm{gen}}$), both without the $H\to\gamma\gamma$ branching ratio. The ggF cross sections are at the generation point $(\at,\bt)=(1,1)$,
  and the matched cross section at the SM point is $12.58\pb$. The $\gamma j$ sample models photon+jet fakes, and the
  $\gamma\gamma+0,1$\,jet sample is used only in the pileup study (section~\ref{sec:res-pileup}). MLM stands for $k_T$-MLM jet merging
  (section~\ref{sec:sim-signal}), and heavy-top EFT for the effective interaction of eq.~\eqref{eq:leff}.}
  \label{tab:sim-samples}
  \small
  \resizebox{\textwidth}{!}{%
  \begin{tabular}{llrrr}
   \toprule
   Process & Matrix elements & $N_{\mathrm{LHE}}$ & $\sigma_{\mathrm{gen}}$ & $\sigma_{\mathrm{matched}}$ \\
   \midrule
   ggF $H$ (signal)        & $H+0,1,2,3$ partons, MLM, heavy-top EFT & $5.2\times10^7$ & $199.8\pb$ & $40.69\pb$ \\
   QCD $\gamma\gamma$      & $\gamma\gamma+2,3$ partons, MLM          & $9.99\times10^7$ & $11.58\pb$ & $2.691\pb$ \\
   QCD $\gamma\gamma$ (pileup study) & $\gamma\gamma+0,1$ partons, MLM & $5.0\times10^6$ & $11.33\pb$ & $5.594\pb$ \\
   $\gamma j$ (fakes)      & $\gamma+2,3$ partons, MLM                & $4.99\times10^7$ & $52.7\,\mathrm{nb}$ & $13.0\,\mathrm{nb}$ \\
   EW $\gamma\gamma jj$    & $\gamma\gamma jj$, $\mathcal{O}(\alpha^4)$ & $1.17\times10^5$ & $0.054\pb$ & \\
   VBF $H$                 & $Hjj$, no $s$-channel $V$, QCD $=0$       & $2.0\times10^6$ & $3.514\pb$ & \\
   $VH$                    & $HV+0,1$ partons, MLM, $V$ inclusive     & $2.0\times10^6$ & $3.749\pb$ & $2.025\pb$ \\
   $t\bar{t}H$             & $t\bar{t}H$                               & $2.0\times10^5$ & $0.418\pb$ & \\
   \bottomrule
  \end{tabular}}
\end{table}

\subsection{Event selection and validation}\label{sec:selection}
At reconstruction level, the event selection follows that of the ATLAS $H\to\gamma\gamma$ measurements~\cite{th:atlas_yy_139_diff}.
Events must contain two photons that pass the identification and isolation criteria of the detector simulation, with $|\eta|<2.37$
excluding the transition region $1.37<|\eta|<1.52$ between the barrel and endcap calorimeters, and with $\pT/\myy>0.35$ for the
leading and $\pT/\myy>0.25$ for the subleading photon.
Each event must also contain at least two jets with $\pT>30\GeV$ and $|y|<4.4$, each separated from both photons by
$\Delta R>0.4$. These are the jets from which the CP-sensitive observables of section~\ref{sec:multijet} are built.
The diphoton invariant mass must lie in the range $105<\myy<160\GeV$, and the fit to the CP-sensitive observables uses the window
$120<\myy<130\GeV$.
This window is wide compared with the simulated mass resolution of section~\ref{sec:sim-detector}, but the simulation has no photon
conversions and therefore underestimates the width of the signal peak. In the two-jet analysis the narrower window $122<\myy<128\GeV$ keeps $94\%$ of the signal and $60\%$ of the
continuum and narrows the expected interval on $\bt$ by $12\%$, part of which comes from the simulated mass resolution.
Within the window, we expect with $300\ifb$ 1563 gluon-fusion signal events in the SM, 479 VBF events, 259 $VH$ and $t\bar tH$
events, and $7.57\times10^4$ continuum events (table~\ref{tab:yields}).
The continuum thus exceeds the signal about fifty-fold, which motivates the S/B categories of
section~\ref{sec:categories}.

The simulated SM signal agrees with the unfolded $H\to\gamma\gamma$ differential cross sections measured by ATLAS at $13\TeV$ and by CMS at $13.6\TeV$ within the measurement uncertainties, including the signed $\dphijj$. Our total Higgs boson cross section in the ATLAS diphoton fiducial region, scaled to $13\TeV$, is $62.8\fb$, against the measured $66.6\pm6.3\fb$.

\section{CP-sensitive observables}\label{sec:multijet}

\subsection{Two-jet observables}\label{sec:dphi-corr}\label{sec:oo}

The interference term $w_I$ in eq.~\eqref{eq:morph} is odd under CP and integrates to zero, so it does not contribute to the rate, as reflected in eq.~\eqref{eq:ring}. Observables that are even under CP, such as the dijet invariant mass or any transverse momentum, likewise carry no information on $\bt$ at linear order. Sensitivity to the sign of $\bt$, and hence to CP violation, requires CP-odd observables.

For $p(k_a)p(k_b)\to H\,j_1(p_1)\,j_2(p_2)$ there are four independent momenta, and the only pseudoscalar that can be constructed from them is the Levi-Civita contraction,
\begin{equation}
  w_I \propto \epsilon_{\mu\nu\rho\sigma}\,k_a^\mu k_b^\nu p_1^\rho p_2^\sigma = \pm\frac{\hat s}{2}\,p_{\mathrm{T}1}p_{\mathrm{T}2}\sin(\phi_1-\phi_2),
  \label{eq:th-eps}
\end{equation}
with $k_a$ along the $+z$ direction. The interference is therefore an azimuthal correlation between the two jets, and its sign is meaningful only for a definite ordering of the jets. We denote by $P$ the reflection $\phi\to-\phi$ of all momenta in azimuth ($p_y\to-p_y$), that is, parity followed by a rotation by $\pi$ about the $y$ axis, perpendicular to the beam. It reverses the sign of eq.~\eqref{eq:th-eps} and of $w_I$ and leaves $w_E$ and $w_O$ unchanged. We call observables that change sign under $P$ P-odd and the others P-even. With three partons further pseudoscalars can be constructed (section~\ref{sec:mj-invariants}).

When the two jets are widely separated in rapidity, the amplitude is dominated by $t$-channel gluon exchange, and the squared matrix element takes the simple form~\cite{th:prz,th:hankele,th:klamke_zeppenfeld,th:delduca_hjj_2006}
\begin{equation}
  |\mathcal{M}|^2 \propto \left(\at\cos\dphijj + \tfrac32\bt\sin\dphijj\right)^2
  = \frac{\at^2+\frac94\bt^2}{2}\Bigl[1+\cos\!\left(2(\dphijj-\alpha_g)\right)\Bigr],
  \label{eq:eikonal}
\end{equation}
in our conventions. The CP-even and CP-odd couplings produce opposite $\cos2\dphijj$ modulations, and a mixture of the two shifts the modulation by the angle $\alpha_g$, with $\tan\alpha_g=3\bt/(2\at)$, so that the sign of $\bt$ appears as the direction of the shift.
To make this shift observable, the jets must be assigned to the two beams~\cite{th:hankele}. Following the ATLAS measurement~\cite{th:atlas_yy_139_diff}, we order the two jets with the highest transverse momentum in rapidity and define the signed angle $\dphijj=\phi_{\mathrm{fwd}}-\phi_{\mathrm{bwd}}$, wrapped to $(-\pi,\pi]$. Alternatively, the momenta of all jets on either side of the Higgs boson can be summed before the azimuthal difference is taken; this defines the grouped angle $\dphigrp$~\cite{th:andersen_arnold_zeppenfeld}.

At small rapidity separations the modulation is diluted by $s$-channel configurations and by additional radiation. Figure~\ref{fig:dphi} shows both angles for our simulated signal at particle level, after the NLO reweighting, for the SM, a pure CP-odd coupling and the mixed point $(\at,\bt)=(1,1)$, without a cut on the rapidity separation of the jets. The grouped angle is shown for the 49\% of the events with at least one jet on each side of the Higgs boson in rapidity, for which it is defined. Relative to the SM, the CP-odd distributions are enhanced near $\pm\pi/2$ and suppressed near $0$ and $\pm\pi$, and the distributions at the mixed point are shifted towards positive angles, as expected from eq.~\eqref{eq:eikonal} for $\bt/\at>0$. The modulation of the grouped angle is much stronger: its $\langle\cos2\Delta\phi\rangle$ is $0.148$ for the SM and $-0.125$ for the CP-odd coupling, compared with $0.118$ and $-0.015$ for $\dphijj$. At the mixed point, the second harmonic of the grouped angle is shifted by $0.93$~rad, close to the value $\alpha_g=\arctan(3/2)=0.98$~rad of eq.~\eqref{eq:eikonal}, while that of $\dphijj$ is shifted by only $0.57$~rad.

\begin{figure}[tbp]
  \centering
  \includegraphics[width=\linewidth]{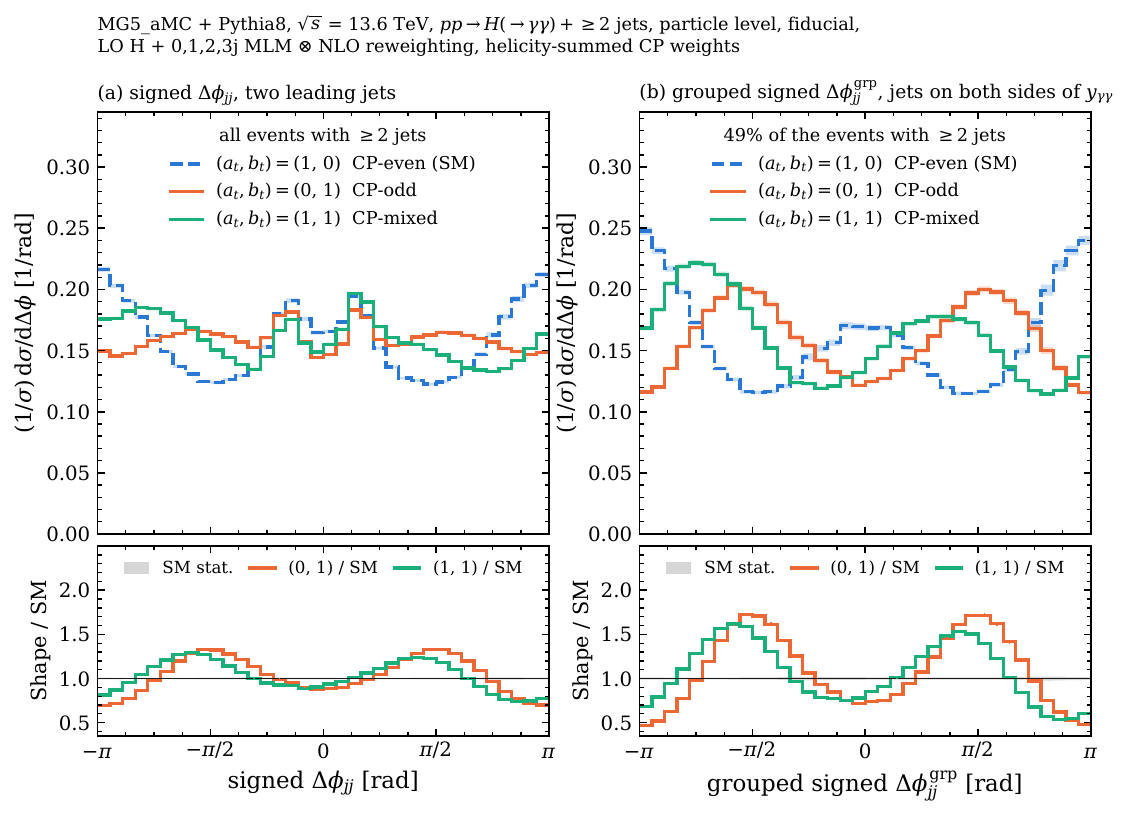}
  \caption{Normalized distributions of (a) the signed $\dphijj$ of the two leading jets and (b) the grouped signed angle $\dphigrp$ of
  gluon-fusion $H(\to\gamma\gamma)+\geq2$\,jets at particle level, after the NLO reweighting of eq.~\eqref{eq:nlo}, for
  $(\at,\bt)=(1,0)$, $(0,1)$ and $(1,1)$, without a cut on $|\Delta y_{jj}|$, computed with the helicity-summed weights.
  Panel (b) contains only the 49\% of the events with at least one jet on each side of the Higgs boson in rapidity and is normalized
  to them. In the fits the other events enter with $\dphijj$. The lower panels show the ratio to the SM shape, whose statistical
  uncertainty (gray band) is at most $1.3\%$ in (a) and $2.3\%$ in (b) per bin. The narrow peaks of $\dphijj$ near $\pm0.4$ and the dip
  at $0$, present for all couplings, come from pairs of nearby jets at the resolution limit $\Delta R=0.4$ of the jet algorithm. They
  are absent in the grouped angle, whose two jet groups lie on opposite sides of the Higgs boson.}
  \label{fig:dphi}
\end{figure}

The signed $\dphijj$ captures the interference fully only in the eikonal limit, whereas optimal observables~\cite{th:atwood_soni,th:davier,th:diehl_nachtmann} use the full kinematics of the event. For a likelihood that is linear in a parameter $\theta$, $p(x|\theta)\propto f_0(x)+\theta f_1(x)$, the ratio $f_1/f_0$ carries all the information on $\theta$ at $\theta=0$ and saturates the Cram\'er--Rao bound~\cite{th:cramer,th:rao} to first order. With eq.~\eqref{eq:morph}, the likelihood ratio to the SM depends on the event only through $w_I/w_E$ and $w_O/w_E$, which are therefore the optimal observables. For $gg\to Hgg$ they are
\begin{equation}
  \OOone = \frac{2\,\mathrm{Re}(\mathcal{M}_E\mathcal{M}_O^*)}{|\mathcal{M}_E|^2},\qquad
  \OOtwo = \frac{|\mathcal{M}_O|^2}{|\mathcal{M}_E|^2} .
  \label{eq:oodef}
\end{equation}
The interference ratio $\OOone$ is P-odd and carries the first-order information on $\bt$, while $\OOtwo$ is P-even and constrains $\bt^2$ through the shape of its distribution. Together they form a sufficient statistic for $(\at,\bt)$ at the level of the hard process, and we therefore fit their two-dimensional distribution. Since $|\at\mathcal{M}_E+\bt\mathcal{M}_O|^2\geq0$, the weights satisfy
\begin{equation}
  w_I^2\leq 4\,w_E\,w_O
  \label{eq:th-bound}
\end{equation}
event by event, so the physical region lies inside the parabola $\OOone^2\leq4\,\OOtwo$. In the eikonal limit $\OOone\to3\tan\dphijj$, so the signed $\dphijj$ is optimal there and $\OOone$ generalizes it to the full phase space. We refer to $\OOone$ and $\OOtwo$, evaluated with the heavy-top $gg\to Hgg$ matrix element on two jets, either the two leading jets or the grouped jets, as the two-jet optimal observables. The discriminants $D_{0-}$ and $D_{CP}$ that CMS constructs from the $H+2$ parton matrix elements~\cite{th:gritsan_2016} are the bounded combinations $1/(1+c\,\OOtwo)$ and, up to a constant factor, $\sqrt{c}\,\OOone/(1+c\,\OOtwo)$, where the constant $c$ fixes the normalization of the CP-odd amplitude. They therefore carry the same information as the two-jet optimal observables. With a third parton none of these observables is exact any more, which is the subject of the following subsections.

\subsection{Fisher information}\label{sec:fisher}

To compare the observables of this section on an equal footing, we quantify the information that each of them carries on $\bt$
through the Fisher information~\cite{res:brehmer_infogeo,th:brehmer_cp_infogeo}.
For an observable divided into bins $k$, the expected number of signal events in bin $k$ follows from eq.~\eqref{eq:morph} as
\begin{equation}
  \nu_k(\at,\bt) = L\left(\at^2E_k+\at\bt\,I_k+\bt^2O_k\right),
  \label{eq:fisher-nu}
\end{equation}
where $L$ is the integrated luminosity and $E_k$, $I_k$ and $O_k$ are the contributions of $w_E$, $w_I$ and $w_O$ to the cross
section in that bin.
For Poisson-distributed bin contents, the Fisher information matrix is
$F_{ij}=\sum_k\partial_i\nu_k\,\partial_j\nu_k/\nu_k$, with $i,j\in\{\at,\bt\}$.
At the SM point $(\at,\bt)=(1,0)$ the derivatives are $\partial_{\bt}\nu_k=L\,I_k$ and $\partial_{\at}\nu_k=2L\,E_k$, so that
\begin{equation}
  F_{\bt\bt}=L\,F,\qquad F=\sum_k\frac{I_k^2}{E_k},\qquad F_{\at\bt}=2L\sum_kI_k=0,
  \label{eq:fisher}
\end{equation}
where the off-diagonal element vanishes because the interference integrates to zero.
The information on $\bt$ at the SM is therefore the same whether $\at$ is fixed or profiled, and in the asymptotic
limit~\cite{cpfit:asimov} the expected 68\% CL half-width on $\bt$, which we denote by $\sigma(\bt)$ throughout, is
$\sigma(\bt)=(LF)^{-1/2}$.

The largest information that any observable can achieve follows from the limit of infinitely fine bins in the full phase space of
the event, which corresponds to an unbinned analysis in which every event enters the likelihood with its own weights,
\begin{equation}
  F_{\mathrm{exact}}=\sum_{\mathrm{events}}w_E\left(\frac{w_I}{w_E}\right)^2 ,
  \label{eq:fisher-exact}
\end{equation}
where the sum runs over the simulated events with their weights.
Writing $I_k=\sum_{i\in k}w_{E,i}\,(w_I/w_E)_i$ and applying the Cauchy--Schwarz inequality within each bin gives
$I_k^2\le E_k\sum_{i\in k}w_{E,i}\,(w_I/w_E)_i^2$, and hence $F\le F_{\mathrm{exact}}$, with equality only if the ratio $w_I/w_E$ is
constant within every bin.
An observable that bins the events in $w_I/w_E$ itself therefore retains all the information on $\bt$ at the SM, up to the finite
width of its bins. This is the sense in which $\OOone$ is optimal, and the ratio $F/F_{\mathrm{exact}}$ measures the fraction of the
information that a given observable retains.
We use binned templates because the backgrounds and the systematic uncertainties are described by them (section~\ref{sec:stat}). The
$16\times8$ binning used throughout retains $0.888$ of the unbinned information of the exact ratios, and an unbinned fit of the learned
observables gives an interval only $4\%$ narrower than the binned one.
With finite simulated samples we subtract the statistical noise in each bin, $I_k^2\to I_k^2-\sum_{i\in k}w_{I,i}^2$. In section~\ref{sec:truth} two-dimensional observables are binned in 16 signed
quantile bins of the first and 8 quantile bins of the second, events in which an observable is undefined go to one extra bin, and
uncertainties come from Poisson bootstrap replicas of the events, with every ratio between observables formed per replica.

The Fisher information at the SM measures only the linear dependence on $\bt$, since $O_k$ enters eq.~\eqref{eq:fisher-nu} at order
$\bt^2$.
The information carried by the CP-even ratio $\OOtwo$, which separates the CP-odd from the CP-even shape along the ellipse of
eq.~\eqref{eq:ring}, is therefore assessed with fits in which $\at$ is free.
In section~\ref{sec:truth} the observables are compared through $F/F_{\mathrm{exact}}$ for the signal alone, while the results of
section~\ref{sec:results} include the backgrounds, which enter the denominator of eq.~\eqref{eq:fisher} as $E_k+B_k$. The off-diagonal element still vanishes, because the bins are symmetric under $P$, $I_k$ is odd, and $E_k$ and $B_k$ are even.

\subsection{Optimal observables for Higgs plus three jets}\label{sec:mj-invariants}\label{sec:mj-oo3}

For events with three partons ($ab\to H\,1\,2\,3$) there are six external momenta and, after momentum conservation, five independent four-vectors. In contrast to the two-parton process, where the interference is determined by a single Levi-Civita contraction, the three-parton interference involves five CP-odd Levi-Civita contractions. Defining $ \epsilon(i,j,k,l)\equiv \epsilon_{\mu\nu\rho\sigma} p_i^\mu p_j^\nu p_k^\rho p_l^\sigma $ and denoting the incoming momenta by $a$ and $b$, the interference term can be written as 
\begin{equation} w_I^{(3)} = \sum_{i<j} c_{ij}\,\epsilon(a,b,i,j) + c_a\,\epsilon(a,1,2,3) + c_b\,\epsilon(b,1,2,3), 
	\label{eq:mj-eps3} 
\end{equation} 
where $i,j\in\{1,2,3\}$ and the coefficients $c$ are parity-even scalar functions determined by the matrix element. Contractions containing the Higgs momentum introduce no new structures, since they can be eliminated with momentum conservation. 
The five contractions are related by Schouten identities, so that in four dimensions they can all be written as one contraction multiplied by rational functions of the scalar products. These functions become singular in parts of phase space, so we keep all five structures. 
In general, a process with $n$ final-state partons admits $\binom{n+2}{4}$ such contractions: one for two partons, five for three and fifteen for four. 

The two-jet optimal observable evaluated on the two leading jets retains only the single invariant $\epsilon(a,b,1,2)$.
The grouped observable of ref.~\cite{th:andersen_arnold_zeppenfeld} constructs two pseudo-jets, $q_A=\sum_{j\in\mathcal{A}}p_j$ and $q_B=\sum_{j\in\mathcal{B}}p_j$, from the sets $\mathcal{A}$ and $\mathcal{B}$ of jets with $y_j<y_H$ and $y_j>y_H$, and evaluates the two-jet observable on these momenta. Its CP-odd invariant is 
\begin{equation} 
	\epsilon(a,b,q_A,q_B) = \sum_{i\in\mathcal{A},\,j\in\mathcal{B}} \epsilon(a,b,i,j). 
	\label{eq:mj-grp} 
\end{equation} 
Compared with eq.~\eqref{eq:mj-eps3}, this construction retains only one particular combination of the five invariants. Contractions involving two jets on the same side of the Higgs boson are discarded, as are the terms $\epsilon(a,1,2,3)$ and $\epsilon(b,1,2,3)$, and the individual coefficients $c_{ij}$ are replaced by an effective coefficient that depends only on the grouped momenta.

The motivation for the grouping is the high-energy limit of the amplitude, where the CP-sensitive correlation is determined by the azimuthal angle between the transverse-momentum sums on the two sides of the Higgs boson~\cite{th:andersen_arnold_zeppenfeld}. Away from that limit, the kinematics of the individual jets can carry additional CP information that the grouping does not retain. The loss is expected to be largest for configurations with a hard central third jet, small rapidity separation, or multiple jets on the same side of the Higgs boson. Section~\ref{sec:truth} quantifies the resulting reduction in sensitivity.

The natural way to recover this information is to evaluate the optimal observables with the matrix element of the three-parton process. For events with a third jet we define
\begin{equation}
  \OOone^{(3)}=\frac{2\,\mathrm{Re}(\mathcal{M}_E\mathcal{M}_O^*)}{|\mathcal{M}_E|^2}\bigg|_{gg\to Hggg},\qquad
  \OOtwo^{(3)}=\frac{|\mathcal{M}_O|^2}{|\mathcal{M}_E|^2}\bigg|_{gg\to Hggg}.
  \label{eq:mj-oo3}
\end{equation}
Unlike the two-jet observables, it contains all five invariants of eq.~\eqref{eq:mj-eps3}, each weighted with its coefficient from the matrix element, and therefore retains the additional CP information carried by the third jet. We evaluate it with the color- and helicity-summed heavy-top amplitudes of the effective interaction of eq.~\eqref{eq:leff}, the same limit in which the signal is generated, using our own implementation of the $gg\to Hgg$ and $gg\to Hggg$ amplitudes (appendix~\ref{app:obs-psp}, and refs.~\cite{DelDuca:2004wt,Dixon:2004za,Badger:2004ty} for the $H+3$ parton amplitudes). The multiplicity-matched optimal observable uses the $gg\to Hgg$ matrix element for events with two jets and eq.~\eqref{eq:mj-oo3} on the three leading jets otherwise. Its construction at reconstruction level is described in section~\ref{sec:obs-oo3}.

\subsection{Optimal observables from reconstructed jets}\label{sec:observables}\label{sec:obs-oo2}\label{sec:obs-oo3}

The optimal observables of sections~\ref{sec:dphi-corr} and~\ref{sec:mj-oo3} are exact functions of the momenta of the hard partons,
whereas the analysis observes photons and jets at the detector level.
To evaluate them on reconstructed events, we take the diphoton four-momentum and the jets with $\pT>30\GeV$ and $|y|<4.4$ of the
event selection (section~\ref{sec:selection}), boost the final state to zero transverse momentum, reconstruct the incoming momenta
from energy and longitudinal-momentum conservation, and treat all partons as gluons in the heavy-top matrix elements.
The number of jets above this threshold, however, is in general not the number of hard partons, as figure~\ref{fig:obs-migration}
shows for the selected signal events.
Events generated with three hard partons make up $56\%$ of all selected signal events. Of these, $47\%$ are reconstructed with two jets,
$36\%$ with three jets and $17\%$ with four or more, the additional jets coming from radiation in the parton shower. Events
generated with two hard partons are reconstructed with two jets in $96\%$ of the cases.
Overall, $68\%$ of the selected events contain exactly two jets, $22\%$ contain three jets and $10\%$ contain four or more.
Mapping the jets onto the two or three partons of a matrix element therefore has no unique solution, and we compare three approaches.
\begin{itemize}
  \item In the first approach, the two-jet optimal observables are evaluated with the two jets of highest transverse momentum, and all other jets are ignored.
  ATLAS selects jets in the same way for its $\dphijj$ measurement~\cite{th:atlas_yy_139_diff}, and this
  appears to be the choice of CMS for its discriminants constructed from the $H+2$ parton matrix elements of
  JHUGen~\cite{th:gritsan_2016} as well.\footnote{CMS uses the two jets with the highest transverse momentum in the $H\to4\ell$
  analysis~\cite{th:cms_4l_anom} (section~4.1 of that paper). In the $H\to\tau\tau$ and $H\to\gamma\gamma$
  analyses~\cite{th:cms_tautau_anom,th:cms_yy_anom}, CMS does not state explicitly which jets are used to compute the optimal
  observable.}
  \item In the second approach, the two-jet optimal observables are evaluated with two pseudo-jets, constructed as the momentum sums of the jets on either side of the diphoton rapidity~\cite{th:andersen_arnold_zeppenfeld}, with the two leading jets used when one side is empty, so that all jets contribute. These grouped two-jet optimal observables are more sensitive than those on the two leading jets (sections~\ref{sec:truth} and~\ref{sec:results}).
  \item In the third approach, events with at least three jets are evaluated with the three-jet optimal observable of eq.~\eqref{eq:mj-oo3} on the three leading jets, while events with exactly two jets keep the grouped two-jet observables. Each event is thus described by a matrix element of its own jet multiplicity up to three, and jets beyond the third are ignored.
  Among the jet assignments compared in appendix~\ref{app:obs-jets} with the same $30\GeV$ jet threshold, no other rule improves on it by more than about $0.2\%$ in $\sigma(\bt)$.
\end{itemize}

\begin{figure}[tbp]
  \centering
  \includegraphics[width=0.62\linewidth]{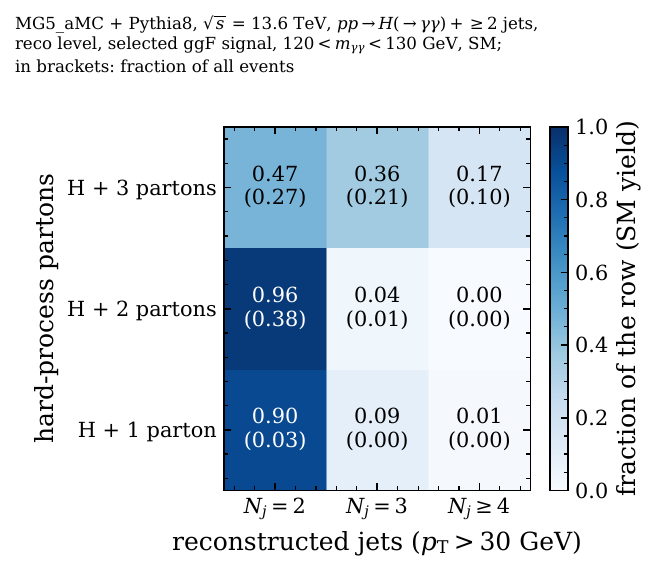}
  \caption{Number of hard-process partons of the MLM-merged signal sample against the number of reconstructed jets with
  $\pT>30\GeV$, for the selected ggF signal at the SM ($120<\myy<130\GeV$). Each row is normalized to one, and the numbers in brackets are
  the fractions of the selected events with at least one hard parton. Events with no hard parton ($1\%$ of the selected signal) are not shown.}
  \label{fig:obs-migration}
\end{figure}

At particle and detector level, however, none of these observables is optimal, since the jets differ from the partons whose
momenta enter the matrix elements.
The parton shower changes the jet momenta and their correspondence to the partons, and the flavors of the partons are unknown,
although subprocesses with at least one quark carry about four fifths of the information of the three-parton events
(section~\ref{sec:truth-flavour}).
In addition, the three-parton events reconstructed with two jets, $27\%$ of all selected events, lose their third parton, and
lowering the jet threshold to $20\GeV$ reduces this migration from $47\%$ to $21\%$. The information carried by such a parton is not
accessible to any of the three approaches. The larger loss, however, comes from the difference between jets and partons, which
affects the events with a third jet as well, as section~\ref{sec:truth} quantifies.
Because the templates of the fit are built with the same procedure from the full simulation, these approximations reduce the
sensitivity but do not bias the measurement, as long as the simulation describes the jets in data.

The observable that is optimal for the observed event is instead the ratio of the expectation values of the weights $w_I$ and
$w_O$ to that of $w_E$, given the observed jets and photons, where the expectation runs over all hard-parton configurations that lead
to the same observed event.
These expectation values cannot be computed from a matrix element, but they can be learned from the simulated events, which is the
subject of the next section.

\section{Learning optimal observables with transformers}\label{sec:mj-learned}\label{sec:obs-learned}

\subsection{Motivation}\label{sec:ml-motivation}

In section~\ref{sec:observables} we argued that, for an observed event, the optimal observables are the ratios of the weights $w_I$
and $w_O$ to $w_E$, each weight averaged over all hard-parton configurations that could have produced the observed jets and photons.
For example, when the third parton falls below the jet threshold, its direction is unknown, and the average runs over all the
directions it could have had.
Computing these averages directly would require integrating over the parton shower, hadronization and detector response, as in the
matrix element method~\cite{Kondo:1988yd,Artoisenet:2010cn,Brochet:2018pqf}. Its transfer functions, however, assume a one-to-one match
of jets and partons and therefore miss jets lost below the threshold or created by the shower.
The simulation contains this information implicitly: for every simulated event we know both the reconstructed jets and photons and the
hard partons that produced them, together with their matrix-element weights.
These ratios of averaged weights can therefore be learned from the simulated events as a function of the reconstructed event, in the same way
that a fit through many scattered data points learns their average.

A function of this kind must accept a varying number of jets, must give the same result irrespective of the order in which the jets
are listed, and must be able to construct from the jet momenta the CP-odd Levi-Civita contractions of eq.~\eqref{eq:mj-eps3}
that carry the interference.
Transformer networks satisfy the first two requirements by construction, since they act on unordered sets of particles of variable
size~\cite{Qu:2022mxj,Spinner:2024hjm}. How the two networks used in this work satisfy the third is described below.

\subsection{Symmetries and learning objectives for the network}\label{sec:ml-objectives}

For each reconstructed event $x$, the networks return two numbers.
The first, $f_E(x)$, estimates the fraction of the sum of the $\at^2$ and $\bt^2$ terms of eq.~\eqref{eq:morph} that arises from the
$\at^2$ term, so that $1-f_E(x)$ is the corresponding fraction of the $\bt^2$ term.
The second, $g(x)$, estimates half the $\at\bt$ interference term relative to the same sum; its sign indicates in which direction the
interference distorts the event.
Both are trained on the simulated events to reproduce targets computed from the matrix-element weights of the hard partons of the
same event,
\begin{equation}
  t_E=\frac{w_E}{w_E+w_O}\in[0,1],\qquad t_I=\frac{w_I}{2(w_E+w_O)}\in[-\tfrac12,\tfrac12],
  \label{eq:mj-targets}
\end{equation}
by minimizing the squared difference between output and target, summed over the simulated events with the weight $w_E+w_O$ of each
event.
Up to the normalization of the CP-odd amplitude, the targets coincide with the discriminants $D_{0-}$ and $D_{CP}$ that CMS evaluates with the
$H+2$ parton matrix elements on two jets~\cite{th:gritsan_2016}, so that the networks learn a weighted average of these discriminants over
all hard-parton configurations consistent with the reconstructed event.
Since many hard-parton configurations lead to similar reconstructed events, the targets scatter for a given $x$, and the output that
minimizes the squared difference is their weighted average~\cite{Brehmer:2018kdj,Brehmer:2018eca}.
The learned optimal observables then follow as
\begin{equation}
  \rhoI(x)=\frac{2g(x)}{f_E(x)}=\frac{\langle w_I\rangle_x}{\langle w_E\rangle_x},\qquad
  \rhoO(x)=\frac{1-f_E(x)}{f_E(x)}=\frac{\langle w_O\rangle_x}{\langle w_E\rangle_x},
  \label{eq:mj-rho}
\end{equation}
where $\langle\cdot\rangle_x$ denotes the average over all hard-parton configurations that lead to the reconstructed event $x$.
At the level of the hard partons, where $x$, including the parton flavors, determines the configuration, $\rhoI$ and $\rhoO$ reduce to the
ratios $w_I/w_E$ and $w_O/w_E$ of section~\ref{sec:oo}.
The ratios could also be learned directly with the weight $w_E$, but they become very large where the SM amplitude nearly vanishes,
whereas the weight $w_E+w_O$ treats the $\at^2$ and $\bt^2$ terms on an equal footing and keeps both targets bounded
(appendix~\ref{app:nets-design}).

Because $\rhoI$ and $\rhoO$ fix the relative size of the three terms of eq.~\eqref{eq:morph}, they determine the probability of an
event at every point of the $(\at,\bt)$ plane, so that a single training suffices for the full parameter space.
This is the morphing-aware construction of ref.~\cite{Brehmer:2018eca}, here with bounded targets and an exactly symmetric
parametrization. Parametrized estimators of the quadratic coefficients of an EFT likelihood ratio were also studied in
refs.~\cite{Chen:2020mev,Chen:2023ind}.
Methods that instead learn the derivative of the likelihood with respect to the couplings at a single reference
point~\cite{Brehmer:2018kdj,Brehmer:2018eca} would, at the SM, provide only $\rhoI$, and therefore only the information linear in
$\bt$.

The exact ratios $w_I/w_E$ and $w_O/w_E$ have two properties that the learned observables should share.
Under the reflection $P$ of all momenta in azimuth, $\phi\to-\phi$, defined in section~\ref{sec:dphi-corr}, the interference
weight $w_I$ changes sign while $w_E$ and $w_O$ are unchanged, so that $\rhoI$ is odd and $\rhoO$ is even under $P$.
In addition, the bound of eq.~\eqref{eq:th-bound} implies $\rhoI^2\le4\rhoO$, which guarantees that the predicted number of events
is positive for every value of $(\at,\bt)$.
A trained network satisfies these properties only approximately, and even a small violation of the first would mimic a CP-odd signal
in the SM.
We therefore impose both exactly, by evaluating the network on the event $x$ and on its mirror image $Px$ and combining its two raw
outputs $h_E$ and $h_I$ as
\begin{equation}
  f_E=\mathrm{sigmoid}\!\left(\tfrac12[h_E(x)+h_E(Px)]\right),\qquad
  g=\tanh\!\left(\tfrac12[h_I(x)-h_I(Px)]\right)\sqrt{f_E(1-f_E)} .
  \label{eq:mj-param}
\end{equation}
The sum over $x$ and $Px$ makes $f_E$, and hence $\rhoO$, exactly even, and the difference makes $g$, and hence $\rhoI$, exactly odd.
The factor $\sqrt{f_E(1-f_E)}$ enforces $\rhoI^2\le4\rhoO$, and the sigmoid and hyperbolic tangent keep both outputs within the
ranges of their targets in eq.~\eqref{eq:mj-targets}.
For a detector that is symmetric under the reflection, the ratios of averaged weights of eq.~\eqref{eq:mj-rho} have both properties as well, so
that imposing them does not restrict what the networks can learn. Residual asymmetries of a real detector are discussed in
section~\ref{sec:st-zsym}.

\subsection{Network architectures}\label{sec:ml-arch}

Each event is presented to the networks as a set of objects, referred to as tokens, which consists of the two photons, the diphoton
system and up to six jets with $\pT>30\GeV$ and $|y|<4.4$.
Every token is described by its four-momentum and its type, without further information on the object, so that the networks have
access to the same kinematic information as the analytic optimal observables, extended to all jets of the event.

The networks are trained on signal events, whose diphoton mass is close to $125\GeV$, but are applied to all selected events with
$105<\myy<160\GeV$.
At evaluation we therefore rescale the photon four-momenta of every event to $\myy=125\GeV$, keeping their directions and the
ratios $\pT/\myy$, so that the learned observables do not depend on $\myy$ directly.
The kinematics of the continuum still change with $\myy$, and the shape of its $\rhoO$ distribution differs between the sidebands
and the fit window, while that of $|\rhoI|$ does not.
The continuum templates are therefore taken from the simulation, and the sidebands fix only the continuum normalization of each
category; in an experiment they would also measure any residual P-odd asymmetry of the detector (section~\ref{sec:stat}).

\paragraph{Particle Transformer.}
The first network is the Particle Transformer (ParT)~\cite{Qu:2022mxj}, which was developed for the identification of jets from
their constituents.
In a transformer, the description of each token is updated repeatedly with information from all other tokens, weighted by attention
scores that the network computes from the tokens themselves.
In ParT, these scores receive an additional contribution computed from physical features of each pair of tokens, such as their
separation in rapidity and azimuth and their invariant mass.
Among these pair features is $\sin\Delta\phi$, which changes sign under the reflection $P$.
The contractions $\epsilon(a,b,i,j)$ of eq.~\eqref{eq:mj-eps3} are proportional to $p_{\mathrm{T},i}\,p_{\mathrm{T},j}\sin\Delta\phi_{ij}$,
and the remaining ones can be built from it together with the other pair features.
Figure~\ref{fig:obs-arch} (top) summarizes the network.
The features of each token are mapped to a vector of 64 numbers, three attention blocks with the pair-feature contribution update these vectors, a final attention step combines them into
a single vector that summarizes the event, and a small fully connected network maps this vector to the two outputs $h_E$ and $h_I$ of
eq.~\eqref{eq:mj-param}.
Before being passed to the network, each event is rotated about the beam axis so that the diphoton system has $\phi=0$. The
observables do not depend on the overall azimuthal orientation of the event, and the network then does not have to learn this
invariance.
The token features and the size of the network are given in appendix~\ref{app:nets-design}.

\paragraph{Lorentz-equivariant geometric-algebra transformer.}
The second network is the Lorentz-equivariant geometric-algebra transformer (L-GATr)~\cite{Spinner:2024hjm,Brehmer:2024yqw}.
Whereas ParT receives features chosen by hand, L-GATr is constructed so that every operation of the network respects Lorentz
symmetry.
Each token is represented by a multivector of the spacetime geometric algebra, an object that combines a scalar, a four-vector,
antisymmetric tensors and a pseudoscalar, with the four-momentum of the token as its vector part.
The layers of the network combine these objects only in Lorentz-covariant ways, and products of four vectors yield their Levi-Civita
contraction $\epsilon(p_i,p_j,p_k,p_l)$ as a pseudoscalar.
L-GATr therefore constructs the CP-odd structures of eq.~\eqref{eq:mj-eps3} directly from the momenta, whereas ParT has to
assemble them from its pair features.
We use the version of L-GATr that also respects parity~\cite{Brehmer:2024yqw}, so that scalar and pseudoscalar quantities are never
mixed inside the network.

As summarized in figure~\ref{fig:obs-arch} (bottom), the scalar and pseudoscalar parts of the output multivectors of each token,
together with a set of scalar output channels, are averaged over the tokens and passed to a small fully connected network, which
returns $h_E$ and $h_I$.
This network combines scalar and pseudoscalar inputs in both outputs, and, as for ParT, the exact parity properties of $\rhoI$ and
$\rhoO$ follow from the symmetrization of eq.~\eqref{eq:mj-param}.
Since the colliding beams single out a direction in space, three fixed reference vectors, the two beam directions and the time
direction, are added to every token.
They are unchanged by rotations about the beam axis and by the reflection $P$, so that, unlike ParT, the network requires no rotation
of the event, and its pseudoscalar outputs change sign under $P$ exactly.
The settings of the network are given in appendix~\ref{app:nets-design}.

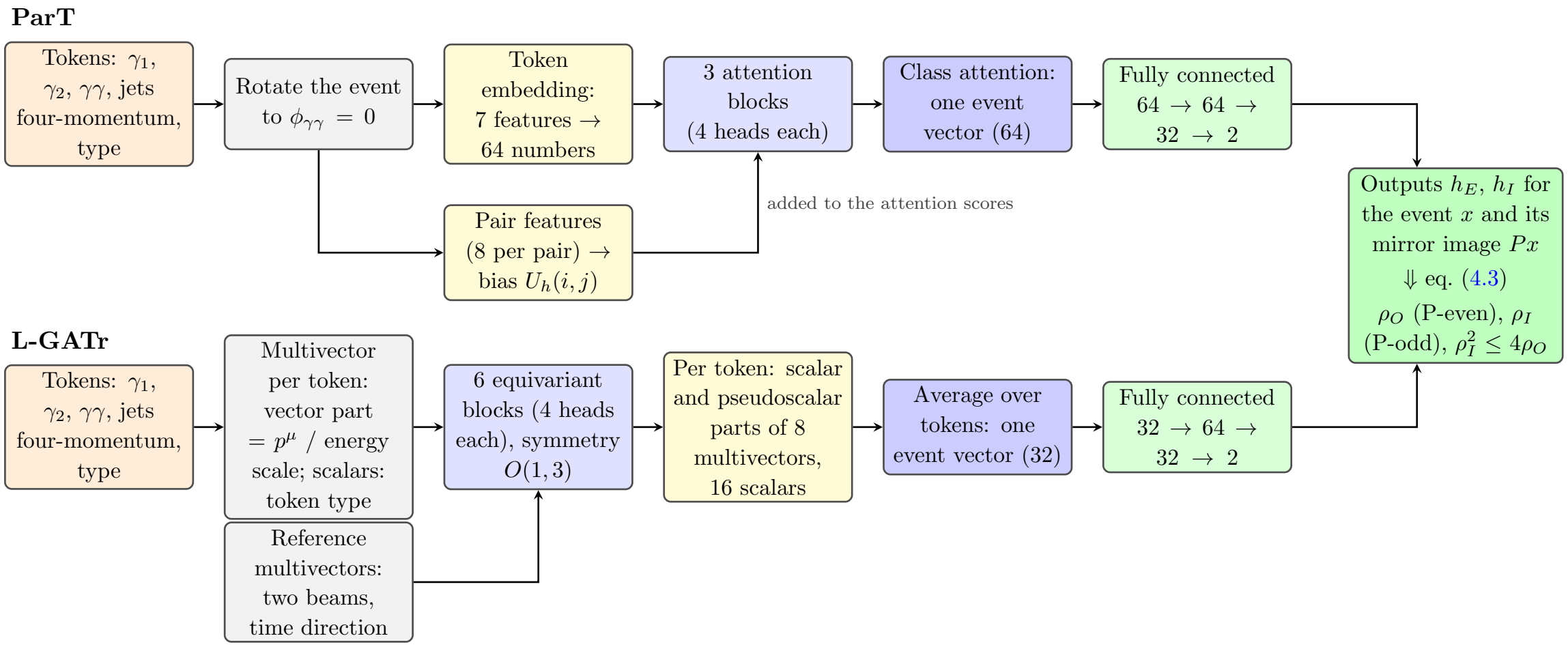
\begin{figure}[tbp]
  \centering
  \resizebox{\linewidth}{!}{%
  \begin{tikzpicture}[box/.style={rectangle,rounded corners=3pt,draw=black!70,minimum height=14mm,text width=27mm,align=center,
                      font=\small,inner sep=3pt,execute at begin node={\hyphenpenalty=10000\exhyphenpenalty=10000}},lab/.style={font=\scriptsize,text=black!75},>=stealth,thick]
  \node[font=\bfseries,anchor=west] at (-1.5,1.35) {ParT};
  \node[box,fill=orange!15] (in) at (0,0) {Tokens: $\gamma_1$, $\gamma_2$, $\gamma\gamma$, jets\\ four-momentum, type};
  \node[box,fill=black!5] (rot) at (3.4,0) {Rotate the event to $\phi_{\gamma\gamma}=0$};
  \node[box,fill=yellow!20] (tf) at (6.8,0) {Token embedding:\\ 7 features $\to$ 64 numbers};
  \node[box,fill=yellow!20] (pf) at (6.8,-2.3) {Pair features (8 per pair) $\to$ bias $U_h(i,j)$};
  \node[box,fill=blue!12] (blk) at (10.2,0) {3 attention blocks\\ (4 heads each)};
  \node[box,fill=blue!20] (cls) at (13.6,0) {Class attention: one event vector (64)};
  \node[box,fill=green!15] (head) at (17.0,0) {Fully connected\\ $64\to64\to32\to2$};
  \draw[->] (in) -- (rot); \draw[->] (rot) -- (tf); \draw[->] (tf) -- (blk); \draw[->] (blk) -- (cls); \draw[->] (cls) -- (head);
  \draw[->] (rot.south) |- (pf.west);
  \draw[->] (pf.east) -| node[lab,pos=0.75,right]{added to the attention scores} (blk.south);
  \node[font=\bfseries,anchor=west] at (-1.5,-3.65) {L-GATr};
  \node[box,fill=orange!15] (lin) at (0,-5) {Tokens: $\gamma_1$, $\gamma_2$, $\gamma\gamma$, jets\\ four-momentum, type};
  \node[box,fill=black!5] (emb) at (3.4,-5) {Multivector per token: vector part $=p^\mu$ / energy scale; scalars: token type};
  \node[box,fill=black!5] (spur) at (3.4,-7.4) {Reference multivectors: two beams, time direction};
  \node[box,fill=blue!12] (lblk) at (6.8,-5) {6 equivariant blocks (4 heads each), symmetry $O(1,3)$};
  \node[box,fill=yellow!20] (feat) at (10.2,-5) {Per token: scalar and pseudoscalar parts of 8 multivectors, 16 scalars};
  \node[box,fill=blue!20] (pool) at (13.6,-5) {Average over tokens: one event vector (32)};
  \node[box,fill=green!15] (lhead) at (17.0,-5) {Fully connected\\ $32\to64\to32\to2$};
  \draw[->] (lin) -- (emb); \draw[->] (emb) -- (lblk); \draw[->] (spur.east) -| (lblk.south);
  \draw[->] (lblk) -- (feat); \draw[->] (feat) -- (pool); \draw[->] (pool) -- (lhead);
  \node[box,fill=green!25,text width=31mm,minimum height=24mm] (out) at (21.0,-2.5)
       {Outputs $h_E$, $h_I$ for the event $x$ and its mirror image $Px$\\[2pt] $\Downarrow$ eq.~\eqref{eq:mj-param}\\[2pt]
        $\rhoO$ (P-even), $\rhoI$ (P-odd), $\rhoI^2\le4\rhoO$};
  \draw[->] (head.east) -| ([xshift=-6mm]out.north) ;
  \draw[->] (lhead.east) -| ([xshift=-6mm]out.south);
  \end{tikzpicture}}
  \caption{Architectures of the two networks that learn the optimal observables. The upper row shows the Particle Transformer (ParT),
  in which eight features of each pair of tokens enter the attention scores as a bias $U_h(i,j)$. The lower row shows L-GATr, with
  reference multivectors for the two beams and the time direction attached to every token. Both networks are evaluated on $x$ and on
  its mirror image $Px$, and eq.~\eqref{eq:mj-param} combines the outputs into an exactly P-even $\rhoO$ and an exactly P-odd $\rhoI$.
  The settings are given in appendix~\ref{app:nets-design}.}
  \label{fig:obs-arch}
\end{figure}

We use two networks to test whether building the symmetries of the problem into the architecture improves the learned
observables.
Parity alone cannot single out an architecture, since eq.~\eqref{eq:mj-param} makes the outputs of any network exactly odd or even
under $P$.
L-GATr adds Lorentz equivariance, whose benefit in the regression of $q\bar q\to Z+n\,g$ amplitudes grows with the number of
gluons~\cite{Spinner:2024hjm}.
Trained with identical inputs, targets and events (a few hundred thousand), the two networks reach the same $\sigma(\bt)$ within
1--2\% at every level (sections~\ref{sec:truth} and~\ref{sec:results}).
On the hard partons both reach the binned exact ratios, and at particle and reconstruction level both are limited by the information
that survives the shower and the detector.

\subsection{Training}\label{sec:ml-training}

The networks are trained on gluon-fusion signal events that pass the diphoton selection and contain at least two jets, with the targets of section~\ref{sec:ml-objectives}, computed from
the helicity-summed weights of each event.
A network evaluated on its own training events describes them too well, and using the same events for the templates of the fit would
overestimate the sensitivity.
We therefore use a five-fold scheme, as in $k$-fold cross-validation.
The signal events are divided by generation job into five
subsets of equal size, and for each subset five networks with different initializations are trained on three of the other subsets,
with the fourth used to stop the training.
Every signal event is evaluated with the average of the five networks of its own subset, none of which saw it in training, so that
all signal events enter the templates. The background samples are evaluated with the average of all twenty-five networks.
The two averages differ per event by about 6\% of the spread of $\rhoI$, small compared with the bin widths of the templates
(appendix~\ref{app:nets-design}).
As for the analytic observables, an imperfect network reduces the sensitivity but does not bias the measurement, since the templates
are constructed with the same networks from the full simulation, provided that the simulation describes the data.
The training settings, the training curves, the calibration and closure tests, the spread between ensemble members and folds, and
the numerical checks of the exact symmetries are given in appendix~\ref{app:nets}.

\subsection{Distributions of the learned observables}\label{sec:ml-dist}

Figure~\ref{fig:rho2d-ptcl} shows the joint distribution of $\rhoI$ and $\rhoO$ at particle level, obtained with ParT networks
trained at that level and evaluated on events not used in their training.
All events lie above the parabola $\rhoO=\rhoI^2/4$, which follows from the bound of eq.~\eqref{eq:th-bound} and is built into the
networks through eq.~\eqref{eq:mj-param}.
In the SM, $49.9\%$ of the cross section lies at $\rhoI>0$, and $\rhoI$ is uncorrelated with $\rhoO$ (correlation coefficient $0.01$).
A pure CP-odd coupling, $(\at,\bt)=(0,1)$, moves events to large $\rhoO$, raising the fraction of the cross section at $\rhoO>3$ from
$7\%$ to $15\%$, but leaves the distribution symmetric in $\rhoI$; only the interference makes it asymmetric.
At $(1,1)$, $55\%$ of the cross section lies at $\rhoI>0$ and the correlation coefficient is $0.10$.
The two amplitudes add for events at $\rhoI>0$ and subtract for events at $\rhoI<0$, and on the negative branch of the parabola, where
they are fully aligned, they cancel completely at $\rhoO=1$ (figure~\ref{fig:rho2d-ptcl}(b)).
The point $(1,-1)$ gives approximately the mirror image ($45\%$ at $\rhoI>0$).
The sign of $\bt$ therefore appears as an asymmetry in $\rhoI$ and its size as an excess of events at large $\rhoO$, and the fit uses
both observables.
The corresponding distributions for the hard partons and at reconstruction level show the same pattern.

\begin{figure}[tbp]
  \centering
  \includegraphics[width=0.96\linewidth]{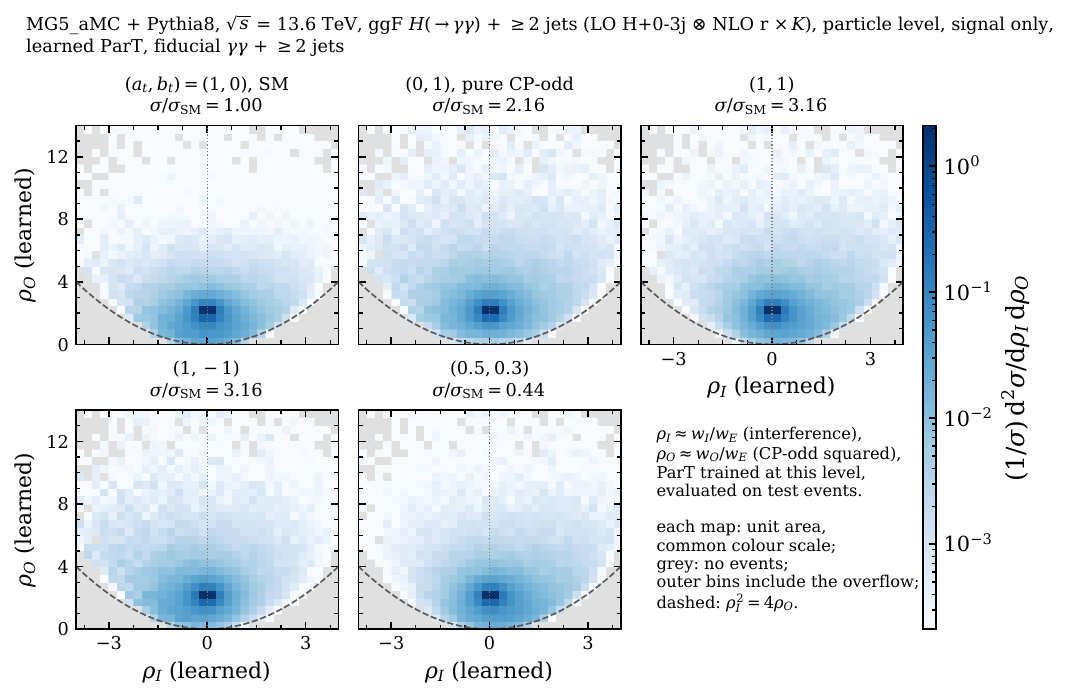}\\[2pt]
  (a)\\[4pt]
  \includegraphics[width=\linewidth]{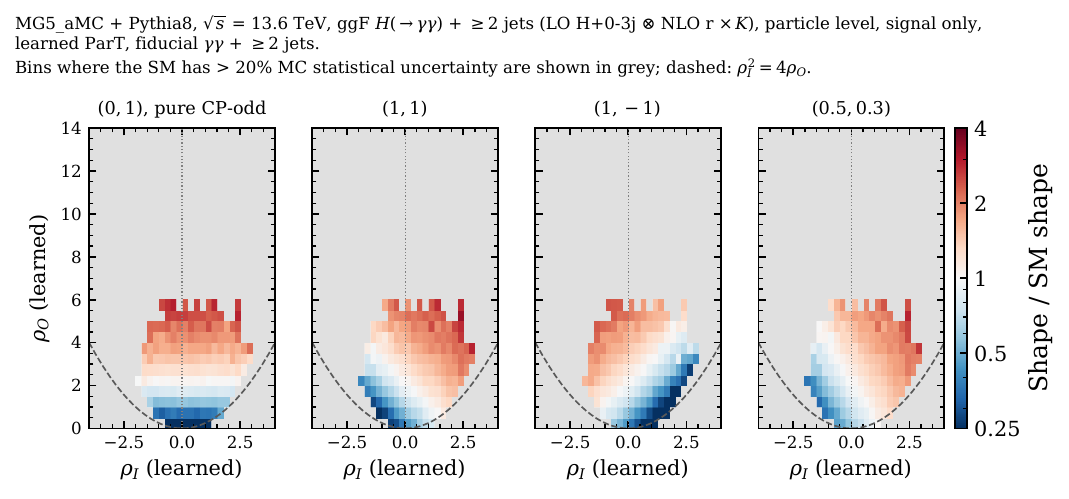}\\[2pt]
  (b)
  \caption{Learned observables $\rhoI$ and $\rhoO$ of ParT for gluon-fusion $H(\to\gamma\gamma)+\geq2$\,jets in the particle-level
  fiducial region, with networks trained at particle level and evaluated on events not used in their training. (a) Normalized joint
  distribution at $(\at,\bt)=(1,0)$, $(0,1)$, $(1,1)$, $(1,-1)$ and $(0.5,0.3)$, on a logarithmic color scale common to all panels.
  The panel titles give the fiducial cross section relative to the SM. The outer bins include the overflow, and gray bins contain no
  events. (b) Ratio of the normalized distribution to that of the SM for the four non-SM points, on a logarithmic color scale from
  $1/4$ to $4$. Gray bins have a statistical uncertainty of the simulated SM prediction above 20\%. In both rows the dashed line is the
  bound $\rhoI^2=4\rhoO$ and the dotted line marks $\rhoI=0$.}
  \label{fig:rho2d-ptcl}\label{fig:rho2d-ptcl-ratio}
\end{figure}

\section{Signal-purity categories}\label{sec:categories}

The observables of sections~\ref{sec:multijet} and~\ref{sec:mj-learned} were constructed and compared for the signal alone.
In the selected events, however, the continuum diphoton background exceeds the signal by a factor of about fifty
(section~\ref{sec:selection}), so the analysis must separate the signal from the background.
Any such separation acts on the same events from which the CP-sensitive observables are computed. A classifier trained on the SM
signal could prefer particular regions of these observables and thereby distort, or sculpt, the distribution of the CP-odd signal.
We therefore separate the signal from the background with classifiers whose output is even under the azimuthal reflection $P$, so
that they cannot select on the sign of any P-odd quantity.
We call them the signal-to-background (S/B) classifiers, and the networks of section~\ref{sec:mj-learned} that learn the optimal
observables the CP networks.
The nominal S/B classifier, the S/B transformer, has the L-GATr architecture of section~\ref{sec:ml-arch} but is trained to separate
the signal from the backgrounds.

\subsection{S/B classifiers}\label{sec:st-classifiers}

In the presence of background, the expected event density at the SM acquires an additional term $w_B$ from the background, and the
optimal observable for $\bt$ becomes $w_I/(w_E+w_B)$, the signal-only ratio $w_I/w_E$ multiplied by the signal purity
$w_E/(w_E+w_B)$ of the event.
Rather than suppress the continuum with further cuts, we therefore divide all selected events into categories of signal purity, which
approximates this product, and fit the CP-sensitive observables in each category (section~\ref{sec:stat}).

The nominal classifier is trained to separate three classes: the gluon-fusion signal, the diphoton continuum including the
electroweak $\gamma\gamma jj$ production and photon-jet fakes, and the other Higgs boson production modes.
Like the CP networks, the classifier receives the photons, the diphoton system and the jets of each event.
In addition, it receives the substructure variables of each jet, which separate quark from gluon jets; the $b$-tagging of each jet and
the missing transverse momentum, which help to identify $t\bar tH$ and $VH$ production; and the leading lepton, which identifies
leptonic $VH$ decays.
It uses the same rescaling of the photon momenta to $\myy=125\GeV$ (section~\ref{sec:ml-arch}), and the continuum $\myy$ spectrum
remains smooth in every category, without a peak at $125\GeV$.
The signal class is weighted with $w_E+w_O$, in which the interference is absent, so that the training is not tied to either CP
hypothesis.

From the class probabilities we form the discriminant $D$, an estimate of the signal purity $(w_E+w_O)/(w_E+w_O+w_B)$, where $w_B$
includes the continuum and the other Higgs boson production modes, and symmetrize it under the reflection $P$ as in
eq.~\eqref{eq:mj-param}, $D_{\mathrm{sym}}(x)=\tfrac12[D(x)+D(Px)]$, so that the categories cannot select on the sign of any P-odd
quantity.
The network is trained with the same five-fold scheme as the CP networks (section~\ref{sec:ml-training}).
Against the continuum, the area under the receiver operating characteristic curve is $0.800$, compared with $0.798$ for the S/B BDT
described below, so the two classifiers separate the signal from the continuum equally well.
Its categories keep the SM interference at zero within the statistics of the simulation, and a pure CP-odd signal differs from the
SM, apart from its $2.16$ times larger total rate, only in how the signal is shared among the categories, by up to $8\%$, which
the fit describes through the $\bt^2$ term of the templates.

As an independent check we also use a boosted decision tree (BDT) trained to separate the signal from the continuum.
This S/B BDT, labeled ``CP-safe BDT'' in the figures, is even under $P$ by construction, because all its inputs are even.
It is a chain of two gradient-boosted decision trees, the first separating the CP-mixed ggF signal from the continuum and the second
ggF from VBF, and only the first score, $s_1$, defines the categories. Its 35 inputs carry no interference information ($0.000\pm0.001$
of that of the signed $\dphijj$), and no combination of them reconstructs $\myy$.
Both classifiers were trained without pileup. Since both are exactly even under $P$, the choice between them cannot bias $\bt$.
With pileup the S/B BDT categories give slightly narrower intervals and are the nominal ones there (section~\ref{sec:res-pileup}).

\subsection{Categories}\label{sec:st-categories}

The output of each classifier is divided into three categories of increasing signal purity, with boundaries chosen to minimize the
expected uncertainty on $\bt$ rather than to maximize the significance of the signal.
Three categories provide about $90\%$ of the gain of six. The BDT boundaries are $s_1=0.66$ and $0.89$, and re-optimizing them for
the learned observables changes $\sigma(\bt)$ by less than $1\%$.
The transformer boundaries, $D_{\mathrm{sym}}=0.0222$ and $0.1028$, are optimized for the learned observables and then kept fixed, so that all observables in
section~\ref{sec:results} are compared with identical events and categories.
With boundaries optimized for them instead, the expected $\sigma(\bt)$ of the grouped two-jet optimal observables would be about
$1\%$ smaller.

Every selected event is assigned to exactly one category according to its classifier output, so that no event is discarded.
Within each category, the events are binned in a CP-sensitive observable, such as the learned or the optimal observables. This
yields for each category $c$ and bin $k$ the signal templates $E_{c,k}$, $I_{c,k}$ and
$O_{c,k}$ of eq.~\eqref{eq:fisher-nu} and the background templates $V_{c,k}$ (VBF), $X_{c,k}$ ($VH$ and $t\bar{t}H$) and $C_{c,k}$
(continuum), whose sum we denote by $B_{c,k}$.
The likelihood of section~\ref{sec:stat} is the product of the Poisson probabilities of all bins in all three categories, so that
$\bt$ is determined from all categories at once.

Table~\ref{tab:yields} lists the category boundaries and the expected numbers of events with $300\ifb$ in each category, for the SM
signal, a pure CP-odd signal, the other Higgs boson production modes and the continuum.
The ratio of the SM signal to the continuum increases from $0.011$ in the first category to about $0.3$ in the third.
The two purer categories carry almost all of the information on $\bt$: about half in the middle category ($53\%$ for ParT) and
$40\%$ in the purest one, which gains in purity what it loses in signal.
The least pure category holds $47\%$ of the signal but only $8\%$ of the information, because the large background there dilutes the
asymmetry.

\begin{table}[tbp]
  \centering
  \caption{Expected numbers of events with $300\ifb$ in the window $120<\myy<130\GeV$ in the three categories of the S/B transformer
  (discriminant $D_{\mathrm{sym}}$), which are nominal without pileup, and of the S/B BDT (first-stage score $s_1$),
  which are nominal with pileup. The column ggF SM gives the gluon-fusion signal
  at $(\at,\bt)=(1,0)$ and ggF$(0,1)$ the pure CP-odd signal. The continuum includes the QCD and electroweak production of
  $\gamma\gamma jj$ and the photon-jet fakes.}
  \label{tab:yields}
  \begin{fitwidth}%
  \begin{tabular}{lrrrrr}
   \toprule
   Category & ggF SM & ggF$(0,1)$ & VBF & $VH$, $t\bar{t}H$ & Continuum \\
   \midrule
   \multicolumn{6}{l}{\emph{S/B transformer}}\\
   $D_{\mathrm{sym}}<0.0222$          & 727.3 & 1453.4 & 180.8 & 203.2 & 64\,640 \\
   $0.0222$--$0.1028$                 & 689.0 & 1588.5 & 264.0 &  48.7 & 10\,536 \\
   $D_{\mathrm{sym}}\geq0.1028$       & 146.5 &  336.9 &  33.9 &   6.7 &     512 \\
   \midrule
   \multicolumn{6}{l}{\emph{S/B BDT}}\\
   $s_1<0.66$         & 727.1 & 1445.5 & 185.1 & 184.4 & 64\,638 \\
   $0.66$--$0.89$     & 648.2 & 1502.1 & 205.7 &  58.0 & 10\,280 \\
   $s_1\geq0.89$      & 187.5 &  431.2 &  87.8 &  16.1 &     769 \\
   \midrule
   Total (either classifier) & 1562.8 & 3378.7 & 478.7 & 258.5 & 75\,688 \\
   \bottomrule
  \end{tabular}
  \end{fitwidth}
\end{table}

\section{Statistical method}\label{sec:stat}

\subsection{Templates and binning}\label{sec:st-templates}\label{sec:st-zsym}

The fit compares the expected numbers of events in bins of the CP-sensitive observable, separately in each category.
For the optimal observables, which each consist of a P-odd quantity, $\OOone$ or $\rhoI$, and a P-even quantity, $\OOtwo$ or
$\rhoO$, we use two-dimensional histograms with 16 bins in the P-odd and 8 bins in the P-even quantity.
The bin edges are chosen so that every bin contains a similar number of simulated events in each category, which keeps the
statistical fluctuations of the simulation small in every bin.
In the P-odd quantity, the bins are placed symmetrically about zero, so that every bin at positive values has a mirror bin at
negative values.
The reflection $P$ of section~\ref{sec:dphi-corr} reverses the sign of the interference and leaves all other contributions
unchanged. In principle, the interference therefore enters the two bins of each pair with opposite signs, while all other
contributions are the same in both, which separates the interference from the rest of the prediction.
This holds because every step of the analysis is symmetric under $P$: the initial state, the detector simulation, the event
selection, the reweighting of the continuum to \sherpa{}, which uses only unsigned azimuthal information (section~\ref{sec:sim}), the
S/B classifiers (section~\ref{sec:st-classifiers}) and the CP-sensitive observables (section~\ref{sec:ml-objectives}).
The P-odd asymmetries of the background templates are consistent with the statistics of the simulation, the largest being 2.5
standard deviations in the continuum of the least pure category; mirroring the background templates in $\rhoI$ would move the fitted $\bt$
at the SM by $-0.034$, about one sixth of its statistical uncertainty.
The reflection $z\to-z$, which exchanges the beams, is not used by the fit: symmetrizing the continuum reweighting under it changes
the expected half-width at $300\ifb$ by $0.003$ for ParT, within the effect of the template statistics.
A real detector is only approximately symmetric under $P$, and a residual asymmetry would mimic interference. Such an asymmetry can be
measured in the $\myy$ sidebands, which contain almost only background, and corrected for or included as an uncertainty.
Our simulated detector is symmetric, so this effect is not included in our results.
The signed angles $\dphijj$ and $\dphigrp$ are binned in 24 bins of equal width between $-\pi$ and $\pi$.
The $16\times8$ binning passes the convergence and stability criteria of appendix~\ref{app:fitval-binning}, and for the angles
36 instead of 24 bins change the expected uncertainty on $\bt$ by only about $1\%$.

\subsection{Likelihood}\label{sec:st-likelihood}

The expected number of events in bin $k$ of category $c$ is
\begin{equation}
  \nu_{c,k}(\at,\bt) = \at^2E_{c,k}+\at\bt\,I_{c,k}+\bt^2O_{c,k} + B_{c,k},\qquad B_{c,k}=V_{c,k}+X_{c,k}+C_{c,k},
  \label{eq:nu}
\end{equation}
where $E$, $I$ and $O$ are the gluon-fusion templates of eq.~\eqref{eq:fisher-nu}, expressed as expected numbers of events for the
integrated luminosity considered, and the background $B$ consists of the VBF templates $V$, the $VH$ and $t\bar tH$ templates $X$,
and the continuum templates $C$.
The number of events in each bin follows a Poisson distribution, and the likelihood $\mathcal{L}$ is the product of the Poisson
probabilities over all bins and categories.
Since we determine the expected sensitivity, we evaluate the likelihood on the SM Asimov data set~\cite{cpfit:asimov}, in which the
number of events in every bin equals its SM expectation.
To constrain $\bt$ without an assumption on the CP-even coupling, we profile $\at$: for every value of $\bt$ we set $\at$ to the
value that maximizes the likelihood.
The prediction depends on the couplings only through $\at^2$, $\at\bt$ and $\bt^2$, so reversing the signs of both couplings leaves
it unchanged, and we take $\at\geq0$.
The 68\% and 95\% CL intervals on $\bt$ are the regions in which $-2\Delta\ln\mathcal{L}$ is below $1$ and $3.84$, and the
corresponding regions in the $(\at,\bt)$ plane are bounded by $2.30$ and $5.99$.
In this statistics-only fit the templates are exact and the background normalizations fixed.
Our nominal results add the systematic uncertainties of section~\ref{sec:st-syst}; the statistical uncertainties of the simulated
templates are discussed in section~\ref{sec:st-mcstat}.

\subsection{Systematic uncertainties}\label{sec:st-syst}

Systematic uncertainties are included through nuisance parameters $\bm\theta$, each constrained by a Gaussian distribution.
A nuisance parameter can change the normalization of a process, the shape of its template, or both, and the expected number of
events becomes
\begin{equation}
  \nu_{c,k}(\at,\bt,\bm\theta) = \sum_p f_p(\at,\bt)\,T_{p,c,k}(\bm\theta)\,\prod_j(1+\delta_{pj})^{\theta_j}\;\mu_{C,c}^{[p=C]} ,
  \label{eq:nusyst}
\end{equation}
where $p$ runs over the six components $E$, $I$, $O$, $V$, $X$ and $C$ of eq.~\eqref{eq:nu}, with the coefficients
$f_p\in\{\at^2,\at\bt,\bt^2,1,1,1\}$, since the couplings are varied only in gluon-fusion production (section~\ref{sec:theory}) and
the backgrounds are taken with their SM prediction.
For a shape uncertainty, the template $T_{p,c,k}$ is interpolated linearly between the nominal template at $\theta=0$ and the
templates obtained with the variation shifted up or down by one standard deviation at $\theta=\pm1$. The normalization
uncertainties $\delta_{pj}$ scale a process as a whole.
Each shape variation is propagated through the full analysis, including the event selection, the S/B classifier and the
CP-sensitive observables, and is applied to all processes and categories at once, so that the fit includes the resulting migration
of events between categories.
The normalization of the continuum is not taken from the simulation but is a free parameter $\mu_{C,c}$ in each category $c$
(the exponent $[p=C]$ is one for the continuum and zero otherwise). It is constrained by the number $n_{\mathrm{sb}}$ of continuum
events in the sidebands $105<\myy<120\GeV$ and $130<\myy<160\GeV$ of that category, with a relative uncertainty
$\sqrt{1/n_{\mathrm{sb}}+0.01^2}$, where $0.01$ accounts for the extrapolation from the sidebands into the window.
With all nuisance parameters fixed to zero, the fit reduces to that of section~\ref{sec:st-likelihood}.

The nuisance parameters and their impact on the expected uncertainty on $\bt$ are listed in table~\ref{tab:st-nuisances}.
For every observable of table~\ref{tab:st-nuisances} the largest impact, between $0.04$ and $0.05$ at $300\ifb$, comes from the
reduction of the SM azimuthal modulation at NLO by the factor $0.74$ of section~\ref{sec:sim-nlo},\footnote{Campbell et al.\ find almost no change of the $\dphijj$ distribution at NLO in a fixed-order calculation with a large
rapidity separation of the jets, $|\Delta\eta_{jj}|>4.2$~\cite{Campbell:2006xx}. Our reduction is obtained with less stringent cuts,
without a cut on the rapidity separation and with the photon selection of the $H\to\gamma\gamma$ analysis, and becomes smaller for
$|\Delta y_{jj}|>2$, where it is $0.85\pm0.09$ (section~\ref{sec:sim-nlo}).}
followed by the shape of the continuum.
A shift of this nuisance parameter by one standard deviation scales the interference and the azimuthal modulation of $E$ and $O$ by
$0.74$, and the opposite shift scales them by $1.26$.
Scaling the interference is conservative, since the NLO samples generated at the CP points show no reduction of the interference
term (section~\ref{sec:sim-nlo}).
For the grouped signed angle, which is not in the table, the jet energy scale and resolution ($0.081$) and the continuum shape
($0.078$) have larger impacts than the NLO modulation ($0.044$).
The photon energy scale and resolution enter as changes of the acceptance of the $\myy$ window for all Higgs boson processes.
The VBF background is taken with SM couplings, since a mismodeling of its P-even shape cannot mimic the interference.
No uncertainty is assigned to the jet directions, since the azimuthal resolution of jets above $30\GeV$ is small compared with the
period $\pi$ of the $\cos2\dphijj$ modulation.

\begin{table}[tbp]
  \centering
  \caption{Nuisance parameters of the fit with systematic uncertainties. In the column Type, ``N'' refers to a normalization
  uncertainty, whereas ``S'' refers to a shape uncertainty. The impact of a nuisance parameter is its contribution to the expected
  uncertainty on $\bt$, given by the 68\% CL half-width of the interval on $\bt$. It is obtained as the difference in quadrature
  between the half-widths with all nuisance parameters free and with this nuisance parameter fixed, in the categories of the S/B
  transformer. The impact is given at $300\ifb$ for the two-jet optimal observables on the two leading-$\pT$ jets (the reference) and on the
  grouped jets, the multiplicity-matched optimal observables and the learned ParT observables, and at $3000\ifb$ for ParT. Some
  nuisance parameters are evaluated together, and a single impact is given for each such group, namely for the jet energy scale and
  resolution, for the photon efficiency, energy scale and resolution and the luminosity, and for the continuum shape, the photon-jet
  fakes and the electroweak $\gamma\gamma jj$ production. The last row gives the impact of all nuisance parameters together, which is
  smaller than the quadrature sum of the rows because the nuisance parameters are correlated in the fit. The ggF cross-section
  uncertainties are those of ref.~\cite{Karlberg:2024zxx}.%
  }
  \label{tab:st-nuisances}
  \small
  \begin{fitwidth}%
  \begin{tabular}{>{\raggedright\arraybackslash}p{3.2cm}p{5.6cm}cccccc}
   \toprule
   Nuisance & Variation & Type & \multicolumn{4}{c}{Impact, $300\ifb$} & $3000\ifb$ \\
            &           &      & leading & grouped & matched & ParT & ParT \\
   \midrule
   Jet energy scale & $\pm\sqrt{0.01^2+(1\GeV/\pT)^2}$, $\oplus2\%$ for $|\eta|>2.5$, correlated over jets and events,
     propagated through selection, classifier and observables & S & \multirow{6}{*}{0.018} & \multirow{6}{*}{0.016} & \multirow{6}{*}{0.012} & \multirow{6}{*}{0.008} & \multirow{6}{*}{0.001} \\
   Jet energy resolution & extra smearing, $1.1\%$ at $30\GeV$ to $0.5\%$ at high $\pT$ & S & & & & & \\
   \cmidrule(lr){1-8}
   Photon efficiency & $3\%$ (Higgs processes) & N & \multirow{6}{*}{0.005} & \multirow{6}{*}{0.005} & \multirow{6}{*}{0.004} & \multirow{6}{*}{0.004} & \multirow{6}{*}{0.001} \\
   Photon energy scale & $0.3\%$, applied as a change of the window acceptance in each category & N & & & & & \\
   Photon energy resolution & $10\%$, applied as a change of the window acceptance in each category & N & & & & & \\
   Luminosity & $2\%$ (Higgs processes) & N & & & & & \\
   \cmidrule(lr){1-8}
   ggF cross section & scale $^{+4.6}_{-6.7}\%$, PDF and $\alpha_s$ $3.2\%$ & N & 0.010 & 0.009 & 0.008 & 0.008 & 0.002 \\
   ggF jet-bin migration & $\pm20\%$ for $\geq3$ jets, 2 jets compensating (ggF yield fixed) & S & 0.001 & 0.006 & 0.006 & 0.006 & 0.003 \\
   NLO kinematic reweighting & $r(x)\to1$, shape only & S & 0.025 & 0.015 & 0.015 & 0.015 & 0.004 \\
   NLO azimuthal modulation & interference and modulation of $E,O$ scaled by $0.74$ at one standard deviation & S & 0.050 & 0.048 & 0.046 & 0.042 & 0.012 \\
   VBF, $VH$ and $t\bar{t}H$ & VBF $20\%$, $VH$ and $t\bar{t}H$ $7\%$ & N & 0.003 & 0.003 & 0.001 & 0.002 & 0.001 \\
   Continuum normalization & free per category, sideband constraint & N & 0.020 & 0.015 & 0.015 & 0.017 & 0.005 \\
   \cmidrule(lr){1-8}
   Continuum shape & \sherpa\ reweighting $r\to1$, shape only & S & \multirow{4}{*}{0.038} & \multirow{4}{*}{0.029} & \multirow{4}{*}{0.027} & \multirow{4}{*}{0.024} & \multirow{4}{*}{0.006} \\
   Photon-jet fakes & $\pm25\%$ of the fake contribution & N & & & & & \\
   Electroweak $\gamma\gamma jj$ & $\pm20\%$ & N & & & & & \\
   \midrule
   All nuisances & & & 0.067 & 0.059 & 0.056 & 0.051 & 0.013 \\
   \bottomrule
  \end{tabular}
  \end{fitwidth}
\end{table}

\paragraph{Parton distributions.}
The simulated samples use the NNPDF2.3 parton distributions~\cite{sim:nnpdf23}. Reweighting every event to the CT25 NNLO
set~\cite{sim:ct25} makes the expected $\sigma(\bt)$ of the learned ParT observables, the grouped two-jet optimal observables and the
grouped angle $3$--$3.5\%$ smaller, because CT25 predicts slightly more quark-initiated events, which carry more information on $\bt$.
The CT25 uncertainties, included as shape nuisance parameters, change $\sigma(\bt)$ by less than $0.1\%$, and a fit to CT25 Asimov data
returns $|\hat b_t|\le0.006$. The choice of parton distributions therefore does not bias $\bt$, and its uncertainty is negligible.

\subsection{Uncertainty from limited statistics of the simulation}\label{sec:st-mcstat}

The templates are obtained from a finite number of simulated events and therefore fluctuate statistically, the continuum templates
most of all.
Because the interference is extracted from the difference between mirror bins (section~\ref{sec:st-templates}), a statistical
fluctuation of the simulated templates between two mirror bins looks like interference.
We compare three treatments of this effect.
\begin{itemize}
\item \emph{Method A.} The templates are used as generated by the simulation, without considering their statistical uncertainty, as
described in section~\ref{sec:st-likelihood}.
\item \emph{Method B, bootstrap.} The fit is repeated 200 times with templates in which every simulated event receives a random
weight drawn from a Poisson distribution with mean one, and the spread of the fitted $\bt$ is added in quadrature to the statistical
uncertainty.
\item \emph{Method C, Barlow--Beeston-lite.} One nuisance parameter per bin describes the statistical uncertainty of the total
prediction~\cite{Conway:2011in}.
\end{itemize}
The background templates are not symmetrized in the P-odd observable, since this would remove the very effect that these methods
estimate.

At $300\ifb$, method~B widens the 68\% interval by 4--5\% for the learned, matched and grouped two-jet observables, for example
from $0.204$ to $0.213$ for ParT.
The variance from the simulated templates is about 9\% of that from the data and comes mainly from the simulated $\gamma\gamma$+jets sample, and less from
the photon-jet fakes.
At $3000\ifb$ the two variances are about equal, and the intervals widen by about 40\%, for example from $0.064$ to $0.089$ for ParT.
The $3000\ifb$ intervals quoted in this paper, obtained with method~A, therefore assume about ten times more simulated continuum than
is available here.
Method~B leaves the ordering of these observables unchanged; the two-jet optimal observables on the two leading jets were not tested
with methods B and C.
Injected Asimov data at three non-SM points, $(\at,\bt)=(1,0.3)$, $(0.9,-0.3)$ and $(0.8,0.45)$, are recovered exactly, the 68\% (95\%) intervals of the grouped two-jet optimal
observables in the BDT categories cover in 66\% (95\%) of 1000 SM pseudo-experiments at $300\ifb$, and samples generated directly
at $(\at,\bt)=(1,\pm1)$ confirm the sign of the interference.

\section{The CP information from hard partons to reconstructed events}\label{sec:truth}

How much information on $\bt$ do the jets beyond the two leading-$\pT$ jets carry, and how much of it survives the parton shower,
hadronization and detector response?
To answer this question, we compare, for the signal alone, the analytic optimal observables $(\OOone,\OOtwo)$ of
section~\ref{sec:multijet}, the learned observables $(\rhoI,\rhoO)$ of the transformer networks of section~\ref{sec:mj-learned}, and
the signed angles $\dphijj$ and $\dphigrp$. The analytic observables are evaluated with the two leading-$\pT$ jets, with the grouped
jets and with the multiplicity-matched matrix element.
Each observable is evaluated separately on the hard partons, on jets at particle level and on reconstructed jets, and the learned
observables are also trained separately at each level.
At each level we determine the fraction $F/F_{\mathrm{exact}}$ of the information on $\bt$ that the observable retains, where $F$ is
the Fisher information of eq.~\eqref{eq:fisher} for the binned observable and $F_{\mathrm{exact}}$ that of the exact ratios,
eq.~\eqref{eq:fisher-exact}.
The information is evaluated for SM couplings, $(\at,\bt)=(1,0)$, and therefore measures how well each observable can detect a small
CP-odd coupling in addition to the SM one.
The pairs of observables are binned in $16\times8$ bins as in the fit, and the signed angles in 16 bins.
Binned in the same way, the exact ratios $w_I/w_E$ and $w_O/w_E$ retain $0.888$ of the unbinned information, the most that any of
these observables can reach.
Our reference is the two-jet optimal observables on the two leading-$\pT$ jets, which correspond to current experimental practice
(section~\ref{sec:observables}).

\subsection{The information ladder}\label{sec:truth-ladder}

The hard partons, whose matrix-element weights are known exactly, carry the maximum information on $\bt$, which the parton shower,
hadronization and detector response can only reduce.
We follow the fraction retained by each observable through these three levels, which we call the information ladder.

\label{sec:truth-location}
The signal sample is generated by merging matrix elements with up to three hard partons (section~\ref{sec:sim}), so that the number
of partons in the hard process of every simulated event is known.
These hard partons are the matrix-element partons above the generation cut $x_{\mathrm{qcut}}=15\GeV$.
Since this cut lies below the jet threshold of $30\GeV$, some third partons give no jet even before the shower, which lowers the
fraction of three-jet events (quantified below).
The parton shower lowers it further, mainly because final-state radiation outside the jet radius $R=0.4$ makes jets softer than
their partons, which the underlying event only partly compensates: $57\%$ of the three-parton events in the particle-level fiducial
region have at least three jets above $30\GeV$, compared with $61\%$ without final-state radiation and $55\%$ without the underlying
event.
Since switching off final-state radiation also changes the acceptance of the MLM merging, these numbers are indicative.
We use the number of hard partons to divide the events into classes, independently of the number of jets into which the partons are later
reconstructed. An event thus keeps its class at all three levels of the ladder, even when its third parton is not reconstructed as
a jet.
Events generated with three hard partons make up $56\%$ of the selected signal but carry $68\%$ of the information, while events with
two hard partons carry the remaining $32\%$.
Events with zero or one hard parton, which pass the selection only because the parton shower produces additional jets, make up $5\%$
of the selected signal and carry no information on $\bt$ at linear order, since their interference term vanishes
(section~\ref{sec:sim}).
The information is therefore concentrated in the events with a third hard parton, precisely those that the two-jet observables
describe only approximately, and we show the ladder separately for events generated with two and with three hard partons.
After reconstruction, however, events with at least three jets make up only $32\%$ of the signal and carry $39\%$ of the information,
since almost half of the three-parton events are reconstructed with two jets (figure~\ref{fig:obs-migration}). This limits how much
of the information of the third parton can be recovered.
These shares count events by the number of reconstructed jets, whereas the shares above (56\% of the events and 68\% of the
information) count them by the number of hard partons. The difference between the two is the migration of three-parton
events into the two-jet category. In both counts the share of the information is larger than the share of events: an event with a
third hard parton, or with a third jet, carries on average about 1.2 times the information of an average selected event.

Figure~\ref{fig:truth-ladder} shows the fraction $F/F_{\mathrm{exact}}$ retained by each observable at the three levels, and
table~\ref{tab:truth-ladder} gives the numbers.
The comparison uses selected signal events in the window $120<\myy<130\GeV$ that were not used in the training of the networks and
that also pass the particle-level fiducial selection of section~\ref{sec:sim}, so that every event can be evaluated at all three
levels and all observables are compared on the same events.
For these events the learned observables come from a single set of networks, not from the average over the five-fold scheme of
section~\ref{sec:ml-training} that is used in the fit.

\paragraph{Hard partons.}
For events with two hard partons, the two-jet optimal observables on the two leading-$\pT$ jets and on the grouped jets coincide.
They are exact up to the treatment of all partons as gluons and retain $0.75$ of the information, compared with $0.82$ for the binned
exact ratios.
For events with three hard partons, the two-jet optimal observables of the two leading-$\pT$ jets retain only $0.41$ of the
information, against $0.90$ for the binned exact ratios.
More than half of the information accessible in these events is therefore lost.
These observables are sensitive to a single invariant, $\epsilon(a,b,1,2)$ of eq.~\eqref{eq:mj-eps3} (section~\ref{sec:mj-oo3}), whereas the
invariant of the grouped jets, eq.~\eqref{eq:mj-grp}, also contains the contractions that involve the third parton.
Grouping the jets therefore recovers part of the information ($0.57$), and the multiplicity-matched optimal observables, which
contain all five invariants, recover most of it ($0.82$).
The learned observables reach the binned exact ratios in both classes of events ($0.82$ and $0.90$), which is the closure test of
the regression. At this level the networks, like the exact ratios, receive the parton types (gluon, quark or antiquark); at particle
and reconstruction level they receive only the photons and the jet momenta.

\paragraph{Shower and detector.}
Most of the information is lost between the hard partons and the jets at particle level.
Part of this loss is the parton flavor, which is no longer available; since the flavor-blind multiplicity-matched optimal
observables retain $0.82$ of the information of three-parton events at the level of the hard partons, against $0.90$ for the exact
ratios, it accounts for at most about $0.08$ of it.
For three-parton events, the learned ParT observables retain $0.50$ at particle level, the multiplicity-matched optimal observables
$0.44$, and the two-jet optimal observables $0.34$ with the grouped jets and $0.30$ with the two leading-$\pT$ jets.
The parametrized detector response of section~\ref{sec:sim} reduces these fractions by about another 10\%, to $0.46$, $0.39$, $0.31$
and $0.27$.
Grouping the jets therefore helps less after the shower: $0.31$ against $0.27$ at reconstruction level, compared with $0.57$ against
$0.41$ for the hard partons.
A comparison of the two classes of events in figure~\ref{fig:truth-ladder} shows that the information lost between the hard
partons and reconstruction level is considerably larger for events with three hard partons than for those with two.
For two hard partons the learned observables lose about $20\%$ of the information, whereas for three hard partons about half of
it is lost.
We attribute this larger loss to two effects.
The first is that almost half of the three-parton events are reconstructed with only two jets, either because the third parton is
generated below the jet threshold, which is possible since the generation cut of the merging, $x_{\mathrm{qcut}}=15\GeV$, lies
below $30\GeV$, or because the shower moves its jet below the threshold, so that the information carried by the third parton is
no longer available at reconstruction level.
The second is that, even when the third jet is reconstructed, the shower and the detector alter its momentum as well as those of
the other two jets.
Since the observables depend on the momenta of all jets, the deviation of each jet from its parton propagates into them, and the
third jet introduces an additional source of error compared to the two-jet case.
The loss of information from the change of the jet momenta is therefore expected to be larger for three-parton events than for
two-parton events, which is what we observe.
Lowering the threshold for the jets used by the learned observables from $30$ to $20\GeV$, so that softer third jets are also
included, recovers only a small part of this loss, from $0.500$ to $0.563$ for ParT at particle level and from $0.457$ to $0.505$ at
reconstruction level.
In the presence of pileup, discussed in section~\ref{sec:res-pileup}, even this small gain largely disappears
(table~\ref{tab:res-main}).

In three-parton events the ordering of the observables is the same at every level.
At reconstruction level, in these events, the learned observables retain $1.70$ times the information of the two-jet optimal
observables with the two leading-$\pT$ jets, the multiplicity-matched optimal observables $1.43$ times, and the two-jet optimal
observables with the grouped jets $1.13$ times.
For events with two hard partons the learned observables retain only $1.07$ times the information of the grouped two-jet optimal
observables. This is expected, since there the two-jet optimal observables differ from the exact ratios only by using the $gg\to Hgg$
matrix element for every parton flavor.
The multiplicity-matched optimal observables coincide with the two-jet ones in these events, except in the rare events with a third
reconstructed jet.

\begin{figure}[tbp]
  \centering
  \includegraphics[width=\linewidth]{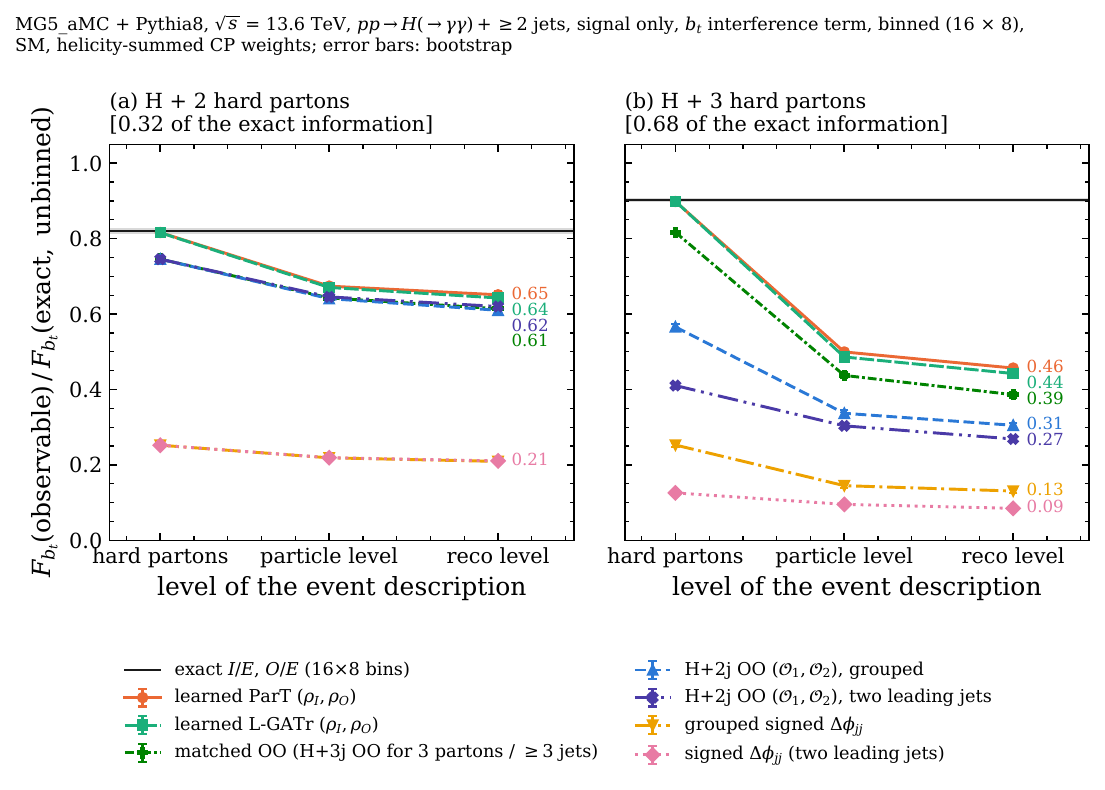}
  \caption{Fraction $F/F_{\mathrm{exact}}$ of the information on $\bt$ retained by each observable, for SM signal events generated
  with (a) two and (b) three hard partons. $F$ is the Fisher information of the binned observable, eq.~\eqref{eq:fisher}, and
  $F_{\mathrm{exact}}$ that of the unbinned exact ratios, eq.~\eqref{eq:fisher-exact}. The observables are evaluated on the hard
  partons, on particle-level jets and on reconstructed jets (``reco level''), and the learned observables are trained separately at
  each level. The pairs of optimal and learned observables are binned in $16\times8$ bins as in the fit, and the signed angles in 16
  bins. The multiplicity-matched optimal observables use the three-parton matrix element for events with three hard partons, or with
  at least three jets at particle or reconstruction level. The horizontal line and gray band show the exact ratios binned in
  $16\times8$ bins and their uncertainty. Error bars come from 100 bootstrap replicas of the events. The panel titles give the share of
  the exact information carried by each class among all selected events ($0.284$ and $0.716$ within the ladder). In (a) the two signed
  angles nearly coincide, as do the two two-jet and the multiplicity-matched optimal observables.}
  \label{fig:truth-ladder}
\end{figure}

\begin{table}[tbp]
  \centering
  \caption{Fraction $F/F_{\mathrm{exact}}$ of the information on $\bt$ retained by each observable, eqs.~\eqref{eq:fisher}
  and~\eqref{eq:fisher-exact}, on the hard partons (``hard''), on jets at particle level (``particle'') and on reconstructed jets
  (``reco''), for all events of the ladder, which have two or three hard partons, and for those with three hard partons (signal only, SM couplings, binned as in
  figure~\ref{fig:truth-ladder}). The binned exact ratios retain $0.888$ of the information for all events, $0.820$ for events with
  two and $0.902$ for events with three hard partons. The statistical uncertainties are between $0.002$ and $0.008$.}
  \label{tab:truth-ladder}
  \small
  \begin{tabular}{lccc@{\hspace{1.4em}}ccc}
   \toprule
    & \multicolumn{3}{c}{all events} & \multicolumn{3}{c}{three hard partons} \\
   Observable & hard & particle & reco & hard & particle & reco \\
   \midrule
   signed $\dphijj$                     & 0.161 & 0.128 & 0.118 & 0.126 & 0.096 & 0.085 \\
   grouped $\dphigrp$                   & 0.251 & 0.166 & 0.153 & 0.253 & 0.145 & 0.131 \\
   two-jet $(\OOone,\OOtwo)$, grouped jets & 0.611 & 0.409 & 0.375 & 0.566 & 0.337 & 0.305 \\
   two-jet $(\OOone,\OOtwo)$, two leading-$\pT$ jets & 0.502 & 0.393 & 0.358 & 0.411 & 0.304 & 0.269 \\
   matched $(\OOone,\OOtwo)$             & 0.802 & 0.495 & 0.449 & 0.817 & 0.437 & 0.386 \\
   learned ParT $(\rhoI,\rhoO)$          & 0.884 & 0.552 & 0.514 & 0.899 & 0.500 & 0.457 \\
   learned L-GATr $(\rhoI,\rhoO)$        & 0.884 & 0.539 & 0.498 & 0.899 & 0.486 & 0.442 \\
   \bottomrule
  \end{tabular}
\end{table}

\subsection{Limitations of the analytic observables}\label{sec:truth-flavour}\label{sec:truth-gain}

Two effects limit the multiplicity-matched optimal observables relative to the learned ones.
First, the analytic observables treat every parton as a gluon.
To study this, we use all selected events at the level of the hard partons, binned in 16 bins per class rather than in the
$16\times8$ bins and on the event subset of the ladder. In this binning the multiplicity-matched optimal observables retain $0.76$ of
the information of all three-parton events, compared with $0.82$ in table~\ref{tab:truth-ladder}.
Three-parton events with at least one quark carry four fifths of the information of the three-parton events.
For these events the multiplicity-matched optimal observables retain $0.77$ of the information, compared with $0.88$ for the binned
exact ratios, whereas for all-gluon events they are exact.
Closing this gap would require the matrix elements of the quark-initiated $H+3$ parton processes, which we leave for future work
(section~\ref{sec:outlook}).

Second, the multiplicity-matched optimal observables use at most three jets.
At reconstruction level, the learned observables improve on the grouped two-jet optimal observables, the baseline of this comparison,
for every number of jets, and the improvement grows with the number of jets. In this signal-only comparison, $\sigma(\bt)$ is reduced to
$0.93$, $0.82$ and $0.73$ of its value for events with two, three, and four or more jets.
With three reconstructed jets, the multiplicity-matched optimal observables recover about 70\% of the difference between the grouped
two-jet and the learned observables, but only 40\% with four or more jets, where they ignore the jets beyond the third.
A fourth jet, present in 10\% of the selected events, which carry 12\% of the exact information, is used only by the learned
observables.
Since the signal is merged only up to three hard partons, a fourth jet comes from the parton shower alone and carries information on
$\bt$ only through its correlation with the three hard partons; a full study requires a sample merged with up to four hard
partons (section~\ref{sec:outlook}).

More than half of the information on $\bt$ in events with a third hard parton is thus not accessible to the two-jet optimal
observables evaluated with the two leading-$\pT$ jets.
After the parton shower, hadronization and detector response, the learned observables still retain $1.70$ times the information of
these observables in three-parton events, and $1.44$ times when all selected events are included.
The analytic multiplicity-matched optimal observables recover a large part of this gain, and the remainder comes mainly from
quark-initiated events and from events with more than three jets.

\section{Results}\label{sec:results}

\subsection{Expected constraints}\label{sec:res-expected}

Table~\ref{tab:res-main} and figure~\ref{fig:res-hierarchy} give the expected 68\% CL half-widths $\sigma(\bt)$ on $\bt$, with $\at$
profiled, for all observables in the same three categories of the S/B transformer.
As in section~\ref{sec:truth}, we take as the reference the two-jet optimal observables evaluated with the two leading-$\pT$ jets,
which appear to correspond to the choice of CMS (section~\ref{sec:observables}).
At $300\ifb$, with statistical uncertainties only, the interval shrinks as the observable uses more of the event.
It is $0.345$ for the grouped signed angle and $0.254$ for the reference, $0.234$ when the jets are grouped, $0.220$ for the
multiplicity-matched optimal observables, and $0.204$ and $0.206$ for the learned observables of ParT and L-GATr.
Relative to the reference, grouping the jets reduces $\sigma(\bt)$ by 8\%, the multiplicity-matched optimal observables by 14\%, and
the learned observables by 20\%.
The reference would need $1.18$, $1.34$ and $1.56$ times the integrated luminosity to reach the same sensitivity.
The systematic uncertainties widen the intervals by 3\% for the optimal and learned observables and by 5\% for the grouped angle;
with them, the reference reaches $0.263$ and ParT $0.210$.
At $3000\ifb$ we find the same ordering and a similar gain, $0.081$ against $0.065$.
Appendix~\ref{sec:interp} translates these intervals into other parametrizations, such as the Wilson coefficient of the CP-odd
Higgs-gluon operator and the CP-mixing angle of the top Yukawa coupling, and compares them with current measurements and indirect
constraints.

To test whether the gain depends on a particular training or a particular set of events, we repeat the comparison in each of the five
subsets of the five-fold scheme of section~\ref{sec:ml-training}, evaluating each subset with networks and category boundaries
obtained from the other four.
Comparing ParT with the grouped two-jet optimal observables on the same events in each subset, ParT gives a $13.7\%$ smaller $\sigma(\bt)$ (ratio $0.863\pm0.004$, mean and standard error over the five subsets), and $0.881\pm0.003$ in the S/B BDT
categories. The reduction is many times the uncertainty of the comparison.
Tuning the category boundaries on independent subsets or on all events changes this ratio by only 1\% ($0.863$ against $0.871$).

Lowering the threshold for the jets used by the observables from $30$ to $20\GeV$, with the event selection unchanged, brings the
multiplicity-matched optimal observables to $\sigma(\bt)=0.210$ and the networks to $0.197$ (ParT) and $0.199$ (L-GATr) without
pileup.
With pileup, in the S/B BDT categories, the gain shrinks to 1\% ($0.306\to0.303$ for the multiplicity-matched optimal observables and
$0.283\to0.280$ for ParT), and we therefore keep $30\GeV$ as the nominal threshold.

\begin{table}[tbp]
  \centering
  \caption{Expected 68\% CL half-width $\sigma(\bt)$ of the interval on $\bt$ for SM Asimov data, with $\at$ profiled, in the window
  $120<\myy<130\GeV$. The first five columns use the categories of the S/B transformer and give $\sigma(\bt)$ with statistical
  uncertainties only (``stat.''), with the systematic uncertainties of table~\ref{tab:st-nuisances} (``+syst.''), and with pileup and
  the emulation of the jet-vertex tagger (JVT), with statistical uncertainties only (``PU'', section~\ref{sec:res-pileup}). The last two
  columns use the categories of the S/B BDT, which are nominal in the presence of pileup, and their ``stat.'' values are obtained from
  the fits without pileup of the pileup study. The multiplicity-matched optimal observables use the three-parton matrix element for
  events with a third jet above $30\GeV$ (nominal) or above $20\GeV$, and the learned observables use the jets above $30\GeV$. The
  networks that also use the jets above $20\GeV$ reach $0.197$ (ParT) and $0.199$ (L-GATr) without pileup. The
  $3000\ifb$ results assume templates without
  statistical uncertainty from the simulation, and with the simulated samples available here they would widen by about 40\%, for ParT
  from $0.064$ to $0.089$ (section~\ref{sec:st-mcstat}). The rows are ordered from the least to the most constraining observable at
  $300\ifb$ with statistical uncertainties only.}
  \label{tab:res-main}
  \small
  \begin{fitwidth}
  \begin{tabular}{lccccccc}
   \toprule
    & \multicolumn{3}{c}{$300\ifb$} & \multicolumn{2}{c}{$3000\ifb$} & \multicolumn{2}{c}{S/B BDT categories, $300\ifb$} \\
   Observable & stat. & +syst. & PU & stat. & +syst. & stat. & PU \\
   \midrule
   grouped signed $\dphigrp$             & 0.345 & 0.362 & 0.474 & 0.109 & 0.111 & 0.337 & 0.420 \\
   two-jet $(\OOone,\OOtwo)$, two leading-$\pT$ jets & 0.254 & 0.263 & 0.352 & 0.079 & 0.081 & 0.254 & 0.337 \\
   two-jet $(\OOone,\OOtwo)$, grouped jets & 0.234 & 0.241 & 0.337 & 0.073 & 0.075 & 0.235 & 0.320 \\
   matched $(\OOone,\OOtwo)$             & 0.220 & 0.227 & 0.313 & 0.069 & 0.071 & 0.224 & 0.306 \\
   matched, $\pT^{j_3}>20\GeV$           & 0.210 & 0.216 & 0.310 & 0.066 & 0.067 & 0.213 & 0.303 \\
   learned L-GATr $(\rhoI,\rhoO)$        & 0.206 & 0.212 & 0.289 & 0.064 & 0.066 & 0.211 & 0.287 \\
   learned ParT $(\rhoI,\rhoO)$          & 0.204 & 0.210 & 0.289 & 0.064 & 0.065 & 0.209 & 0.283 \\
   \bottomrule
  \end{tabular}
  \end{fitwidth}
\end{table}

\begin{figure}[tbp]
  \centering
  \includegraphics[width=\linewidth]{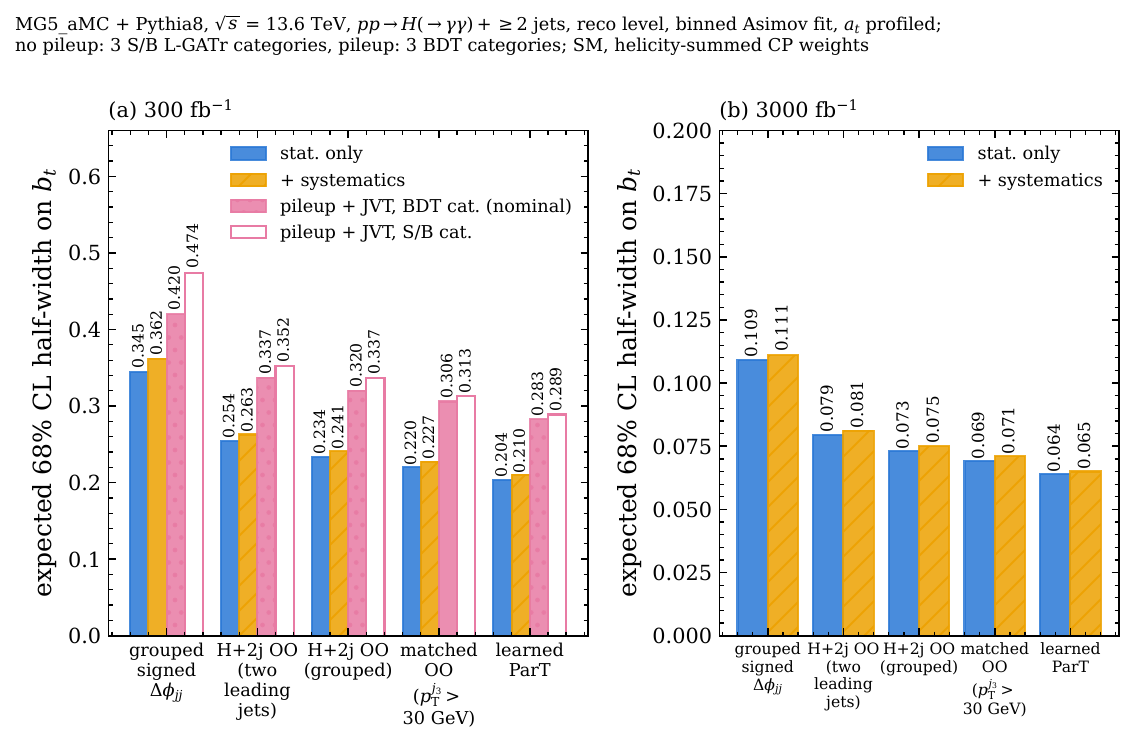}
  \caption{Expected 68\% CL half-width $\sigma(\bt)$ for SM Asimov data with $\at$ profiled, at (a) $300\ifb$ and (b) $3000\ifb$,
  in the S/B-transformer categories with $120<\myy<130\GeV$. The bars give $\sigma(\bt)$ with statistical uncertainties only, with
  systematic uncertainties added, and at $300\ifb$ with pileup and JVT. The filled pileup bars use the S/B BDT categories, which are
  nominal with pileup, and the open ones the S/B-transformer categories. ``OO'' stands for optimal observables and ``H+2j OO'' for the
  two-jet ones, built with the two leading-$\pT$ jets or with the grouped jets as indicated in parentheses. L-GATr, within 1\% of ParT,
  and the variants with jets above $20\GeV$ are given in the text and in table~\ref{tab:res-main}.}
  \label{fig:res-hierarchy}
\end{figure}

Figure~\ref{fig:res-contours} shows the expected two-dimensional 68\% and 95\% CL regions in the $(\bt,\at)$ plane.
Since the total rate is CP-even, it constrains only the combination $\at^2+2.16\,\bt^2$ of eq.~\eqref{eq:ring}, where $2.16$ is the
ratio of the CP-odd to the SM yield of the selected events (table~\ref{tab:yields}).
All points on the ellipse $\at^2+2.16\,\bt^2=1$ therefore predict the SM rate, including a pure CP-odd coupling,
$(\at,\bt)=(0,\pm0.68)$, and only the shapes of the CP-sensitive observables can distinguish them.
At $300\ifb$ the two-parameter 95\% CL regions of the multiplicity-matched optimal observables and of the learned observables are
closed, while those of the grouped angle and of the two-jet optimal observables of the two leading-$\pT$ jets extend along the
ellipse to $\at=0$.
Tested on its own, with $\bt$ profiled (a one-parameter test), a pure CP-odd coupling is disfavored by $-2\Delta\ln\mathcal{L}=5.7$
with the two-jet optimal observables of the two leading-$\pT$ jets, $7.6$ with the multiplicity-matched and $8.5$ with the ParT
observables, and by only $3.6$ with the grouped angle.
It is therefore excluded at 95\% CL by every observable except the grouped angle, and by the two-jet optimal observables of the two
leading-$\pT$ jets at $2.4\sigma$.
The learned and multiplicity-matched optimal observables also distinguish a CP-odd coupling from a reduced CP-even one better than the
two-jet optimal observables, so their 95\% regions along the ellipse are shorter.
The regions of the learned observables are smaller than those of the two-jet optimal observables with the two leading-$\pT$ jets by
about $1.25$ in their extent along $\bt$, consistent with the one-dimensional intervals with $\at$ profiled.
The intervals are slightly asymmetric, for example $[-0.206,+0.201]$ for ParT, because of statistical fluctuations of the simulated
templates, mainly of the continuum, which are covered by the treatment of section~\ref{sec:st-mcstat}.

\begin{figure}[tbp]
  \centering
  \includegraphics[width=\linewidth]{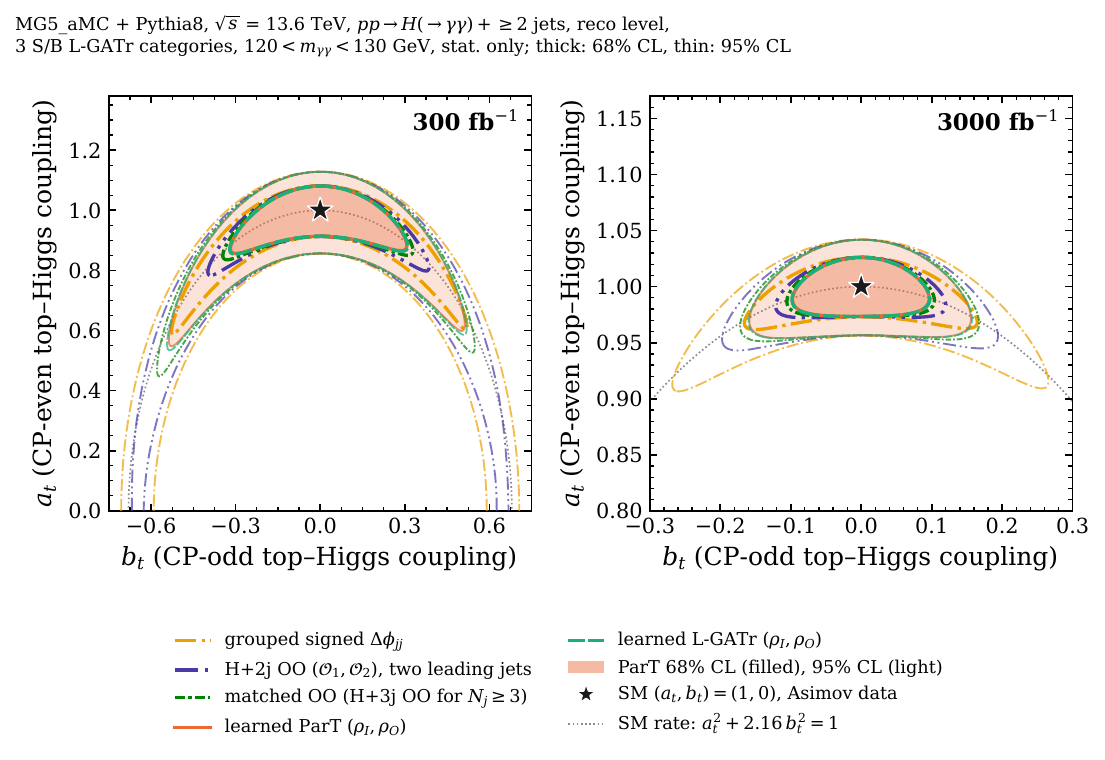}
  \caption{Expected two-parameter 68\% (thick) and 95\% (thin) CL regions, $-2\Delta\ln\mathcal{L}=2.30$ and $5.99$, in the
  $(\bt,\at)$ plane for SM Asimov data at $300\ifb$ (left) and $3000\ifb$ (right), with statistical uncertainties only, in the
  S/B-transformer categories with $120<\myy<130\GeV$. The regions are shown for the grouped signed angle, the two-jet optimal
  observables of the two leading-$\pT$ jets, the multiplicity-matched optimal observables and the learned observables of ParT (filled)
  and L-GATr. The dotted line is the ellipse $\at^2+2.16\,\bt^2=1$ of eq.~\eqref{eq:ring}, on which the SM rate is reproduced.
  ``H+2j OO'' and ``H+3j OO'' denote the two- and three-jet optimal observables and $N_j$ the number of jets.}
  \label{fig:res-contours}
\end{figure}

\paragraph{Gain by jet multiplicity.}\label{sec:res-njet}
To see where the gain arises in the full analysis with backgrounds, we repeat the comparison separately for events with $N_j=2$ and
with $N_j\geq3$, where $N_j$ is the number of reconstructed jets, with the categories and the binning determined within each subset
(figure~\ref{fig:res-njet}).
Relative to the grouped two-jet optimal observables, the baseline of this comparison, the learned observables reduce $\sigma(\bt)$ by
4\% in events with two jets ($0.288$ against $0.300$) but by 22\% in events with at least three jets ($0.277$ against $0.358$), which
make up a third of the signal.
The gain therefore comes from the additional jets, as for the signal alone in section~\ref{sec:truth}.
In events with two jets the multiplicity-matched optimal observables coincide with the two-jet ones.
In events with at least three jets they reach $0.329$ and recover about 36\% of the gain of the learned observables over the grouped
two-jet optimal observables, a smaller fraction than in section~\ref{sec:truth-gain}, where the comparison is made for the signal
alone and in terms of information.
These values are obtained from the Fisher information with backgrounds and agree with the profile-likelihood intervals of
table~\ref{tab:res-main} to within 2\%, and the same holds in the S/B BDT categories.

\begin{figure}[tbp]
  \centering
  \includegraphics[width=\linewidth]{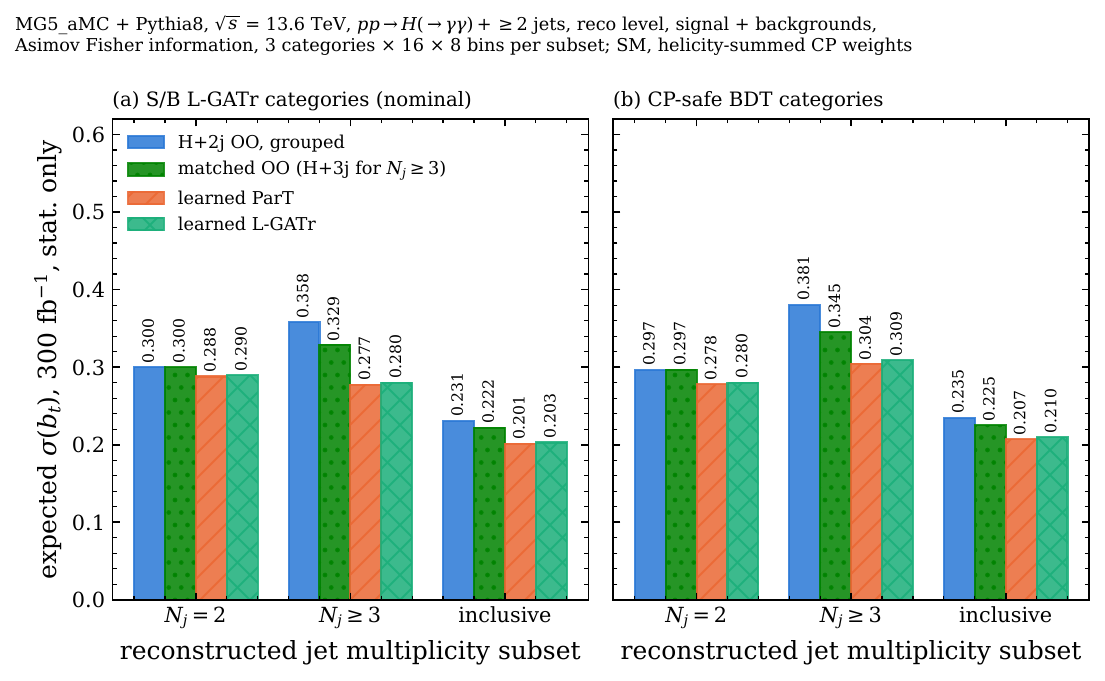}
  \caption{Expected $\sigma(\bt)$ at $300\ifb$, with statistical uncertainties only, from the Fisher information of signal and
  backgrounds for SM Asimov data, for events with two reconstructed jets ($N_j=2$), with at least three ($N_j\geq3$), and for all
  events. The three categories and the $16\times8$ bins are determined within each subset, using (a) the S/B transformer (labeled
  ``S/B L-GATr'' in the figure) and (b) the S/B BDT (labeled ``CP-safe BDT''). Shown are the grouped two-jet optimal observables
  (``H+2j OO''), the multiplicity-matched optimal observables (``H+3j'') and the learned observables of ParT and L-GATr.}
  \label{fig:res-njet}
\end{figure}

\subsection{Stability of the results}\label{sec:res-robustness}

We now ask whether the gain of the learned observables persists under less idealized conditions: with systematic uncertainties, with
a more conservative view of the NLO corrections to the azimuthal modulation, and with pileup.

\paragraph{Systematic uncertainties.}\label{sec:syst}
The 3\% widening by the systematic uncertainties of table~\ref{tab:st-nuisances} (section~\ref{sec:st-syst}) is the same for every
optimal and learned observable, so it leaves the gain of the learned observables essentially unchanged.
Two variations not included in the table change the intervals by at most 1\%, for the grouped angle only: an uncertainty of
$\pm30\%$ instead of $\pm20\%$ on the migration of gluon-fusion events into $N_j\geq3$, and a reduction of the gluon-fusion events with
$\Delta R_{jj}<1$ by the factor $9.0/13.4$, which corrects the excess of close jet pairs at LO.
Not covered are the modeling of the parton shower and hadronization and a merging-scale variation beyond the SM rate
(section~\ref{sec:sim}). Since the gain comes from events with three or more jets, these are the most important missing uncertainties.

\paragraph{NLO reduction of the azimuthal modulation.}
The normalization of the interference template sets the conversion of the observed asymmetry into $\bt$. The NLO-modulation nuisance
of section~\ref{sec:st-syst}, which scales this template, is therefore the largest systematic uncertainty (impact $0.042$ for ParT), and
the SM Asimov data constrain it only weakly, through the CP-even modulation.
Scaling the interference with it is conservative: in the NLO samples generated at the CP points (appendix~\ref{app:merging-nlo}) the
$\cos2\dphijj$ term is somewhat smaller than at LO ($0.89\pm0.10$), but the NLO/LO ratios of the interference moments in reconstructed
events lie above one, $1.11\pm0.10$ for $\sin2\dphijj$, $1.10\pm0.06$ for ParT and $1.08\pm0.06$ for L-GATr.
Applying the nuisance to the modulation of $E$ and $O$ only, as these samples suggest, gives half-widths of $0.206$ (ParT) and $0.237$
(grouped two-jet optimal observables) with all systematic uncertainties, instead of the nominal $0.210$ and $0.241$ quoted throughout
this paper, whereas a reduction of the interference fixed at the full factor $0.74$ would widen all intervals by about 30\%  (appendix~\ref{sec:int-limits} gives the corresponding bound on $\alpha_t$).

\paragraph{Pileup.}\label{sec:res-pileup}
For the pileup study we overlay on a subset of the simulated events a Poisson-distributed number of additional proton-proton
interactions with $\langle\mu\rangle=55$, as in Run~3, drawn at random from a library of simulated minimum-bias events.
As in ATLAS, the pileup contribution is then subtracted: charged particles from pileup vertices are removed and jets are
corrected by the median pileup density times the jet area. We suppress pileup jets with an emulation of the ATLAS jet-vertex
taggers~\cite{det:jvt,det:fjvt}, and add the diphoton events with zero or one hard parton that pass the two-jet selection through
pileup jets.
The networks, classifiers and binning are kept from the training without pileup, which is conservative, and the S/B BDT categories
are used, which with pileup give intervals 1--5\% narrower than those of the S/B transformer for the optimal and learned
observables and 11\% narrower for the grouped angle (section~\ref{sec:st-classifiers}).
Pileup widens the $300\ifb$ intervals by a factor $1.25$ for the grouped angle, $1.33$ for the two-jet optimal observables of the two
leading-$\pT$ jets and $1.36$ for the other optimal and learned observables.
Since the classifier is trained without pileup, pileup moves events into the most signal-like category and lowers its S/B from
$0.21$ to $0.09$.
The grouped angle widens less, in part because more of its information lies in this category, which loses the least.
The learned observables keep their advantage, with $\sigma(\bt)=0.283$ for ParT against $0.337$ for the reference.
The $3000\ifb$ results do not include the pileup of about 200 interactions expected at the HL-LHC.

\section{Conclusions and outlook}\label{sec:conclusions}\label{sec:outlook}

In gluon-fusion Higgs boson production, the interference between the CP-even and CP-odd Higgs-gluon couplings, which is sensitive to
the sign of $\bt$, becomes observable only through the kinematics of the accompanying jets.
Current measurements build their CP-sensitive observables from two jets, and the experimental practice, which we have taken as our
reference, appears to be to evaluate the two-jet optimal observables with the two leading-$\pT$ jets.
In this paper we have asked how much information on $\bt$ the additional jets carry, in particular a third jet, and how much of it can
be recovered in a realistic $H\to\gamma\gamma$ analysis at $13.6\TeV$.
We have compared three strategies: grouping the jets on either side of the Higgs boson, an analytic three-jet optimal observable
constructed from the $gg\to Hggg$ matrix element, and learned observables from two transformer networks, ParT and L-GATr, trained by
weighted regression on the matrix-element weights. For each of them we have tracked the information on $\bt$ from the hard partons
to the reconstructed events.

Events generated with three hard partons make up $56\%$ of the selected signal but carry $68\%$ of the information on $\bt$.
With three partons the interference is no longer described by a single Levi-Civita contraction but by five, and the two-jet optimal
observables evaluated with the two leading-$\pT$ jets, which are sensitive to only one of them, retain only $41\%$ of the information
of these events at the level of the hard partons, compared with $90\%$ for the exact ratios in the same binning.
Grouping the jets raises this fraction to $57\%$, the three-jet optimal observable to $82\%$, and the learned observables reach the
binned exact ratios.
Most of this information is lost in the parton shower, which alters the momenta of the jets, so that they no longer correspond
directly to the momenta of the hard partons, even when the third parton is reconstructed as a jet. Almost half of the three-parton
events are reconstructed with only two jets, but this accounts for only about 5 of the 44 percentage points lost between the hard
partons and reconstruction level.
Even after the detector simulation, however, the learned observables retain $1.70$ times the information of the reference in
three-parton events, and the multiplicity-matched optimal observables $1.43$ times.

The full analysis includes the continuum background, the systematic uncertainties, and signal-versus-background (S/B) categories
defined with a transformer classifier that is insensitive to parity.
In it, the expected 68\% CL half-width on $\bt$ at $300\ifb$ is $0.263$ for the two-jet optimal observables with the two
leading-$\pT$ jets, $0.241$ when the jets are grouped, $0.227$ for the multiplicity-matched optimal observables and $0.210$ for the
learned observables.
Reaching the sensitivity of the learned observables with the two-jet optimal observables on the two leading-$\pT$ jets would require
about 56\% more integrated luminosity. The gain is the same at $3000\ifb$, where the half-width decreases from $0.081$ to $0.065$.
It is stable across the five subsets of the simulated events and comes mostly from the events with three or more reconstructed jets.
The multiplicity-matched optimal observables, which require no training, achieve about two thirds of the reduction of the half-width
from the reference to the learned observables. Without pileup, they achieve almost 90\% of it if the third jet may be as soft as
$20\GeV$.
The two networks perform equally well: with our training samples, the Lorentz-equivariant L-GATr offers no advantage over ParT.
Pileup widens the intervals of the optimal and learned observables by 33--36\% without changing their ordering.
If the effective Higgs-gluon coupling arises from the top-quark loop, the expected sensitivity corresponds to a CP-mixing angle of
the top Yukawa coupling $|\alpha_t|<37^\circ$ at 95\% CL with $300\ifb$, about as strong as today's direct $t\bar tH$
measurements, and $8.2^\circ$ with $3000\ifb$ (appendix~\ref{sec:interp}).

These results rely on several approximations.
The multi-jet information is described at leading order, with matrix elements for up to three hard partons, so that a fourth jet is
produced only by the parton shower and carries information on $\bt$ only through its correlation with the three hard partons.
NLO predictions exist for SM Higgs boson production with three jets~\cite{Greiner:2015jha}, whereas our CP-odd and interference terms
for three-parton events are at LO.
The reduction of the azimuthal modulation at NLO is treated conservatively, by allowing it to reduce the interference as well,
although NLO samples generated at CP-odd and mixed couplings show no such reduction; all intervals quoted with systematic
uncertainties include this conservative treatment.
Our intervals do not include an uncertainty in the modeling of the parton shower and hadronization. This is the most important
uncertainty not covered here, since the gain arises in events whose jet multiplicity is set by the parton shower and the jet threshold.
They also do not include the full top-quark mass dependence of three-parton events, the pileup conditions of the HL-LHC, or the
statistical uncertainty of the simulated templates at $3000\ifb$, which would require about ten times more simulated continuum than
is available here.

\paragraph{Outlook.}
Several extensions follow directly from this study.
On the analytic side, including the quark-initiated $H+3$ parton matrix elements would raise the information retained in three-parton
events with quarks from $0.77$ to $0.88$ at the level of the hard partons (section~\ref{sec:truth-flavour}). The one-loop
$gg\to Hggg$ matrix element with the full top-quark mass dependence, and a signal sample merged with up to four hard partons, would
extend the study to the finite top-quark mass and to a fourth jet.
Further open studies include a lower jet threshold combined
with a rejection of soft pileup jets, the dependence on the merging scale, classifiers and networks trained with pileup, and the
separation of the gain into the use of the additional jets and an implicit correction for the parton shower and detector response.

For experimental analyses, these results show that the CP information of the Higgs-gluon coupling is not confined to the two
leading-$\pT$ jets but is also carried by the additional jets of the event, and that a substantial part of it can be recovered from
the reconstructed event.
Since the learned observables are calibrated, exactly odd or even under parity, and enter a standard binned template fit with the same
treatment of systematic uncertainties as the analytic optimal observables, they can be included in existing analyses without changes
to the statistical procedure.
Replacing the two-jet optimal observables evaluated with the two leading-$\pT$ jets by the learned observables would then provide the
same sensitivity to $\bt$ with about two thirds of the integrated luminosity.

\section*{Data availability}
The analysis code is available at \url{https://github.com/prakafle/hyy-jets-cp}. The simulated event samples, the trained
networks and the templates used in this work can be made available by the authors upon reasonable request.

\acknowledgments
The work of K.M., M.Y. and K.X. was supported in part by the U.S. National Science Foundation under awards PHY-2310497 and PHY-2609606.
The work of P.K., N.P. and J.H. was supported in part by the U.S. National Science Foundation under awards PHY-2309950 and
PHY-2608087. The work of K.X. was also supported by the National Natural Science Foundation of China and by the State Key Laboratory of Dark Matter Physics.
M.Y.\ gratefully acknowledges financial support for this publication by the Fulbright U.S. Student Program, which is sponsored by the U.S. Department of State and the Chilean Fulbright Commission.
This work was supported in part through computational resources and services provided by the Institute for Cyber-Enabled Research
at Michigan State University.
The authors used Claude (Anthropic) to assist with the writing of the manuscript and with setting up the numerical simulations;
all results and text were checked by the authors, who take full responsibility for the content of this article.

\appendix
\section{Effective couplings of the signal model}\label{app:signal}\label{app:signal-coupl}

This appendix discusses the coefficients of the effective Lagrangian of eq.~\eqref{eq:leff}, on which the signal model of
sections~\ref{sec:theory} and~\ref{sec:sim-signal} is built, and how they relate to the couplings extracted by the fit. The CP-even coefficient follows from the low-energy theorem, which relates the coupling of a zero-momentum Higgs boson to the heavy-quark contribution to the QCD $\beta$ function~\cite{th:kniehl_spira}, in the same way as the theorem of ref.~\cite{th:svvz} relates its coupling to photons to the heavy-particle contributions to the QED $\beta$ function.
The CP-odd coefficient, $\alpha_s/(8\pi v)=\tfrac32\times\alpha_s/(12\pi v)$, follows from the axial anomaly~\cite{th:adler,th:bell_jackiw}, as derived for a pseudoscalar Higgs boson in refs.~\cite{th:kniehl_spira,th:spira_hgg}.
Neither coefficient depends on $m_t$ at leading power.
Beyond leading order the CP-even coefficient receives a QCD correction, a factor $1+\tfrac{11}{4}\,\alpha_s/\pi$ in the amplitude~\cite{th:dawson,th:djouadi_spira_zerwas}, while the CP-odd one is not renormalized (Adler--Bardeen theorem~\cite{th:adler_bardeen,th:spira_hgg}).
The factor $3/2$ between the two coefficients, and with it the ratio $\sigma(0,1)/\sigma(1,0)=9/4$ of $gg\to H$, is therefore a leading-order relation.
The fit determines $\at$ and $\bt$ as the coefficients of the effective vertex~\eqref{eq:leff}, and the higher-order correction of the signal model is common to the three terms of the cross section; at points with a CP-odd coupling it is checked with NLO samples generated there (appendix~\ref{app:merging-nlo}).
The different corrections to the two Wilson coefficients matter when $(\at,\bt)$ are translated into top Yukawa couplings, the coefficients $\kappa_t$ and $\tilde\kappa_t$ of the top-quark Yukawa
interaction~(\ref{eq:yukawa}). At NLO the matching gives $\tilde\kappa_t/\kappa_t=(1+\tfrac{11}{4}\,\alpha_s/\pi)\,\bt/\at\approx1.10\,\bt/\at$ for $\alpha_s(m_H)\approx0.11$, which loosens the bounds on the mixing angle $\alpha_t$ of appendix~\ref{sec:interp} by about 10\% in $\tan\alpha_t$ (for example from $37^\circ$ to $40^\circ$).

At $(\at,\bt)=(1,1)$ and $|\Delta y_{jj}|>2$, the phase of the modulation of the signed $\dphijj$ is $0.985$ at particle and $0.987$
at reconstruction level, the eikonal value $\arctan(3/2)=0.983$. At parton level it is smaller ($0.84$, rising to $0.95$ for
$|\Delta y_{jj}|>4$), while in $gg\to Hgg$ without a third parton it reaches the eikonal value ($0.98$ for $|\Delta y_{jj}|>3$).

\section{NLO reweighting and the azimuthal modulation}\label{app:mlm}

This appendix describes the NLO reference sample of the reweighting of section~\ref{sec:sim-nlo} and the reduction of the azimuthal
modulation at NLO, which enters the fit as a nuisance parameter.
Unless stated otherwise, all samples are SM, at particle level, in the fiducial region with two photons and at least two jets, with the $H\to\gamma\gamma$ branching ratio included in the cross sections.

\subsection{The NLO reference sample}\label{app:mlm-checks}

The FxFx sample of ggF $H+0,1,2$\,jets at NLO uses the Higgs Characterization model~\cite{th:hc_framework,Demartin:2014fia}
in the heavy-top limit and the NNPDF3.0 NLO parton distributions~\cite{Ball:2014uwa}. It is generated with \mgfive{} 3.5.12, with
the FxFx fixes of version 3.6.4, and showered with \pythia{}. It contains $2.0\times10^6$ events with a cross section of $35.31\pb$ (without the branching ratio), $31\%$ of them
with negative weights. In the fiducial region with two jets its effective number of events is 8170, which limits the precision of every
comparison made with it.
The classifier for $r(x)$ has no azimuthal input, nor any of the variables from which the azimuthal angles could be reconstructed.

\paragraph{Generator settings.} The LO signal uses the default \mgfive{} scales for $k_T$-MLM merging (the scale of the core process
after $k_T$ clustering, and $\alpha_s$ at each emission evaluated at its clustering scale), partons with $|\eta|<4.7$, $x_{\mathrm{qcut}}=15\GeV$
and $Q_{\mathrm{cut}}=22.5\GeV$. The $\gamma\gamma$ and $\gamma$+jets backgrounds are merged in the same way, with photons above $25\GeV$,
$|\eta|<2.5$, $\Delta R_{\gamma\gamma},\Delta R_{\gamma j}>0.2$, $100<\myy<165\GeV$ (diphoton samples only) and Frixione
isolation~\cite{sim:frixione} ($R_0=0.4$, $n=\epsilon=1$). The electroweak $\gamma\gamma jj$ sample has partons with $\pT>20\GeV$,
$|\eta|<5$ and $\Delta R_{jj}>0.4$, VBF Higgs production partons above $15\GeV$ with $|\eta|<5$ and no merging, and $VH+0,1$\,jets uses
$k_T$-MLM. The FxFx sample ($pp\to H+0,1,2$\,jets at NLO, five-flavor scheme, NNPDF3.0 NLO) uses the FxFx scales~\cite{sim:fxfx}, partons
with $\pT>10\GeV$ and the merging scale $\mu_Q=20\GeV$, and is showered with \pythia{} 8.317 (A14) with the global-recoil settings
recommended for FxFx. The classifier for $r(x)$ is an \textsc{XGBoost}~\cite{bkgm_xgboost} model (300 trees of depth 4, learning rate
$0.05$, logistic loss that accepts negative weights) with five-fold cross-fitting and $r\in[0.1,10]$; its eleven inputs are the number of
jets, the $\pT$ of the three leading jets, $|y_{j_1}|$, $|y_{j_2}|$, $\Delta y_{jj}$, $|y_{\gamma\gamma}|$, $H_T$, the absolute
Zeppenfeld variable of the diphoton system and $y_{j_1}y_{j_2}$. The \sherpa{} 2.2.15 continuum merges $\gamma\gamma+0,1$\,jets at NLO
(OpenLoops~2~\cite{sim:openloops2}) and $\gamma\gamma+2,3$\,jets at LO with MEPS@NLO~\cite{sim:mepsnlo} at a merging scale of about
$10\GeV$, with NNPDF3.0 NNLO and the METS scales.

\subsection{Reduction of the azimuthal modulation at NLO}\label{app:merging}\label{app:merging-nlo}

We compare the SM signed $\dphijj$ of the LO merged sample with that of the FxFx NLO sample. Figure~\ref{fig:sim-dilution} gives the coefficient $c$ in three regions, for the nominal LO sample and for a second LO sample with only $H+2,3$ partons, each before and after the NLO reweighting.
The reweighting, which has no azimuthal input, changes $c$ only slightly, and the NLO and LO modulations differ:
at NLO the modulation is smaller by a factor $0.74\pm0.07$ in the fiducial region, $0.85\pm0.09$ for $|\Delta y_{jj}|>2$ and $0.76\pm0.10$ when $m_{jj}>400\GeV$ is also required.
The $H+2,3$ parton sample, which lacks the zero- and one-parton events, gives a slightly smaller gap.

\begin{figure}[tbp]
  \centering
  \includegraphics[width=0.6\linewidth]{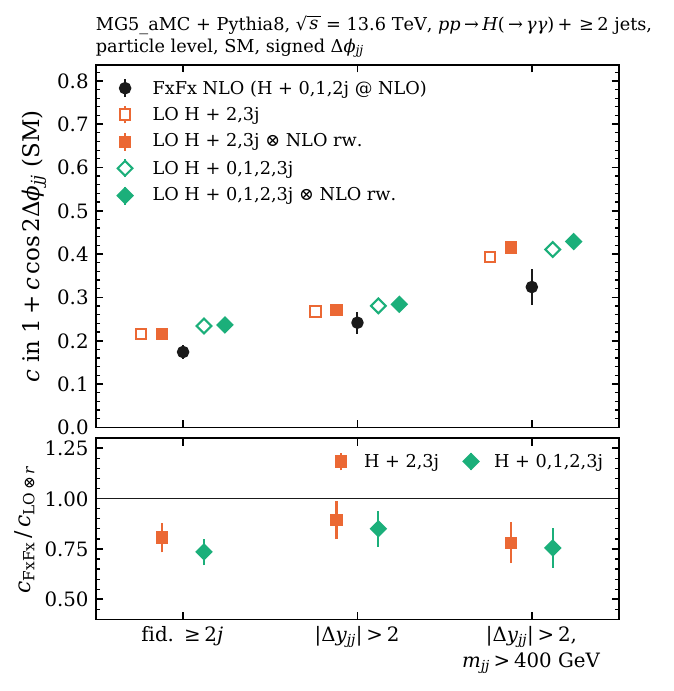}
  \caption{Coefficient $c$ of the SM modulation $1+c\cos2\dphijj$ at particle level, in the fiducial region with at least two jets,
  with $|\Delta y_{jj}|>2$, and with in addition $m_{jj}>400\GeV$. Points show the FxFx NLO $H+0,1,2$\,jet sample. Open and filled
  markers show the LO $H+2,3$ and $H+0,1,2,3$ parton MLM samples before and after the NLO reweighting. The lower panel gives the ratio
  of the NLO to the reweighted LO coefficient. Error bars are statistical.}
  \label{fig:sim-dilution}
\end{figure}

\paragraph{Nuisance parameter.} FxFx samples generated directly at CP points, compared with the LO model,
show the same reduction of the CP-even modulation but no reduction
of the interference. The $1\sigma$ variation of the NLO-modulation nuisance parameter of section~\ref{sec:st-syst} scales the
interference and the modulation of the CP-even and CP-odd shapes by the measured SM ratio, $0.74$, and the opposite variation is its
mirror image (a factor $1.26$). In practice, the variation flattens the binned signed-$\dphijj$ shapes of $E$ and $O$ towards their averages in five
$|\Delta y_{jj}|$ bins and scales $I$ uniformly.
For the interference this goes beyond the measured effect, since NLO is slightly larger than LO.
For the CP-even modulation it approximates an effect that is closer to an additive shift ($-0.041\pm0.007$ in $c$).
For the CP-odd modulation, which NLO does not reduce, the variation is conservative in size but not in direction.

\section{Detector simulation}\label{app:detector}

This appendix describes the \delphes{} card of section~\ref{sec:sim-detector}, that is, the changes to the stock ATLAS card, its validation against published ATLAS performance and the correction of the photon efficiency. The pileup card is summarized in section~\ref{sec:res-pileup}.

\subsection{Changes to the ATLAS card}\label{app:detector-card}

The card starts from the ATLAS card distributed with \delphes{} in \mgfive{} 3.5.5 and is tuned to public ATLAS performance results only.
It uses particle-flow jets with an extra smearing to the published ATLAS resolution~\cite{det:atlas_jes}, the photon efficiencies
and isolation of the ATLAS tight identification and FixedCutLoose isolation working
points~\cite{sim:atlas_photon_id_2015_16,sim:atlas_egamma_eff_run2,sim:atlas_egamma_perf_2015_17}, and the measured electromagnetic
energy resolution~\cite{sim:atlas_egamma_calib_2015_16,sim:atlas_egamma_calib_run2}. The jet
response is corrected with flat factors to within about 2\%, and the jet-shape variables used as quark/gluon discriminants by the
classifiers are computed from the particle-flow constituents of each jet. All changes follow public ATLAS references.

The resulting diphoton mass resolution, jet energy resolution and photon efficiencies are checked against published ATLAS performance.
The signed $\dphijj$ is reconstructed with an effective resolution of $0.042$\,rad and no bias, and the grouped angle with $0.091$\,rad.

\section{Construction of the analytic observables}\label{app:obs}

This appendix gives the construction of the two-jet optimal observables and its extension to three jets.

\subsection{The two-jet optimal observables}\label{app:h2j}

This subsection describes how the two-jet optimal observables $\OOone$ and $\OOtwo$ of section~\ref{sec:oo} are computed from
reconstructed events and compares the heavy-top amplitudes with the one-loop amplitudes with the full top mass.
The remaining approximation studies (parton flavor, grouping of the jets, information from hard partons to reconstructed events) and the
complete two-jet analysis in the categories of the CP-safe BDT, made with the per-helicity CP weights of the earlier analysis, are not shown.

\subsubsection{Definition and distributions}\label{app:h2j-eval}

The observables are evaluated with the construction of section~\ref{sec:obs-oo2}, from the $gg\to Hgg$ amplitudes of the effective
interaction~(\ref{eq:leff}), i.e.\ in the heavy-top limit in which the signal is generated.
As a cross-check we also evaluate them with the one-loop amplitudes of the top Yukawa coupling~(\ref{eq:yukawa}) with the full top mass,
computed with the one-loop integrals of the \textsc{Collier} library~\cite{Denner:2016kdg}.
Treating all partons as gluons is exact in the eikonal limit but not for quark-initiated subprocesses, so the observables are close to,
but not exactly, optimal.
The rest of this subsection deals with the transverse boost and the amplitudes; the cost of the parton-flavor
approximation and of the ignored third parton is quantified in section~\ref{sec:truth}.
None of these approximations breaks the parity property. Under $\phi\to-\phi$ of the input momenta, $\OOone\to-\OOone$ and
$\OOtwo\to\OOtwo$ exactly, because a genuine matrix element is evaluated on the reflected configuration. The approximate $\OOone$ is still a P-odd
observable with zero SM mean, so an asymmetry in it can only come from $\bt\neq0$ or from a P-odd detector or background
effect.
At particle level the
interference shifts $\OOone$ to positive or negative values depending on the sign of $\bt$, while the CP-odd point populates
large $\OOtwo$.

\subsubsection{Phase-space point and amplitudes}\label{app:oo}

Unless stated otherwise, information is the signal-only Fisher information $F$ on $\bt$ at the SM of section~\ref{sec:fisher}
(binned, or over events for the unbinned bound).
The reference is the exact per-event ratio $w_I/w_E$ of the generator, which knows the true parton flavors and all hard partons.

\paragraph{Phase-space point and incoming momenta.}\label{app:oo-psp}

The matrix element needs a $gg\to Hgg$ phase-space point. We take the Higgs boson momentum from the diphoton system, keeping its
mass, and the two outgoing gluons from the two grouped jets (section~\ref{sec:obs-oo2}), made massless ($E=|\vec p|$). The reconstructed $Hjj$ system
has a transverse momentum from additional jets, initial-state radiation and resolution. Relative to its invariant mass
$\sqrt{\hat s}$, the median is $0.048$ for the grouped jets ($0.127$ at the 90th percentile), $0.069$ for the two leading jets and
$0.035$ for the hard-scatter partons.

We boost the final state transversely to the frame in which its total transverse momentum vanishes and then fix the incoming
gluons by four-momentum conservation,
\begin{equation}
  k_{a,b}=\frac{E\pm p_z^{\mathrm{tot}}}{2}\,(1,0,0,\pm1),
  \label{eq:th-kin}
\end{equation}
where $(E,0,0,p_z^{\mathrm{tot}})$ is the total final-state momentum after the boost. Momentum and invariant mass are then conserved to double
precision ($|k_a+k_b-\sum p|/\sqrt{\hat s}\leq2.3\times10^{-15}$).
Since $\OOone$ and $\OOtwo$ are ratios of Lorentz-invariant squared amplitudes, the result is unchanged under a boost of the
whole event or a transverse boost of the final state. It does
change, at the level of $4\times10^{-3}$ of the spread of $\OOone$, under a longitudinal boost of the final state alone, because the
incoming energies then change, as they physically should.

The construction used in earlier work applies eq.~(\ref{eq:th-kin}) to the unboosted final state. It violates transverse-momentum
conservation by the transverse momentum of the $Hjj$ system and keeps only about half of the binned information of the boosted
construction ($0.180$ against $0.337$ of the exact information for reconstructed grouped jets). In the full analysis, made with the
per-helicity weights (the ratio does not depend on them), the boost narrows the 68\% interval of $\OOone\times\OOtwo$ by a factor
$1.26$ at $300\ifb$ and $1.30$ at $3000\ifb$.

\subsection{Three-jet observables at reconstruction level}\label{app:obs-psp}\label{app:obs-chain}\label{app:obs-flavour}\label{app:obs-jets}

The three-jet observables extend the two-jet construction of appendix~\ref{app:h2j} to the three leading jets.
Events with exactly two jets keep the grouped two-jet observables.

\paragraph{Phase-space point and matrix element.}
For every event with at least three jets ($\pT>30\GeV$, $|y|<4.4$) the $gg\to Hggg$ phase-space point is built from the reconstructed
diphoton four-momentum, which keeps its mass, and the three $\pT$-leading jets, made massless ($E=|\vec p|$).
As for two jets, the final state is boosted transversely to zero total transverse momentum, the jets are kept exactly massless after
the boost, and the incoming gluons follow from eq.~(\ref{eq:th-kin}), $k_{a,b}=\tfrac12(E\pm p_z^{\mathrm{tot}})(1,0,0,\pm1)$.
Energy and momentum are then conserved to $4\times10^{-16}$.
Since $\OOone$ and $\OOtwo$ are ratios of Lorentz-invariant squared amplitudes, the arguments of appendix~\ref{app:oo-psp} for the
transverse boost apply unchanged.
The observables of eq.~(\ref{eq:mj-oo3}) are the ratios of the color- and helicity-summed heavy-top $gg\to Hggg$ squared amplitudes
of the CP-even, interference and CP-odd terms, from our own implementation of the heavy-top amplitudes, written for this paper
following refs.~\cite{DelDuca:2004wt,Dixon:2004za,Badger:2004ty}.
They were evaluated for every reconstructed event with at least three jets, in every simulated sample and every systematic variation.
All values are finite, and $\OOone^2\le4\OOtwo$ holds for every event, as it must for the positive semi-definite $2\times2$ matrix of
the weights.
Under the reflection $P$ of section~\ref{sec:dphi-corr}, $\OOone(Px)=-\OOone(x)$ and $\OOtwo(Px)=\OOtwo(x)$ hold bit by bit, because
a genuine matrix element is evaluated on the reflected configuration.
Only the three leading jets enter: we did not implement the $H+4$ parton matrix element, so a fourth jet is ignored.

\paragraph{All-gluon approximation.}
Every parton is treated as a gluon, as for the two-jet observables, since the quark-initiated $H+3$ parton amplitudes are not
implemented.
The approximation keeps the parity properties above, so it can cost sensitivity but cannot create a CP-odd signal. Its cost in information
is given in section~\ref{sec:truth-flavour}.

\paragraph{Assignment of jets to partons.}
There is no unique way to map the reconstructed jets onto the partons of a matrix element.
We compared thirteen assignment rules and scans of their thresholds on the signal selected in the
BDT categories, through the signal-only information of the $(\OOone,\OOtwo)$ histogram ($16\times8$ bins), with events with two and
with at least three jets binned separately.
The numbers below are the expected $\sigma(\bt)$ relative to that of the grouped two-jet observable (smaller is better).
The two-jet observable on the two leading jets gives $1.037$, and exclusive generalized-$k_t$ clustering ($p=0.5$) of all jets into
two pseudo-jets gives $0.979$.
Using the three-jet matrix element on the three leading jets whenever a third jet exists gives $0.941$, a $5.9\%$ smaller
$\sigma(\bt)$.
Clustering four or more jets into three instead, with the $k_t$, generalized-$k_t$ or Cambridge/Aachen algorithm, gains a further
$0.2\%$ only.
Every rule that falls back to the two-jet observable for soft third jets loses sensitivity ($0.982$ for a fallback below
$\pT^{j_3}=50\GeV$), while lowering the third-jet threshold helps steadily ($0.911$ at $20\GeV$).
An event with a third jet should therefore always be treated as a three-jet event.
For comparison, the learned observables reach $0.874$ (ParT) and $0.884$ (L-GATr), and the exact per-event ratios of the hard process
$0.629$.
In the full fit with systematic uncertainties, the two-jet observables on the two leading jets give a $9\%$ wider interval than on the
grouped jets. This is more than the $3.7\%$ found here, because the full fit adds the backgrounds and uses the S/B categories, in which
the two observables separate signal from background differently.

\section{Training of the learned observables}\label{app:nets}

This appendix gives the design of the regression and the settings of the networks. We checked the training curves, the calibration
and the reweighting closure of every network.

\subsection{Design of the regression and of the networks}\label{app:nets-design}

\paragraph{Learned observables as conditional expectations.} Let $z$ be the hard parton-level event, with the weights $w_E(z)$,
$w_I(z)$, $w_O(z)$ of eq.~(\ref{eq:morph}), and $x$ the event the analysis observes, at particle or reconstruction level, related to $z$ by
the shower and detector transfer density $p(x|z)$. The density of $x$ is $p(x|\at,\bt)\propto\at^2W_E(x)+\at\bt\,W_I(x)+\bt^2W_O(x)$
with $W_c(x)=\int\!dz\,p(x|z)\,w_c(z)\,p_{\mathrm{gen}}(z)$ and $p_{\mathrm{gen}}$ the generation density. The ratios
$W_I/W_E=\mathbb{E}[w_I|x]/\mathbb{E}[w_E|x]$ and $W_O/W_E=\mathbb{E}[w_O|x]/\mathbb{E}[w_E|x]$ are the learned observables $\rhoI$ and
$\rhoO$ of eq.~(\ref{eq:mj-rho}), and the likelihood ratio to the SM is
\begin{equation}
  \frac{p(x|\at,\bt)}{p(x|1,0)} = \frac{\sigma(1,0)}{\sigma(\at,\bt)}\left[\at^2+\at\bt\,\rhoI(x)+\bt^2\rhoO(x)\right] .
  \label{eq:mj-lr}
\end{equation}
The pair $(\rhoI,\rhoO)$ is therefore sufficient for $(\at,\bt)$ at the level of $x$, and $\rhoI$ is the score at the SM.
For regression weights $v(x,z)>0$ and targets $t=n/v$, the functional $\mathcal{R}[f]=\mathbb{E}_{p_{\mathrm{gen}}}[v\,(f(x)-t)^2]$ is
minimized point by point by $f^*(x)=\mathbb{E}[n|x]/\mathbb{E}[v|x]$~\cite{Brehmer:2018kdj,Brehmer:2018eca}, since, conditioned on
$x$, the loss is a parabola in $f$ with its minimum there. The targets are joint $(x,z)$ quantities, but the minimizer is the ratio of
conditional expectations at the level of $x$, so that the regression learns eq.~(\ref{eq:mj-rho}) without unfolding and without
evaluating $p(x|z)$.

\paragraph{Event weight of the regression.} Regressing a ratio gives the ratio of conditional expectations only if each event is
weighted by its denominator.
Regressing $w_O/w_E$ with weight $w_E$ is therefore correct but badly conditioned, because $w_O/w_E$ is unbounded and becomes very
large where the SM amplitude almost vanishes.
The regression therefore uses the CP-democratic weight $v=w_E+w_O$ and the bounded targets of eq.~(\ref{eq:mj-targets}), with the
loss $\mathrm{MSE}_E+4\,\mathrm{MSE}_I$, the sum of the weighted mean squared errors of the two outputs, which does not change the
minimizer. We impose the parity symmetry exactly rather than through a data augmentation, which would impose it only
approximately, and we do not train a classifier between the SM and a fixed CP-odd point, whose output would mix $w_E$, $w_I$ and $w_O$
with weights set by the chosen point.

\paragraph{Particle Transformer.} Seven features per token ($\ln(1+\pT)$, $y$, $p_z/E$, $\sin\phi$, $\cos\phi$, $\ln(1+E)$,
$\ln(1+m)$) are embedded into $d=64$ dimensions together with a learned vector per token type. Three particle-attention blocks with
four heads are followed by one class-attention block, and a fully connected head network ($64\to64\to32\to2$) gives the two outputs
$h_E$, $h_I$ of eq.~(\ref{eq:mj-param}). The feed-forward layers of each block have width 128, and the model has about $1.5\times10^5$
parameters.
Eight features of each pair of tokens ($\Delta y$, $\sin\Delta\phi$, $\cos\Delta\phi$, $\ln\Delta R$, $\ln k_{\mathrm T}$, $\ln\zeta$,
$\ln(1+m^2)$ and a validity flag), standardized without shifting the two antisymmetric ones, are mapped by a small network with two
hidden layers to one attention bias $U_h(i,j)$ per head, which enters all three particle-attention blocks.
Because $\sin\Delta\phi$ is odd under $P$, this bias is the route by which ParT receives the CP-odd
azimuthal information.

\paragraph{L-GATr.} We use six blocks with sixteen multivector and thirty-two scalar channels and four heads. The scalar and
pseudoscalar parts of eight output multivectors and sixteen scalar channels are averaged over the tokens and passed to the same head
and parametrization as for ParT (figure~\ref{fig:obs-arch}, bottom), with a head network $32\to64\to32\to2$. The four-momenta are
divided by one fixed energy scale, the median token energy of the training sample, and the 32 averaged features by fixed constants,
which keeps the equivariance, and the token type enters as scalar channels. The model has about $2.5\times10^5$ parameters. The final
symmetrization of eq.~(\ref{eq:mj-param}) is kept, so parity is exact even if the group setting were lost.

\paragraph{Training and cross-fitting.}
The three training folds hold about $3.1\times10^5$ selected signal events, all of which are used by L-GATr and a random $3\times10^5$
by ParT. ParT is trained for up to 12 epochs with learning rate $10^{-3}$ and L-GATr
for up to 30 epochs with $5\times10^{-4}$, both with a batch size of 1024 and early stopping on a validation set. Both use the AdamW
optimizer with weight decay $10^{-5}$, a linear warm-up over the first 3\% of the steps followed by cosine decay, and gradient
clipping at norm 1. The training weights are the LO generation weights, and the NLO reweighting of eq.~(\ref{eq:nlo}) is applied only
to the templates, so the networks learn the LO conditional expectations, which costs optimality but cannot bias the templates.
The signal jobs
are split into five folds. For each fold an ensemble of five networks with different seeds is trained on three of the other folds
and validated on the fourth, and the observable of an event is the ensemble average of the networks that never saw its job.
Every signal event thus enters the templates, and the template statistics are those of the full sample. The background samples are
scored with the average of all twenty-five networks. The averages are taken over $\rhoI$ and $\rhoO$, which keeps their parity
properties and the bound $\rhoI^2\le4\rhoO$, since the region it defines is convex. Per event, the ensemble of one fold differs from
the average of all twenty-five networks by about 6\% of the spread of $\rhoI$ (fold-to-fold correlation $0.995$), so the signal
templates describe the observable applied to the background to that precision. Training a ParT member takes 1--3 minutes on a GPU.

\section{The diphoton continuum model}\label{app:bkg}

This appendix describes how the \mgfive{} continuum is reweighted to the \sherpa{} MEPS@NLO prediction and normalized to data, and how
the photon-jet fakes are modeled. The background model is the same for every observable.

\subsection{Reweighting to MEPS@NLO}\label{app:bkg-rw}

The target is a \sherpa{} 2.2~\cite{sim:sherpa22} MEPS@NLO~\cite{sim:mepsnlo} prediction for $\gamma\gamma+0,1$\,jets at NLO and
$\gamma\gamma+2,3$\,jets at LO, with OpenLoops~2~\cite{sim:openloops2} one-loop amplitudes, generated at $13.6\TeV$ and passed through the
same \delphes{} card.
Its effective sample size after the selection is about 90 times smaller than that of the \mgfive{} sample, too small to provide the
templates directly.
We therefore reweight every \mgfive{} event with the ratio $r(x)=\sigma_{\textsc{Sherpa}}(x)/\sigma_{\textsc{MG5}}(x)$ at reconstruction
level, learned out of fold by a gradient-boosted classifier~\cite{bkgm_xgboost} with 20 kinematic inputs.
The azimuthal inputs are unsigned angles in $[0,\pi]$, so that $r$ is symmetric in the signed $\dphijj$ and cannot create a P-odd
asymmetry.

We compare the reweighted \mgfive{} sample with \sherpa{} on the
training inputs, on held-out kinematic variables and on the CP observables.
The reweighting closes within the \sherpa{} statistical precision everywhere in the bulk, including the held-out signed $\dphijj$
($\chi^2=5.7$ for 11 bins), $\OOtwo$ ($2.2/11$) and $\OOone$ ($7.7/11$).
The $\cos2\dphijj$ modulation of the continuum, $2\langle\cos2\dphijj\rangle$, is $0.502\pm0.020$ in \sherpa{}, $0.487$ in \mgfive{}
and $0.495$ after the reweighting, and for $|\Delta y_{jj}|>2$ it is $0.429\pm0.035$, $0.390$ and $0.425$.
In the pileup study each pileup event takes $r$ from its twin without pileup (section~\ref{sec:res-pileup}).

\subsection{Normalization to data}\label{app:bkg-norm}

No diphoton measurement at $13.6\TeV$ is public yet, and the $13.6\TeV$ $H\to\gamma\gamma$
results~\cite{CMS:2025hyy136,ATLAS:2023run3hyy} give only Higgs boson cross sections.
We therefore normalize the continuum to the ATLAS measurement of isolated-photon pair production at
$13\TeV$~\cite{bkgm_atlas_diphoton_13tev}, which provides $d\sigma/d\myy$ in a fiducial region with photons above $40$ and $30\GeV$,
$|\eta|<2.37$ outside $1.37$--$1.52$, particle-level isolation $E_T(\Delta R<0.2)<0.09\,\pT$, $\Delta R_{\gamma\gamma}>0.4$ and no jet
requirement.
Our \sherpa{} sample, generated in $90<\myy<175\GeV$, can be compared in the ten published bins within $93<\myy<166.5\GeV$.
It is scaled from $13.6$ to $13\TeV$ with $\sigma(13\TeV)/\sigma(13.6\TeV)=0.945\pm0.022$, obtained from LO \mgfive{}
$pp\to\gamma\gamma j$ in the same region, with the difference to the ratio for $pp\to\gamma\gamma$, $0.967$, as its uncertainty.
Integrated over $93<\myy<166.5\GeV$, our \sherpa{} prediction, $13.20\pm0.30\pb$, is $0.855$ of the measured $15.44\pm1.12\pb$.

Its $\myy$ shape agrees with the data, with a ratio flat to $\pm3\%$ ($\chi^2=20.7$ for 9 degrees of freedom, statistical
uncertainties only).
Our \sherpa{} setup does not include the loop-induced $gg\to\gamma\gamma$ box. With the LO box added, the prediction agrees with the
data within the uncertainty.
We do not generate the box with jets but multiply the whole $13.6\TeV$ continuum, including the fakes, by $1/0.855=1.169$.
This assumes that the ratio measured without a jet requirement holds with two jets and at $13.6\TeV$.
The resulting continuum is about $2.2$ times the LO \mgfive{} prediction.
Normalizing instead to \sherpa{} as it is, or to \sherpa{} plus the box, changes the expected widths on $\bt$ by $-7\%$ and $-2.5\%$
in a simplified, Fisher-information version of the fit of the two-jet analysis.
In the fit with systematic uncertainties the continuum normalization of each category is free and constrained by the $\myy$ sidebands
of that category (section~\ref{sec:st-syst}), so the fit absorbs the normalization, but the assumed continuum level still enters
the expected intervals through the Asimov data.
We study this dependence by scaling the continuum by a factor $k_C$
between 0.5 and 2, with the helicity-summed weights in the BDT categories; $k_C=1.2$ widens the interval by 8\%.

\subsection{Photon-jet fakes and other Higgs boson production}\label{app:bkg-fakes}\label{app:bkg-other}
Every jet within the photon acceptance of a $\gamma+2,3$\,jet event is promoted in turn to a photon candidate, and the fakes are
normalized to the fraction $\gamma j/\gamma\gamma=18.6/78.7$ measured by ATLAS for the inclusive diphoton
selection~\cite{th:atlas_yy_36}.
We assume that this fraction also holds with two jets and vary only its normalization ($\pm25\%$, table~\ref{tab:st-nuisances}), while
the continuum normalization of each category is free in the fit.
The remaining $2.7\%$ of that selection, events with two jets misidentified as photons, are not simulated, and the sideband-constrained
continuum normalization of each category absorbs them.
The Working Group cross sections for VBF and $VH$ would widen the expected intervals by about 1\% and are not applied.

\section{Details of the statistical method}\label{app:fitval}\label{sec:st-validation}

This appendix gives the implementation of the nuisance parameters and the choice of the binning. The closure and coverage tests are summarized in
section~\ref{sec:st-mcstat}.

\paragraph{Implementation of the nuisance parameters.} In the fit with systematic uncertainties of section~\ref{sec:st-syst},
shape variations that are available in one direction only are mirrored to obtain the opposite variation, normalization
uncertainties that are asymmetric have separate upward and downward values, and the change of a template by a shape variation is
limited to five times its nominal content in each bin.

\subsection{Binning and positivity of the templates}\label{app:fitval-binning}

The binning was chosen in a convergence scan in the BDT categories, with the grouped two-jet $(\OOone,\OOtwo)$ as the test
observable.
For a sequence of binnings, from $6\times3$ to $24\times8$, we computed the Fisher uncertainty on $\bt$ with the statistical noise
of the simulated interference template subtracted, and the 68\% and 95\% half-widths of the likelihood fit.
A binning is admissible if every populated bin of every category contains at least two simulated background events and one simulated
signal event, and if the two halves of the simulated sample (split by event number) give Fisher uncertainties within 10\% and 95\%
half-widths within 5\% of each other. Among the admissible binnings, the procedure takes the coarsest one whose 95\% half-width is
within 2\% of the best, $12\times8$.
We use the slightly finer $16\times8$ for every two-dimensional observable.
It passes the same criteria, its 68\% half-width is 1\% narrower ($0.234$ against $0.237$), and the finer binnings $16\times12$ and
$24\times8$ change it by about 1\%.
For the angles, 36 bins instead of 24 change the Fisher uncertainty by about 1\%.
The template statistics, the positivity of the signal model in every bin and the precision of the profile were checked.
All intervals quoted in this paper come from a continuous profile, with $\at$ minimized at every $\bt$ and the crossings
found by bisection.

\section{Interpretation in other parametrizations and comparison with existing constraints}\label{sec:interp}

In this appendix we translate the expected constraints of section~\ref{sec:results} into other parametrizations of the CP-odd
coupling, and compare them with the current ATLAS and CMS measurements in the $H\to\gamma\gamma$ channel and with indirect constraints on
the CP structure of the top Yukawa coupling.

\subsection{Parametrizations}\label{sec:int-maps-short}

If only the top quark runs in the loop, the coefficients of the Yukawa interaction~(\ref{eq:yukawa}), which we denote
$\kappa_t$ and $\tilde\kappa_t$ when they refer to the top Yukawa coupling, equal $\at$ and $\bt$ at leading order, as
we verified for two-parton events against a full-$m_t$ loop model, and the CP-mixing angle of the
top Yukawa coupling is $\tan\alpha_t=\bt/\at$. At NLO $\tilde\kappa_t/\kappa_t$ is about $1.10\,\bt/\at$ (appendix~\ref{app:signal-coupl}),
and the values of $\alpha_t$ below use the leading-order relation. With the gluon couplings normalized to the rate,
$\tan\alpha_g=3\bt/(2\at)$ and $\sin^2\alpha_g=f_{a3}^{ggH}$, the CP-odd fraction used by CMS~\cite{th:cms_tautau_anom}. The Warsaw-basis
coefficient of the operator $H^\dagger H\,\widetilde{G}^a_{\mu\nu}G^{a\,\mu\nu}$ is related to our coupling by
$\bt=+13.0\,\cHGt$ at $\Lambda=1\TeV$, calibrated with SMEFTsim~3.0.2~\cite{Brivio:2020onw} in the conventions of the ATLAS
measurement. 
The expected statistical-only half-widths of the learned ParT observables, $|\bt|<0.204$ ($0.411$) at 68\% (95\%) CL with $300\ifb$
(table~\ref{tab:res-main}), correspond to $|\cHGt|<0.016$ ($0.032$) at $\Lambda=1\TeV$, and with $3000\ifb$ ($|\bt|<0.064$, $0.126$) to
$|\cHGt|<0.0049$ ($0.0096$). These limits hold for any $\at$, since $\at$ (or $c_{HG}$) is profiled.

A mixing angle, in contrast, depends on $\at$ as well. At the edge of the $\bt$ interval the fit moves along the rate ellipse of eq.~\eqref{eq:ring} to $\at<1$
(to $\at\approx0.79$ at the 95\% CL edge with $300\ifb$), so that $\arctan\bt$ underestimates the allowed angle. We therefore scan the
angle itself, writing $(\at,\bt)=\kappa\,(\cos\alpha_t,\sin\alpha_t)$ and profiling the overall strength $\kappa$ together with all
nuisance parameters. This gives $|\alpha_t|<12.3^\circ$ ($29.0^\circ$) at 68\% (95\%) CL with $300\ifb$, and $3.7^\circ$ ($7.4^\circ$)
with $3000\ifb$, statistical uncertainties only. In the rate-normalized angle, which is a monotonic function of $\alpha_t$, this is
$|\alpha_g|<18.1^\circ$ ($39.7^\circ$) and $\sin^2\alpha_g<0.097$ ($0.41$) with $300\ifb$, and $5.5^\circ$ ($11.0^\circ$) and
$0.0092$ ($0.036$) with $3000\ifb$.

\subsection{Comparison with the current data}\label{sec:int-recast}

We recast the published ATLAS $139\ifb$ measurement~\cite{th:atlas_yy_139_diff} with our signal model at particle level.
Its four signed-$\dphijj$ bins see only the interference, so with
$\at$ profiled these data constrain only the rate and allow $|\bt|\lesssim0.7$--$0.8$ along the ellipse of eq.~\eqref{eq:ring}
(observed, 95\% CL), while with the quadratic term and the change of the total width we find $|\cHGt|<0.023$ observed and $0.026$
expected at 95\% CL, within 6\% of the ATLAS expected limit.

CMS has analyzed the $H\to\gamma\gamma$ final state with $138\ifb$ at $\sqrt{s}=13\TeV$, using discriminants built from the $H+2$
jet matrix elements, to measure the CP-odd fraction $f_{a3}^{ggH}$ of ggF production~\cite{th:cms_yy_anom}. Its expected 68\% CL
interval, $f_{a3}^{ggH}\in[-0.65,0.63]$, corresponds to $\tan\alpha_t=\bt/\at\in[-0.88,0.85]$ with the CMS conversion
($[-0.91,0.87]$ with our heavy-top factor $2.25$), and the expected 95\% CL interval covers the whole range. Since CMS constrains
$\bt/\at$, we compare in the same variable, with the angle scan described above. With $138\ifb$, our learned ParT observable gives a
68\% CL interval of $\tan\alpha_t$ with half-width $0.35$ at $13.6\TeV$ and $0.37$--$0.39$ when the cross sections are scaled to
$13\TeV$ (statistical uncertainties only). Our expected interval is thus 2.2--2.4 times narrower than the CMS one. We have checked that, with the CMS binning of the
discriminants and the matrix-element discriminant of the two leading jets, we obtain results similar to those of CMS. With pileup, in the
nominal BDT categories, the half-width becomes $0.58$--$0.63$ and the factor 1.4--1.5. At 95\% CL our projection for this data set
does not exclude a pure CP-odd coupling either, with $-2\Delta\ln\mathcal{L}=3.3$--$3.6$ at $\alpha_t=90^\circ$ without pileup.

This comparison is only indicative. Our intervals are Asimov projections with a fast detector simulation and without the experimental
systematic uncertainties of the CMS analysis. Our pileup ($\langle\mu\rangle=55$) is higher than in Run~2 (about 34), so the Run~2
value lies between our two numbers. Finally, the sign conventions of $f_{a3}^{ggH}$ and $\bt$ have not been matched, which matters
only for asymmetric intervals.

\subsection{Indirect constraints}\label{sec:int-limits}

If the effective $Hgg$ coupling comes only from the top loop, $\tan\alpha_t=\bt/\at$ is the CP-mixing angle of the top Yukawa
coupling, eq.~(\ref{eq:yukawa}), and our projection can be compared with other probes of it. With the angle scan of
appendix~\ref{sec:int-maps-short} and all systematic uncertainties, the learned ParT observables give $|\alpha_t|<37^\circ$ at 95\% CL with
$300\ifb$ and $|\alpha_t|<8.2^\circ$ with $3000\ifb$ ($29^\circ$ and $7.4^\circ$ with statistical uncertainties only). The two-jet
optimal observables give $68^\circ$ and $10.3^\circ$ with the two leading-$\pT$ jets and $47^\circ$ and $9.5^\circ$ with the grouped
jets. The systematic uncertainties widen the angle more than they widen $\bt$, because the normalization uncertainties weaken the
rate constraint along which the fit moves to large angles.

With the NLO relation of appendix~\ref{app:signal-coupl} the $37^\circ$ becomes $40^\circ$. Pileup, evaluated with statistical uncertainties only in the nominal BDT categories, widens the ParT bound to
$46^\circ$ with $300\ifb$ and $10.3^\circ$ with $3000\ifb$ (two leading jets: $77^\circ$ and $12.5^\circ$). A reduction of the interference
by the full NLO/LO factor $0.74$ of the SM modulation, which the NLO samples generated at CP points do not show (section~\ref{sec:syst}),
would give $40^\circ$ instead of $29^\circ$ with statistical uncertainties only. All these are asymptotic intervals: for $\bt$ from the learned observables, the
95\% intervals cover in 93\% of SM pseudo-experiments, and the interval calibrated with the pseudo-experiments is 1--2\% wider.
The $3000\ifb$ values require about ten times more simulated continuum than is available here
(section~\ref{sec:st-mcstat}) and do not include the pileup expected at the HL-LHC.

Table~\ref{tab:alphat-short} compares them with
representative published constraints, read from the papers and converted with the relations of appendix~\ref{sec:int-maps-short}. 

\begin{table}[tbp]
  \centering
  \caption[Constraints on a CP-odd top Yukawa or Higgs-gluon coupling]{Expected sensitivity of this study (learned observables,
  including systematic uncertainties, 95\% CL, angle scan with the overall strength profiled) compared with representative published constraints on a CP-odd top
  Yukawa or Higgs-gluon coupling, at 95\% CL unless stated, observed unless stated (expected in brackets where we quote it). The
  entry marked ``$\at=1$'' is obtained with the CP-even coupling fixed to its SM value, so that it constrains $\bt$ through the rate
  and does not distinguish a CP-odd from a CP-even coupling; with $\at$ profiled, the same data allow $|\bt|\lesssim0.7$--$0.8$
  (section~\ref{sec:int-recast}). In the other collider entries the CP-even coupling or the signal strengths are free. }
  \label{tab:alphat-short}
  \small
  \begin{fitwidth}
  \begin{tabular}{lll}
   \toprule
   Probe & As published & Our $\alpha_t$ or $\bt$ \\
   \midrule
   \emph{This study} (expected), $300\ifb$ / $3000\ifb$ & --- & $|\alpha_t|<37^\circ$ / $8.2^\circ$ \\
   \emph{This study} with pileup (stat.\ only), $300\ifb$ / $3000\ifb$ & --- & $|\alpha_t|<46^\circ$ / $10^\circ$ \\
   ATLAS ggF $\gamma\gamma$, $139\ifb$~\cite{th:atlas_yy_139_diff} & $\cHGt\in[-0.031,0.035]$ ($[-0.028,0.027]$) & $-22^\circ<\alpha_t<24^\circ$ ($-20^\circ<\alpha_t<19^\circ$), $\at=1$ \\
   CMS ggF$+2j$ and $t\bar{t}H$~\cite{th:cms_tautau_anom} & $f_{CP}^{Htt}\in[-0.07,0.51]$ & $-15^\circ<\alpha_t<46^\circ$ (CMS sign) \\
   ATLAS $t\bar{t}H$, $\gamma\gamma$, Run 2+3~\cite{ATLAS:2026ttHyyRun3} & $|\alpha|<38^\circ$ ($48^\circ$) & $|\alpha_t|<38^\circ$ \\
   Global rate fits~\cite{Bahl:2020wee} & $2\sigma$, top-loop dominance & $|\alpha_t|\lesssim22.5^\circ$--$27^\circ$ \\
   Electron EDM, ACME 2018, JILA~\cite{th:acme2018,th:jila_eedm} & rescaled (ours), 90\% CL & $|\tilde\kappa_t|\lesssim1.3\times10^{-3}$, $5\times10^{-4}$ \\
   \bottomrule
  \end{tabular}
  \end{fitwidth}
\end{table}

Briefly, the
ATLAS ggF $+2$\,jets limit on $\cHGt$ comes from the rate with $\at$ fixed, and the CMS shape analyses, with the signal strength
free, exclude only part of the $f_{a3}^{ggH}$ range at 95\% CL. The direct measurements in $t\bar{t}H$ and $tH$ exclude
$|\alpha_t|>38^\circ$ at 95\% CL~\cite{ATLAS:2026ttHyyRun3}. The VBF measurements of CP-odd $HVV$ couplings, through top-quark loops,
allow $|\bt|$ of order 10 to 100. Global fits of the Higgs boson rates give
$|\alpha_t|\lesssim22.5^\circ$--$27^\circ$, assuming that the loops are dominated by the top quark, and the EDM bounds are orders of
magnitude stronger. The electron-EDM bounds, however, assume a top-loop origin of $\bt$ and an SM electron Yukawa coupling and can be weakened by
cancellations, while a CP-odd $Hgg$ vertex from new colored states is constrained by the hadronic (neutron and diamagnetic) EDMs, with
large hadronic uncertainties~\cite{th:chien_edm}.

With $300\ifb$ and without pileup the diphoton channel alone would thus constrain $\alpha_t$ about as well as today's direct
$t\bar{t}H$ measurements, but less well than the global rate fits, which rely on assumptions beyond the top-loop dominance needed to
translate $\bt$ into $\alpha_t$. With pileup it would be weaker than the direct measurements. With $3000\ifb$ it would go well beyond
today's direct measurements and global fits, and combined with $t\bar{t}H$ measurements of similar precision~\cite{Cepeda:2019klc} it
could distinguish a CP-odd top Yukawa coupling from a new contribution to the effective $Hgg$ vertex. The constraints on $\bt$ and
$\cHGt$ themselves do not depend on this interpretation.

\bibliographystyle{JHEP}
\bibliography{refs,refs_theory,refs_sim,refs_interp,refs_ml}

\end{document}